\documentclass[twocolumn]{aastex631}

\newcommand{\kms}{\,\mathrm{km} \, \mathrm{s}^{-1}}

\newcommand{\Msolpc}{\, \mathrm{M}_{\sun} \, \mathrm{pc}^{-2}}

\newcommand{\alphavir}{\alpha_{\mathrm{vir}}}
\newcommand{\cs}{c_{\mathrm{s}}}

\newcommand{\Mcloud}{M_{\mathrm{cloud}}}
\newcommand{\Rcloud}{R_{\mathrm{cloud}}}
\newcommand{\Sigmacloud}{\Sigma_{\mathrm{cloud}}}
\newcommand{\Msun}{\mathrm{M}_{\sun}}

\usepackage[flushleft]{threeparttable}
\usepackage{booktabs}
\usepackage{multirow}
\usepackage{amsmath}

\begin{document}

\title{Bursts of star formation and radiation-driven outflows produce efficient LyC leakage from dense compact star clusters}

\author[0000-0002-0311-2206]{Shyam H. Menon}
\affiliation{Department of Physics and Astronomy, Rutgers University, 136 Frelinghuysen Road, Piscataway, NJ 08854, USA}
\affiliation{Center for Computational Astrophysics, Flatiron Institute, 162 5th Avenue, New York, NY 10010, USA}

\author[0000-0001-5817-5944]{Blakesley Burkhart}
\affiliation{Department of Physics and Astronomy, Rutgers University, 136 Frelinghuysen Road, Piscataway, NJ 08854, USA}
\affiliation{Center for Computational Astrophysics, Flatiron Institute, 162 5th Avenue, New York, NY 10010, USA}

\author[0000-0002-6748-6821]{Rachel S. Somerville}
\affiliation{Center for Computational Astrophysics, Flatiron Institute, 162 5th Avenue, New York, NY 10010, USA}

\author[0000-0003-2377-9574]{Todd A. Thompson}
\affiliation{Center for Cosmology and Astro-Particle Physics, Ohio State University, 140 W. 18th Ave, Columbus, OH 43210, USA}
\affiliation{Department of Physics, Ohio State University, 191 W. Woodruff Ave, Columbus, OH 43210, USA}
\affiliation{Department of Astronomy, Ohio State University, 140 W. 18th Ave, Columbus, OH 43210, USA}

\author[0000-0001-5065-9530]{Amiel Sternberg}
\affiliation{Center for Computational Astrophysics, Flatiron Institute, 162 5th Avenue, New York, NY 10010, USA}
\affiliation{ School of Physics and Astronomy, Tel Aviv University, Ramat Aviv 69978, Israel}
\affiliation{Max-Planck-Institut für extraterrestrische Physik (MPE), Giessenbachstr., 85748 Garching, Germany}



\begin{abstract}
The escape of LyC photons emitted by massive stars from the dense interstellar medium of galaxies is one of the most significant bottlenecks for cosmological reionization. The escape fraction shows significant scatter between galaxies, and anisotropic, spatial variation within them, motivating further study of the underlying physical factors responsible for these trends. We perform numerical radiation hydrodynamic simulations of idealized clouds with different gas surface densities (compactness) $\Sigma \sim 10^2$--$10^5 \, \Msolpc$, meant to emulate star cluster-forming clumps ranging from conditions typical of the local Universe to the high ISM-pressure conditions more frequently encountered at high redshift. Our results indicate that dense compact star clusters with $\Sigma \gtrsim 10^4 \, \Msolpc$ efficiently leak LyC photons, with cloud-scale luminosity-weighted average escape fractions  $\gtrsim 80\%$ as opposed to $\lesssim 10\%$ for $\Sigma \sim 100 \, \Msolpc$. This occurs due to higher star formation efficiencies and shorter dynamical timescales at higher $\Sigma$; the former results in higher intrinsic LyC emission, and the latter implies rapid evolution, with a burst of star formation followed by rapid gas dispersal, permitting high LyC escape well before the intrinsic LyC emission of stellar populations drop ($\sim 4 \, \mathrm{Myr}$). LyC escape in dense clouds is primarily facilitated by highly ionized outflows driven by radiation pressure on dust with velocities $ \sim 3$ times the cloud escape velocity. We also vary the (assumed) dust abundances ($Z_{\mathrm{d}}$) and find a very mild increase ($\sim 10 \%$) in the escape fraction for $\sim 100$ lower $Z_{\mathrm{d}}$. Our results suggest a scenario in which localized compact bursts of star formation in galaxies are disproportionately productive sites of LyC leakage. We briefly discuss possible observational evidence for our predictions and implications for cosmic reionization. 

\end{abstract}

\keywords{Stellar feedback (1602) --- High-redshift galaxies(734) --- Young star clusters(1833) --- Reionization(1383) -- Interstellar medium(847)}


\section{Introduction}

The Epoch of Reionization (EoR) marks the most prominent phase transition of the Universe, when the intergalactic medium (IGM) comprised of neutral Hydrogen (\ion{H}{1}) was (re-)ionized by the first stars, galaxies and quasars \citep{Loeb_2001,Gnedin_2022}. The timing of reionization has been well constrained through spectroscopic studies of high redshift (high-$z$) galaxies and quasars \citep[e.g.,][]{Mcgreer_2015}, absorption lines via the Lyman-$\alpha$ forest \citep{Fan_2006,Mcquinn_2016,Burkhart_2022}, and the integrated optical depth of Thomson scattering to the cosmic microwave background \citep{Planck_2020}, placing the conclusion of the EoR at $z\sim 6$ \citep{Robertson_2015,Bosman_2022}. However, there is still significant debate on the dominant underlying sources of reionization, with data from the recently launched James Webb Space Telescope proving fertile ground for testable predictions \citep{Robertson_2022}.  

Specifically, the relative role of the different sources of UV radiation -- i.e. low-mass and high-mass star-forming galaxies and the accretion disks of active galactic nuclei (AGN) -- is heavily debated. This is primarily quantified through the combination of the relative numbers (UV luminosity functions) of these sources, their intrinsic hydrogen-ionizing Lyman continuum (LyC) photon production efficiencies ($\zeta_{\mathrm{ion}}$), and their LyC escape fraction ($f_{\mathrm{esc},\mathrm{LyC}}$). While AGN have been identified as potential contributors to reionization due to the copious amounts of high-energy radiation they emit and their high escape fractions \citep{Madau_2015}, their inferred abundances at high-$z$ however, have proved difficult to reconcile with the required LyC photon budgets to conclude reionization by $z\sim 6$ \citep{Fan_2006,Qin_2017,Mitra_2018,Parsa_2018,Hassan_2018,Dayal_2024}\footnote{We note that it is possible that the higher number densities of AGN detected by JWST might change this picture -- see \citet{Madau_2024}.}. This suggests that star-forming galaxies are the dominant contributors to reionization \citep{Dayal_2018}. At high-$z$, low-mass galaxies are more numerous, and harbor weaker potential wells permitting stellar feedback to easily eject gas and leak LyC photons \citep[e.g.,][]{Paardekooper_2013,Wise_2014,Xu_2016,Finkelstein_2019,Bera_2023}; the feedback, however, would also maintain low star-formation rates and therefore potentially meager LyC photon budgets \citep{Kimm_2017}. On the other hand, massive galaxies have higher star formation rates, and thereby a larger flux of LyC photons \citep{Sharma_2016,Naidu_2020}; conversely, their large reservoirs of gas and dust, and steep potential wells have been argued as inhibiting factors for LyC escape. Some theoretical studies also find a relatively weak dependence of escape fraction on host galaxy mass \citep{Ma_2015} and/or an evolution of this dependence with redshift \citep{Yeh_2023}

While galaxy-level studies provide key statistical constraints, young massive stars produce LyC photons while embedded in dense clouds with scales of tens of parsecs.
Thus, the most prominent bottleneck to LyC escape is clearing out the dense gas surrounding young stellar populations 
\citep{Paardekooper_2015,Trebitsch_2017}. 
These scales are unresolved in studies probing indirect tracers of LyC escape at high-$z$, and even in (rare) low-redshift analog LyC leakers \citep{Leitet_2013,Leitherer_2016,Izotov_2016,Flury_2022b}. State-of-the-art EoR cosmological simulations only marginally resolve these scales at best \citep[e.g.,][]{Rosdahl_2022} and typically adopt subgrid recipes for LyC escape, with galaxy-averaged $f_{\mathrm{esc},\mathrm{LyC}}$ values highly sensitive to the adopted subgrid escape fraction \citep[e.g.,][]{Kostyuk_2023}. This is highly relevant because LyC escape in resolved galaxy populations and/or lensed galaxies have been found to be highly anisotropic and to show significant variation within the galaxy \citep[e.g.,][]{Rivera_2019,Choi_2020,Ramambason_2020}. This implies that a galaxy-averaged estimate along a given (projected) line of sight might not be representative of the full 3D LyC escape that takes this spatial variation into account -- a subtlety that numerical simulations have been increasingly able to identify with improved numerical resolution \citep{Yeh_2023,Choustikov_2024}. Another point to consider is that the timescales over which stellar populations emit LyC photons are short ($\lesssim 10 \, \mathrm{Myr}$), and the interaction of stellar feedback -- especially pre-supernova mechanisms such as radiation and winds -- with the ISM over these timescales are likely crucial in the context of the instantaneous value of $f_{\mathrm{esc},\mathrm{LyC}}$. These spatial and temporal variations likely contribute to the significant scatter found in identified correlations of $f_{\mathrm{esc},\mathrm{LyC}}$ with host galaxy properties and their (aperture-averaged) emission-line features -- such as the [\ion{O}{3}]/[\ion{O}{2}] equivalent width (EW) ratios and Ly$\alpha$ emission, among others \citep[e.g.][]{Flury_2022b}. All of these factors indicate that probing the physics of massive star (cluster) formation and feedback-regulated LyC escape on small spatial and temporal scales can provide insights relevant for identifying the underlying environmental properties that favor LyC escape. 

In this study we conduct idealized numerical simulations of turbulent self-gravitating clouds designed to emulate clumps/regions of star formation within a galaxy. We model a range of clouds with different initial conditions to infer the physical factors that determine the efficiency of LyC escape on cloud scales. Specifically, we probe a range of initial cloud gas surface densities $\Sigmacloud  \sim 10^2$--$10^5 \, \Msolpc$ to mimic ISM conditions typical of clumps ranging from weakly star-forming galaxies in the local Universe, to the high-pressure gas-rich conditions more frequently encountered in the high redshift Universe. We extend the upper limits of $\Sigmacloud$ probed in similarly designed numerical simulations in the literature \citep{Howard_2018,Kim_2019_escape,He_2020,Kimm_2022}, by $\sim$ 2 orders of magnitude motivated by recent findings from JWST revealing young, dense, compact super star clusters \citep{Pascale_2023,Pascale_2024,Mowla_2024,Adamo_2024} and extremely blue, UV-luminous, compact galaxies \citep{Finkelstein_2023a,Casey_2023,Mcleod_2023,Robertson_2023,Harikane_2023,Morishita_2023} in the EoR with stellar surface densities $ \gtrsim 10^4$--$10^5 \, \Msolpc$. We show below that these ``extreme" high density star formation environments are highly productive sites of LyC escape due to their high star formation efficiencies, short dynamical timescales, and radiatively driven outflows that facilitate the formation of density-bounded ionized channels.

The paper is organized as follows: In Section~\ref{sec:Methods} we describe the numerical methods used, the initial conditions of our clouds and the range of parameter space we explore.
In Section~\ref{sec:Results} we present the evolution of our clouds, compare the temporal trends of $f_{\mathrm{esc},\mathrm{LyC}}$ we obtain over our parameter space, and discuss the physics driving the escape of LyC photons. In Section~\ref{sec:discussion} we discuss some possible implications of our results, compare with observational findings that seem broadly consistent with our results, and identify the caveats with our findings imposed by our numerical setups. Finally, in Section~\ref{sec:summary} we conclude with a brief summary of our results.

\section{Methods}
\label{sec:Methods}

\subsection{Numerical Methods}
We solve the equations of self-gravitating radiation hydrodynamics for all our simulations. We use the \texttt{FLASH} magneto-hydrodynamics code \citep{Fryxell_2000,Dubey_2008} for our simulations, with the explicit Godunov method in the split, five-wave HLL5R (approximate) Riemann solver \citep{Waagan_2011} for the hydrodynamics. The Poisson equation for the self-gravity is solved using a multi-grid algorithm implemented in \texttt{FLASH} \citep{Ricker_2008}. Sink particles are used to follow the evolution of gas at unresolved scales, the formation of which is triggered when gas properties satisfy a series of conditions to test for collapse and star formation \citep{Federrath_2010_Sinks}. Specifically, sinks form when the local gas density exceeds $\rho_{\rm sink} = \pi \cs^2/G \lambda_{\mathrm{J}}^2$ -- where $\cs$ is the thermal sound speed, and $lambda_{\mathrm{J}}$ the Jeans length -- and the gas is gravitationally bound and collapsing over a zone of radius 2.5 times the minimum cell resolution (i.e. the accretion radius $r_{\mathrm{sink}}$). Once created, a sink particle can accrete gas, if computational cells within $r_{\mathrm{sink}}$ exceed $\rho_{\mathrm{sink}}$, and satisfy the additional conditions above. Gravitational interactions of sink particles with gas and other sinks are included, and a second-order leapfrog integrator is used for sink particle dynamics \citep{Federrath_2010_Sinks,FederrathBanerjeeSeifriedClarkKlessen2011}. To model the radiative transfer and the associated energy and momentum transfer to gas, we use the Variable Eddington Tensor-closed Transport on Adaptive Meshes method (\texttt{VETTAM}; \citealt{Menon_2022}). \texttt{VETTAM} uses the VET closure obtained with a time-independent ray-trace solution \citep{Buntemeyer_2016} to close the moment equations and an implicit global temporal update for the radiation moment equations for each band, accounting for the coupling between the bands due to UV radiation being reprocessed to the IR by dust. We solve for the (non-equilibrium) ionisation state of hydrogen set by photoionization-recombination balance in an operator-split fashion with subcycling\footnote{The subcycle timestep is limited such that the change in ionization state in each substep is restricted to less than 10\%.}, similar to the approach presented in \citet{Kim_2017}; the photoionization rate is obtained from the solution for the LyC radiation field from \texttt{VETTAM}, and we use the temperature-dependent case B recombination coefficient reported in \citet{Hui_1997}. The radiative output from sink particles is included as a smoothed source term in the moment equations, where we have tested convergence in the smoothing parameters \citep[see][]{Menon_2022b}. We used a fixed uniform grid resolution of $256^3$ for all our simulations. Although \texttt{VETTAM} fully supports AMR, we chose a fixed modest resolution for simplicity and to enable a more controlled experiment across our parameter space. We demonstrate convergence in the SFE (within $\lesssim 5 \%$) at these resolutions in similar numerical setups in \citet{Menon_2023}. 

\subsection{Simulation setup}

Our simulation setup is very similar to that described in \citet{Menon_2023,Menon_2024} with some additional modifications due to the inclusion of photoionization-recombination balance. We model isolated clouds (primarily) parameterized by their mass surface density \(\Sigmacloud = \Mcloud / (\pi \Rcloud^2)\) \(\Mcloud\) where $\Mcloud$ is the mass of the cloud and $\Rcloud$ its radius; this results in a gas density of \(\rho_{\mathrm{cloud}} = \Mcloud / [(4/3) \pi \Rcloud^3]\). We achieve different target values of $\Sigmacloud = 10^2$--$10^5 \, \Msolpc$ by fixing $\Mcloud = 10^5 \, \Msun$ and varying $\Rcloud$. The lower and upper ranges of $\Sigmacloud$ span the conditions of GMC/star-forming clumps typical of (weakly) star-forming galaxies in the local Universe and high-redshift/starburst conditions respectively. The simulated clouds are placed in an ambient medium with a density \(\rho = \rho_{\mathrm{cloud}} / 100\) within a computational domain of size \(L = 4 \Rcloud\). The fluid is initialized with turbulent velocities following a power spectrum \(E(k) \propto k^{-2}\), with a natural mix of solenoidal and compressive modes for \(k / (2\pi/L) \in \left[2, 64\right]\), generated using methods described in \citet{Federrath_2010} and implemented as in \citet{FederrathEtAl2022ascl}. The cloud's velocity dispersion \(\sigma_v\) is scaled so that the clouds are marginally bound, i.e., \(\alphavir = 2\), where \(\alphavir\) is defined as
\[
    \alphavir = \frac{2E_{\mathrm{kin}}}{E_{\mathrm{grav}}} = \frac{5 \Rcloud \sigma_v^2}{3G \Mcloud}\,,
\]
with \(E_{\mathrm{kin}} = (1/2) \Mcloud \sigma_v^2\) and \(E_{\mathrm{grav}} = (3/5) G \Mcloud^2 / \Rcloud\). Diode boundary conditions are applied to the gas quantities, allowing gas to escape through the boundaries but preventing any inflow. We set the initial gas temperature to be equal to $10 \, \mathrm{K}$ everywhere in the domain implying a turbulent Mach number $\mathcal{M} = \sigma_v/c_{\mathrm{s}}$, where $c_{\mathrm{s}} = \sqrt{k_{\mathrm{B}}T/m_{\mathrm{H}}} = 0.28 \, \kms$ is the sound speed of the gas. We also assume that the fluid is a strongly coupled dust-gas mixture with a dust-to-gas ratio parameterized by the abundance relative to the solar neighborhood value $Z_{\mathrm{d}}^{'} = Z_{\mathrm{d}}/Z_{\mathrm{d},\sun}$, where $Z_{\mathrm{d},\sun} = 0.0081$ \citep{Weingartner_2001b}.

We model radiation feedback in three wavelength bands: Lyman continuum (LyC; $h\nu \geq 13.6 \, \mathrm{eV}$), Far-UV (FUV; $6.6 \leq h\nu < 13.6 \, \mathrm{eV}$), and the infrared (IR). Photons in the LyC band can be absorbed by, and photoionize, atomic hydrogen in our simulations. Dust absorbs both LyC and FUV photons, heating the dust, which re-emits in the IR; dust can also absorb and re-emit the IR photons, i.e. effectively IR scattering, which is treated by our IR band. The sources of LyC and FUV photons are sink particles that form in our simulations that physically represent a population of stars rather than individual stars; their radiative outputs are linearly proportional to the sink particle masses via luminosity-to-mass ratios for both bands. We use the analytical fitting functions presented in \citet{Kim_2016}\footnote{For the LyC band, \citet{Kim_2016} provide a fitting function for the number of ionising photons emitted by the stellar population per second; this is converted to a luminosity by adopting a mean energy per LyC photon of 18 eV, which is the appropriate luminosity-weighted frequency-averaged value for a fully sampled IMF \citep{Kim_2023}.} for the luminosity-to-mass ratios for a given total star cluster mass, which we compute as the total instantaneous stellar mass. The functions in \citet{Kim_2016} were fitted to the median luminosity-to-mass ratios obtained from several realizations of calculations with the \texttt{SLUG} stellar population synthesis code for a range of star cluster masses ($\sim 10^2 \, \Msun$ -- $10^6 \, \Msun$) assuming a Chabrier IMF; this approach therefore takes into account the effect of the luminosity-to-mass decreasing (on average) due to stochastic sampling of massive stars when the mass of the cluster is low ($\lesssim 10^4 \, \Msun$). For cluster masses higher than $\gtrsim 10^4 \Msun$, the luminosity-per-unit mass saturates to values of $\Psi_{\mathrm
FUV} = 943 L_\sun \Msun^{-1}$ and $\Psi_{\mathrm
LyC} = 380 L_\sun \Msun^{-1}$ respectively. However, we do not take into account the temporal evolution of the radiative outputs since most of our simulations are run for relatively short timescales over which these values do not change significantly. 

We use a constant (frequency-averaged) cross section of (neutral) hydrogen atoms to LyC photons of $\sigma_{\mathrm{H}} = 3 \times 10^{18} \, \mathrm{cm}^{2}$. The dust opacities per unit (gas) mass for the LyC and FUV bands are taken to be equal to $\kappa_{\mathrm{FUV}} = 1250 Z_{\mathrm{d}}^{'} \, \mathrm{cm}^2 \, \mathrm{g}^{-1}$ and $\kappa_{\mathrm{FUV}} = 1000 Z_{\mathrm{d}}^{'} \, \mathrm{cm}^2 \, \mathrm{g}^{-1}$ respectively. We use the Planck and Rosseland-mean opacities (scaled by $Z_{\mathrm{d}}^{'}$) in the IR using the dust temperature-dependent model of \citet{Semenov_2003}. We assume the dust temperature instantly comes into radiative equilibrium with the IR radiation field; see \citet{Menon_2024} for the caveats associated with this assumption, none of which should affect the results of this paper. We initialize our clouds with zero radiation energy/flux in the UV and an IR radiation field corresponding to a dust temperature of $T_{\mathrm{d}} = 10 \, \mathrm{K}$. We adopt Marshak boundary conditions \citep{Marshak_1958} for the radiation with the background value set to match the initial conditions. 

We use a quasi-isothermal two-temperature equation of state for the thermal evolution of the gas, which by construction assumes instantaneous thermal equilibrium with two distinct (stable) phases that purely depends on the ionization state of the gas -- i.e. a warm, fully photoionized phase with temperature $T_{\mathrm{ion}} = 10^4 \, \mathrm{K}$, and a cold, fully neutral phase with temperature $T_{\mathrm{cold}} = 10 \, \mathrm{K}$. The temperature is then linearly scaled in between these two limits depending on the ionization fraction for intermediate ionization states. This two-temperature approximation has been adopted by several works in the past \citep{Gritschneder_2009,Kim_2018,Menon_2020}, and is a reasonable approximation to first order to model the photoionized/neutral transition since the cooling timescale is very short compared to the dynamical timescale. However, this approach would not be able to produce warmer neutral gas due to other radiative processes that could act on LyC-shielded gas, such as photodissociation and the photoelectric effect. While this warmer gas would likely affect fragmentation in the cloud -- which we do not focus on in this paper -- it is unlikely to affect the bulk evolution of the cloud and the cloud-scale competition between feedback and gravity to zeroth order; the key physics that is relevant in these contexts are photoionization and radiation pressure on gas/dust, all of which is captured in our simulations. We do not include the feedback pathways of stellar winds, protostellar jets, and supernovae in our simulations. The latter occur on timescales longer than the cloud lifetimes; we discuss the implications of our omission of the former mechanisms in Section~\ref{sec:caveats}.

\subsection{Parameter Space}

\begin{table*}
\caption{Summary of our simulation suite and their initial condition parameters.}
\centering
\label{tab:Simulations}
\begin{threeparttable}
\begin{tabular}{c c c c c c c c c c}
\toprule
\multicolumn{1}{c}{$\Sigma_{\mathrm{cloud}}$}& \multicolumn{1}{c}{$M_{\mathrm{cloud}}$}& \multicolumn{1}{c}{$R_{\mathrm{cloud}}$}&  \multicolumn{1}{c}{$n_{\mathrm{cloud}}$}& \multicolumn{1}{c}{$\sigma_{v}$}& \multicolumn{1}{c}{$\mathcal{M}$}& \multicolumn{1}{c}{$v_{\mathrm{esc}}$}& \multicolumn{1}{c}{$t_{\mathrm{ff},\mathrm{cloud}}$}& \multicolumn{1}{c}{$\mathrm{Z_\mathrm{d}}$}\\
$[\Msun \, \mathrm{pc}^{-2}]$& $[\Msun]$& [pc]& [$\mathrm{cm}^{-3}$]& [km/s]& & [km/s]& [Myr]& [$Z_{\mathrm{d},\sun}$]&\\
\midrule
$10^2$ &$10^5$ &$18$ &$1.7$$ \times 10^{2}$ &$5.4$ &20 &7 &4 &1\\
$10^3$ &$10^5$ &$5.6$ &$5.3$$ \times 10^{3}$ &$10$ &33 &12 &0.7&1\\
$10^4$ &$10^5$ &$1.8$ &$2$$ \times 10^{5}$ &$17$ &60 &22 &0.13&1\\
$10^5$ &$10^5$ &$0.6$ &$5.3$$ \times 10^{6}$ &$30$ &110 &39 &0.02 &1\\
$10^4$ &$10^5$ &$1.8$ &$2$$ \times 10^{5}$ &$17$ &60 &22 &0.13&$0.1$\\
$10^4$ &$10^5$ &$1.8$ &$2$$ \times 10^{5}$ &$17$ &60 &22 &0.13&$0.01$\\
\bottomrule
\end{tabular}
\begin{tablenotes}
\small
\item \textbf{Notes}: Columns in order indicate - Model: $\Sigma_{\mathrm{cloud}}$: mass surface density of the cloud given by $\Sigma_{\mathrm{cloud}} = M_{\mathrm{cloud}}/(\pi R_{\mathrm{cloud}}^2)$, $M_{\mathrm{cloud}}$: mass of cloud, $R_{\mathrm{cloud}}$: radius of cloud, $n_{\mathrm{cloud}}$: number density of the cloud given by $n_{\mathrm{cloud}} = 3M_{\mathrm{cloud}}/(4 \pi R_{\mathrm{cloud}}^3m_{\mathrm{H}})$ where $m_{\mathrm{H}}$ is the mass of atomic hydrogen, $\sigma_{v}$: turbulent velocity dispersion of the cloud, $\mathcal{M}$: turbulent Mach number given by the relation $\sigma_v/c_{\mathrm{s}}$ within the cloud where $c_{\mathrm{s}}$ is the initial thermal sound speed in the cloud ($\sim 0.2 \, \kms$),  $v_{\mathrm{esc}}$: escape velocity of the cloud, $t_{\mathrm{ff}}$: free-fall time of the cloud, $Z_{\mathrm{d}}$: dust-to-gas ratio in units of the corresponding value in the solar neighborhood ($Z_{\mathrm{d},\sun}$). All our simulations use a resolution of $256^3$. 
\end{tablenotes}
\end{threeparttable}
\end{table*}

The suite of simulations that we present in this study is summarized in Table 1. The primary parameter that we explore in our simulation suite is $\Sigmacloud$, where we adopt values $ = 10^2 \, \Msolpc$ -- $10^5 \, \Msolpc$ to progressively mimic conditions typical of GMCs in the star-forming spiral galaxies that are common in the local Universe to those in the very compact, rapidly star forming galaxies expected (and now observed) in the high-redshift Universe, respectively. The velocity dispersions follow our assumption of initial marginal stability ($\alphavir = 2$), which leads to turbulent Mach numbers between 20 and 110 for a fixed initial gas temperature of $T = T_{\mathrm{cold}} = 10 K$. We note that the escape velocities ($v_{\mathrm{esc}}$) are higher and the dynamical timescales ($t_{\mathrm{ff}}$) lower for higher $\Sigmacloud$ -- this feature is very important for the results we will show below. All of the above simulations are run with solar-neighbourhood-like dust abundances ($Z_{\mathrm{d}}^{'} = 1$); for $\Sigmacloud = 10^4 \, \Msolpc$, we also run cases with lower dust abundances of $Z_{\mathrm{d}}^{'} = 0.1$ and $Z_{\mathrm{d}}^{'} = 0.01$. This is meant to mimic the conditions of lower metallicities in the high(er) redshift Universe, especially in low-mass ($M_* \sim 10^7$ -- $10^9 \, \Msun$) galaxies that primarily contribute to reionization. Our explored range of dust-to-gas ratios spans that occupied by high-redshift LyC leaking galaxies in cosmological simulations \citep[e.g.,][]{Choustikov_2024}, observed low-redshift LyC leakers \citep[e.g.,][]{Flury_2022b}, and Epoch of Reionization (EoR) galaxies observed with JWST \citep{Curti_2023,Nakajima_2023}. We note that the lower dust abundances are only one of several physical effects that manifest differently at lower metallicities; the gas temperatures in thermal balance ($T_{\mathrm{ion}}$ and $T_{\mathrm{cold}}$) would be higher due to weaker metal-line cooling\footnote{For primordial metallicities the temperature is 30000K, corresponding to a sound speed of 15 km/s in ionized gas; so will affect $\Sigmacloud \lesssim 10^3 \, \Msolpc$, but likely not denser clouds}, and the UV radiative outputs of lower-metallicity stars (and therefore star clusters) higher. We chose not to include these effects here for a more controlled experiment. In Section~\ref{sec:caveats} we discuss the caveats associated with this assumption in interpreting our $Z_{\mathrm{d}}^{'} < 1$ simulations as low-metallicity counterparts, and conclude that the change to the dust opacity with metallicity is the principal effect to capture, and that the effects of the gas temperature and stellar spectra are relatively minor in the context of our results.

\section{Results}
\label{sec:Results}

\subsection{General evolution of the simulations}

We discuss the general time evolution of our simulations in this section. The initial turbulent fluctuations form filamentary structures that become gravitationally unstable, and go on to collapse until they form sink particles (which represent sub-clusters of stars), the first of which forms by a timescale $\lesssim 0.5 t_{\mathrm{ff},\mathrm{cloud}}$. Star formation continues to occur through both the formation of new sink particles and accretion onto existing sinks. As the sinks become more massive, their UV radiative output proportionately grows, and this starts to photoionize and provide momentum feedback (radiation pressure) on the gas surrounding the sinks. The subsequent evolution differs among runs with different $\Sigmacloud$; lower $\Sigmacloud$ ($\Sigmacloud \lesssim 10^3 \, \Msolpc$) runs photoionize and drive radiation pressure-driven bubbles more successfully, whereas higher $\Sigmacloud$ runs continue to accrete with relatively mild dynamical impact from the radiation. Eventually the bulk of the gas in these runs is photoionized and evacuated by a time $\lesssim 2 t_{\mathrm{ff}}$, which for the low-$\Sigmacloud$ run corresponds to $\lesssim 8 \, \mathrm{Myr}$. The higher $\Sigmacloud$ ($\Sigmacloud > 10^3 \, \Msolpc$) runs on the other hand, still contain a significant amount of dense gas in the central regions, although there are signs of outflows through low-density channels. It is only at much later (dynamical) times ($\gtrsim 7 t_{\mathrm{ff}}$) that the bulk of the gas is driven out in an outflow, consistent with \citet{Menon_2023}. Broadly, the general evolution shows that at higher $\Sigmacloud$, star formation and the dispersal of the remaining gas go on for longer multiples of the cloud free-fall times.

\begin{figure*}
    \centering
    \includegraphics[width=\textwidth]{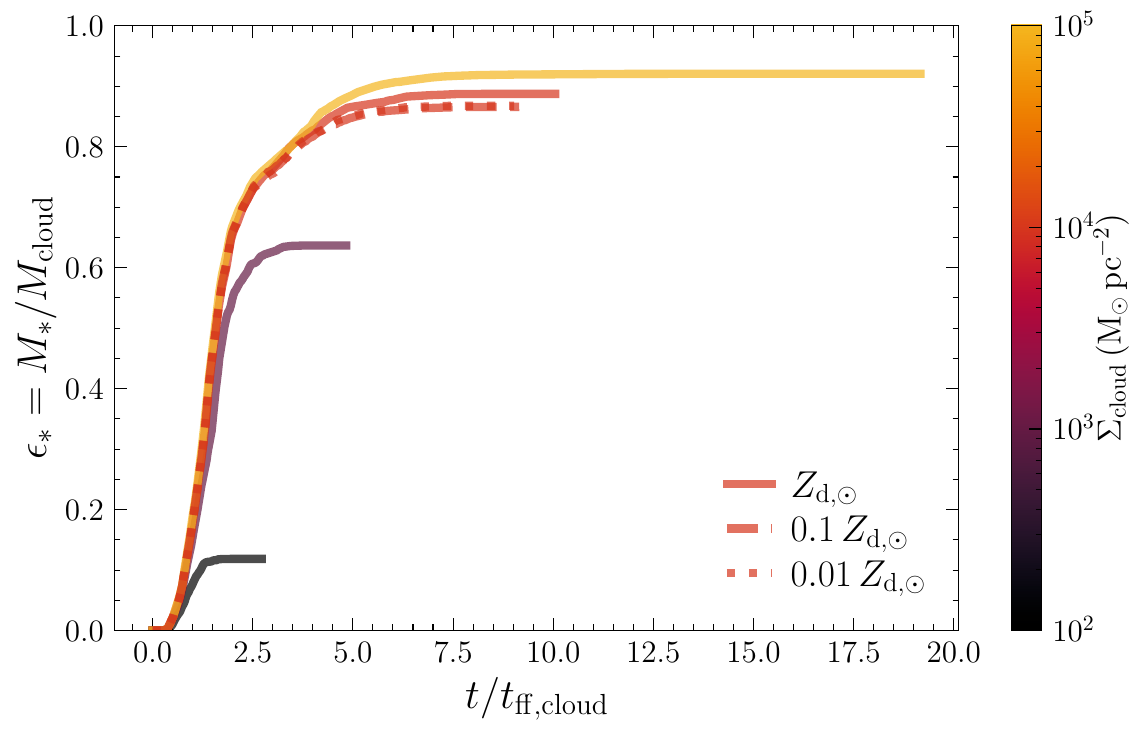}
    \caption{Time evolution of the integrated star formation efficiency $\epsilon_*$ -- i.e. the fraction of initial cloud gas mass converted to stars -- across the simulations in this study, shown as a function of time in units of the cloud free-fall time ($t_{\rm ff, cloud}$; Table~\ref{tab:Simulations}). $\epsilon_*$ increases with time until (radiative) feedback from young stars is able to counteract gravity, at which point the rate of star formation ($\dot{\epsilon_*}$) decreases, and eventually $\epsilon_*$ saturates. The point at which this occurs depends on $\Sigmacloud$, with more efficient star formation in denser clouds.}
    \label{fig:SFE}
\end{figure*}

This behavior is reflected in Figure~\ref{fig:SFE}, which shows the integrated star formation efficiency $\epsilon_*$ -- the fraction of the initial cloud mass that has been converted into stars -- as a function of time. All our simulations show similar trends: $\epsilon_*$ rises rapidly on a timescale ~$t_{\mathrm{ff},\mathrm{cloud}}$, and then saturates due to radiative feedback on a timescale that depends on $\Sigmacloud$. The principal differences between models are the timescale for saturation and the saturated value of $\epsilon_*$ which increases sharply with $\Sigmacloud$, consistent with previous works \citep{Fall_2010,Grudic_2018,Lancaster_2021c,Menon_2023}. If we compare these values with the results of \citet{Menon_2023} where largely similar numerical methods were used, except that photoionization was not included, we find that the runs with $\Sigmacloud \lesssim 10^3 \, \Msolpc$ here have much lower $\epsilon_*$, whereas the values for the higher $\Sigmacloud$ cases are more or less identical. This is consistent with theoretical expectations -- the $\Sigmacloud \lesssim 10^3 \, \Msolpc$ cases have $v_{\mathrm{esc}} \lesssim c_{\mathrm{s},\mathrm{ion}}$, the ionized gas sound speed, where photoionization is expected to be the dominant feedback mechanism \citep{Dale_2012,Kim_2018}. The higher $\Sigmacloud$ cases do not satisfy this condition, and therefore photoionization makes a negligible contribution to the feedback-matter competition, which is rather dominated by radiation pressure on dust and gas. This does not mean that the photoionization does not have any dynamical impact in our simulations; indeed, we will see below that gas needs to be photoionized/photoevaporated to enable escape of LyC photons. Additionally, the ionizing flux also affects the phase structure of radiation pressure-driven outflows. 

\subsection{Escape Fractions}
\label{sec:EscapeFractions}

We now present the escape of UV photons and its dependence on the cloud properties. Since we explicitly solve the radiation moment equations, we can track the local absorption rate of UV photons by dust and gas -- $\dot{E}_{\mathrm{dust}}$ and $\dot{E}_{\mathrm{gas}}$ respectively --and compare this to the net UV radiative output in the domain to obtain the escape fraction. The net UV radiative output is obtained by summing up the luminosities of the individual sinks ($L_*$). We can then define the global escape fraction of the simulation at any time as 
\begin{equation}
    f_{\mathrm{esc},\nu} = 1 - \frac{ \sum_{\mathcal{V}} \left( \dot{E}_{\mathrm{dust},\nu} + \dot{E}_{\mathrm{gas},\nu} \right) dV}{\sum_{*} L_{*,\nu} },
\end{equation}
where the subscript $\nu$ refers either to the LyC or the FUV bands. We note that gas only absorbs LyC photons\footnote{Technically molecular hydrogen gas absorbs photons in the FUV energy range 11.2-13.6 eV, which excites the Lyman-Werner transitions. However, the molecular-atomic phase transition and associated radiative processes are not included in this study.} and therefore $\dot{E}_{\mathrm{gas},\mathrm{FUV}} = 0$, whereas for dust it is non-zero for both bands. We can also define the fraction of radiation that is absorbed by neutral gas ($f_{\mathrm{gas}}$) and by dust ($f_{\mathrm{dust}}$) as --
\begin{equation}
    f_{\mathrm{gas},\nu} = \frac{ \sum_{\mathcal{V}} \dot{E}_{\mathrm{gas},\nu}  dV}{\sum_{*} L_{*,\nu} },
    \label{eq:fgas}
\end{equation}
and 
\begin{equation}
    f_{\mathrm{dust},\nu} = \frac{ \sum_{\mathcal{V}} \dot{E}_{\mathrm{dust},\nu}  dV}{\sum_{*} L_{*,\nu} },
    \label{eq:fdust}
\end{equation}
respectively. This maintains the closure relation $f_{\mathrm{esc},\nu} + f_{\mathrm{gas},\nu} + f_{\mathrm{dust},\nu} = 1$.

\begin{figure*}
    \centering
    \includegraphics[width=0.9\textwidth]{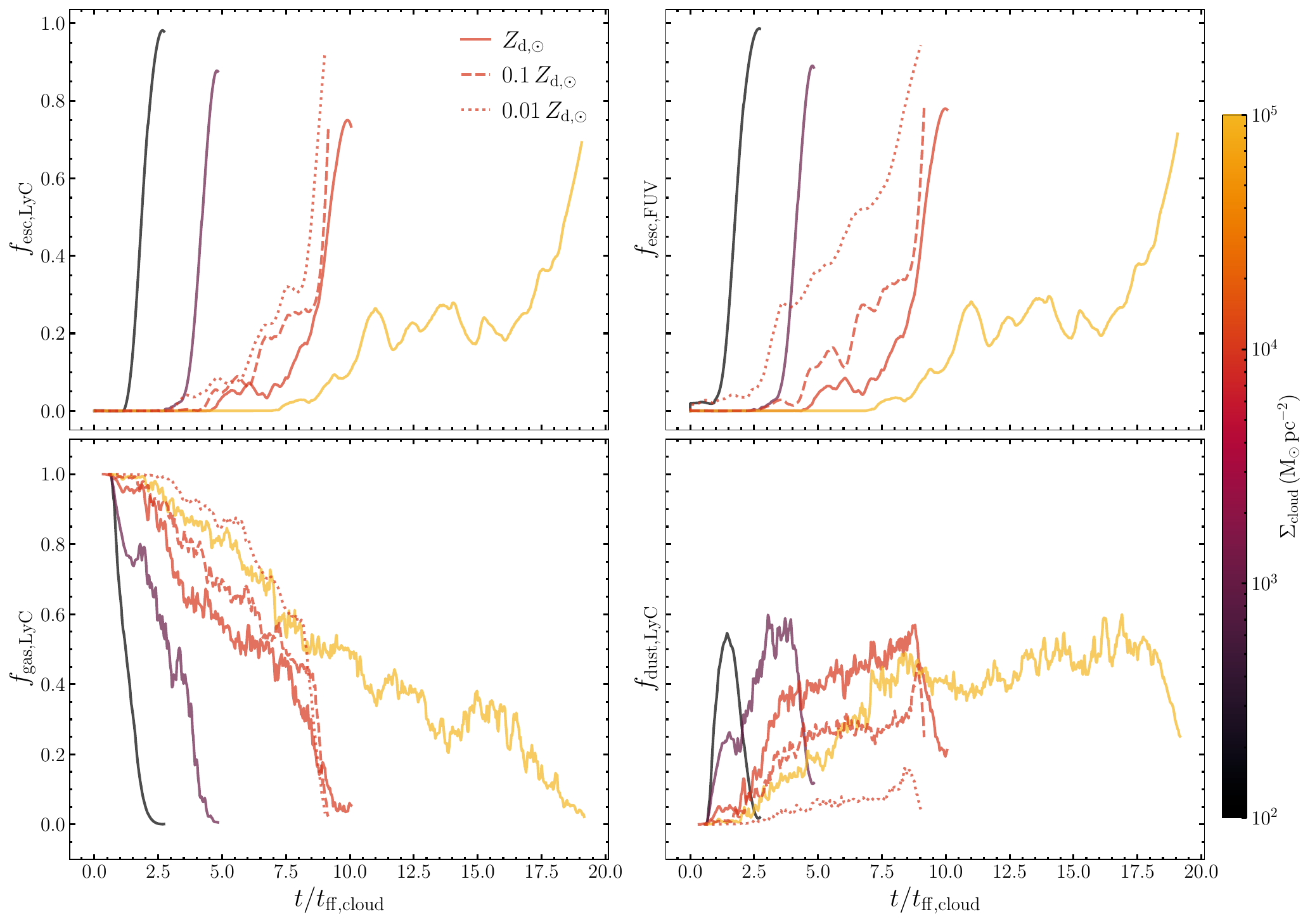}
    \caption{This figure shows the possible fates of photons (see Section~\ref{sec:EscapeFractions}) emitted by the stellar populations across our simulations, shown as a function of time scaled by $t_{\mathrm{ff},\mathrm{cloud}}$. Top-left: escape fraction of LyC photons ($f_{\mathrm{esc},\mathrm{LyC}}$); top-right: escape fraction of FUV photons ($f_{\mathrm{esc},\mathrm{FUV}}$); bottom-left: fraction of LyC photons absorbed by neutral $\mathrm{HI}$ gas ($f_{\mathrm{gas},\mathrm{LyC}}$); bottom-right: fraction of LyC photons absorbed by dust ($f_{\mathrm{dust},\mathrm{LyC}}$). We do not show the latter two for the FUV band since $f_{\mathrm{gas},\mathrm{FUV}} = 0$ and $f_{\mathrm{dust},\mathrm{FUV}} = 1 - f_{\mathrm{esc},\mathrm{FUV}}$. Most LyC photons are absorbed by ionization-bounded gas at earlier times when $f_{\mathrm{esc},\mathrm{LyC}}$ is low, followed by a shorter phase where dust can absorb up to $\sim 50\%$ of photons as radiative feedback clears gas. Denser clouds require significantly more free-fall times to clear their surroundings and leak LyC photons, and have a longer phase where dust absorption is important.}
    \label{fig:LyCEscapeFraction}
\end{figure*}

In Figure~\ref{fig:LyCEscapeFraction} we show $f_{\mathrm{esc},\mathrm{LyC}}$ and $f_{\mathrm{esc},\mathrm{FUV}}$ as a function of the time since star formation began ($t_*$) for all our simulations. The temporal trend is broadly similar across our simulations -- an initial obscured phase followed by a rapid increase in $f_{\mathrm{esc}}$ due to dispersal of gas and dust by feedback. We can see, consistent with expectations from the general evolution discussed above, that lower $\Sigmacloud$ runs achieve higher escape fractions much more rapidly (in units of $t_{\mathrm{ff},\mathrm{cloud}}$), and that they show a more rapid evolution from obscured ($f_{\mathrm{esc}} \sim 0$) to transparent ($f_{\mathrm{esc}} \sim 1$) to UV photons. 

By comparing the top-left and top-right panels in Figure~\ref{fig:LyCEscapeFraction}, we can see that the escape fractions of LyC and FUV photons more or less trace each other, albeit with a slightly shorter timescale for FUV escape evident on careful inspection. 
However, this similar evolution of LyC and FUV escape fractions becomes less pronounced in the simulations with lower dust to gas ratios.
Comparing the dashed and dotted lines with the solid lines in Figure~\ref{fig:LyCEscapeFraction}, we see that there is significantly more FUV photon escape at earlier $t/t_{\mathrm{ff},\mathrm{cloud}}$ than LyC escape. The strong effect of the dust abundance on FUV  escape relative to LyC escape  results from the fact that the only source of opacity for the FUV photons is from dust. On the other hand, LyC photons can be absorbed by both neutral gas and dust. For the LyC band, Figure~\ref{fig:LyCEscapeFraction} shows that different dust abundances yield similar LyC escape fractions at early times, while the curves start to diverge at later times. This suggests that dust absorption is more important at later times. Indeed, in Figure~\ref{fig:LyCEscapeFraction} we show that the fraction of LyC photons absorbed by dust, $f_{\mathrm{dust}}$ (Equation~\ref{eq:fdust}), increases with time, with gas dominating the absorption at earlier times, and dust acting as the main source of opacity later in the evolution. In fact for a brief period -- which depends on $\Sigmacloud$ -- most ($\gtrsim 60\%$) of the LyC photons emitted by the stellar population are consumed by dust rather than by gas or by escaping the computational volume (see lower panel in Figure~\ref{fig:LyCEscapeFraction}). This occurs because most of the gas has already been photoionized and/or photoevaporated by ionizing radiation, and it is the surviving dust in the ionized gas that absorbs photons. However, this dusty ionized gas is also gradually evacuated via thermal pressure (at smaller $\Sigmacloud$), or via radiation pressure on dust grains (at higher $\Sigmacloud$), causing $f_{\mathrm{dust}}$ to decrease and  $f_{\mathrm{esc},\mathrm{LyC}}$ to proportionally increase. The former mechanism likely only occurs for the runs with lower $\Sigmacloud$, which have $v_{\mathrm{esc}} \lesssim c_{\mathrm{s},\mathrm{ion}}$, permitting the ionized gas (and the dust along with it) to escape freely; for higher surface density clouds, radiation pressure on dust acts to evacuate the dusty ionized gas. At earlier times when both sources of opacity exist, the factor $\sim 1000$ higher cross section of neutral hydrogen to ionizing UV photons over dust ensures that it acts as the dominant opacity. For runs with lower dust abundance, Figure~\ref{fig:LyCEscapeFraction} shows that even at later times, dust still plays a relatively minor role as a source of obscuration; for instance for $Z_{\mathrm{d}} = 0.01 Z_{\mathrm{d},\sun}$, $f_{\mathrm{dust}} < 0.2$ at all times. 

\subsection{Evacuation of channels by radiative feedback}
\label{sec:evacuation}

\begin{figure*}
    \centering
    \includegraphics[width=\textwidth]{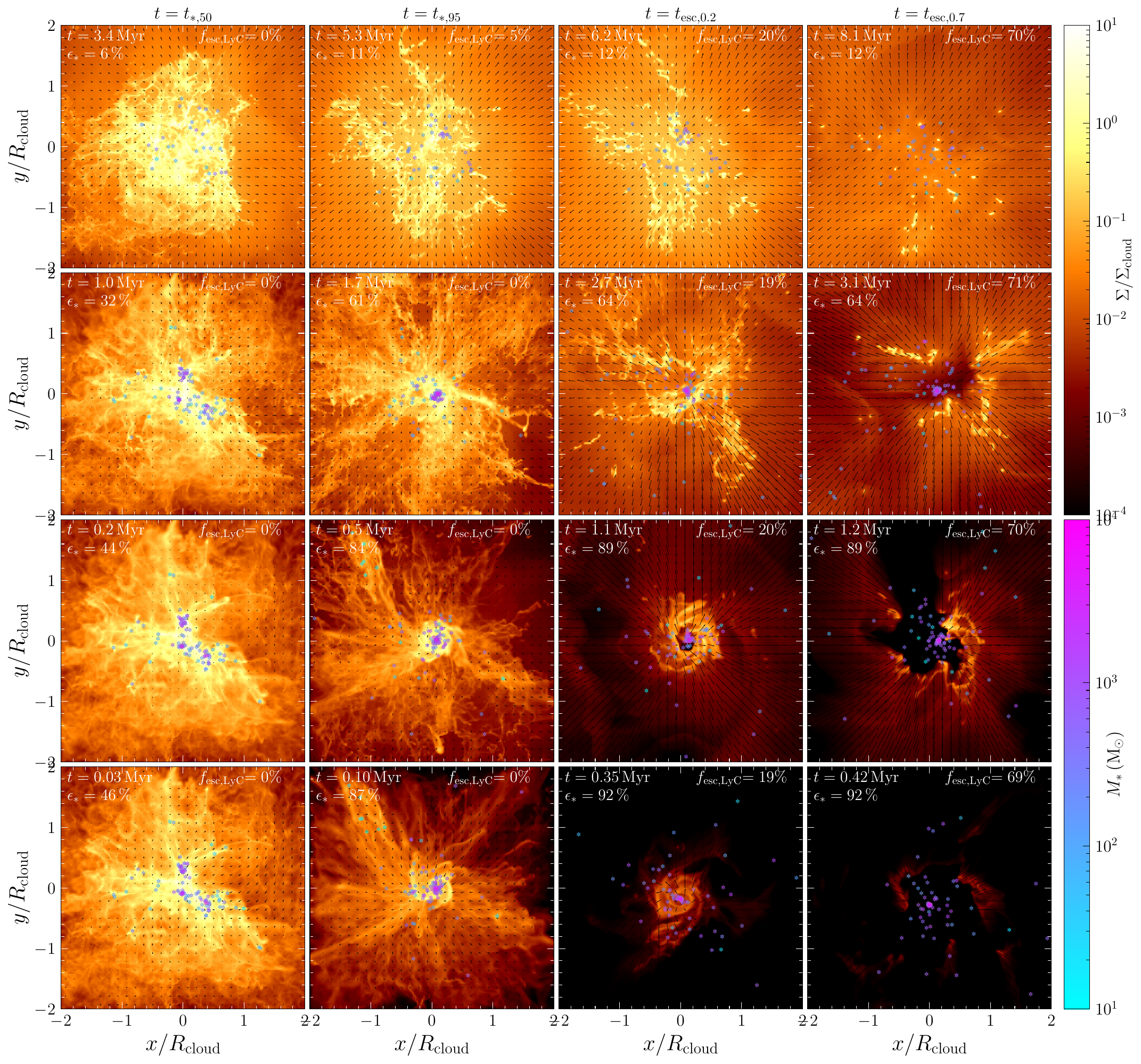}
    \caption{Projected gas density distributions for the different $\Sigmacloud$ runs (rows), plotted at the characteristic timescales defined in Section~\ref{sec:evacuation}: i.e. $t_{*,50}$, $t_{*,95}$, $t_{\mathrm{esc},0.2}$ and $t_{\mathrm{esc},0.7}$ (left to right). Time in physical units ($t$), star formation efficiencies ($\epsilon_*$) and instantaneous escape fractions ($f_{\mathrm{esc}}$) are annotated. Arrows indicate the velocity field.}
    \label{fig:Projection_1}
\end{figure*}

In the previous section we presented the following qualitative scenario for LyC escape: neutral gas is increasingly photoionized by LyC radiation, and the resultant dusty ionized gas subsequently escapes via thermal expansion, or is driven out by radiation pressure on dust in dense clouds. Since the clouds are turbulent, this likely occurs through increasingly wide channels of UV photon escape. While this qualitative picture holds across our parameter space, there is clearly a distinction in the (dynamical) timescales required for this process to play out. We explain the physics behind this behavior in this section.

\begin{figure}
    \centering
    \includegraphics[width=0.48\textwidth]{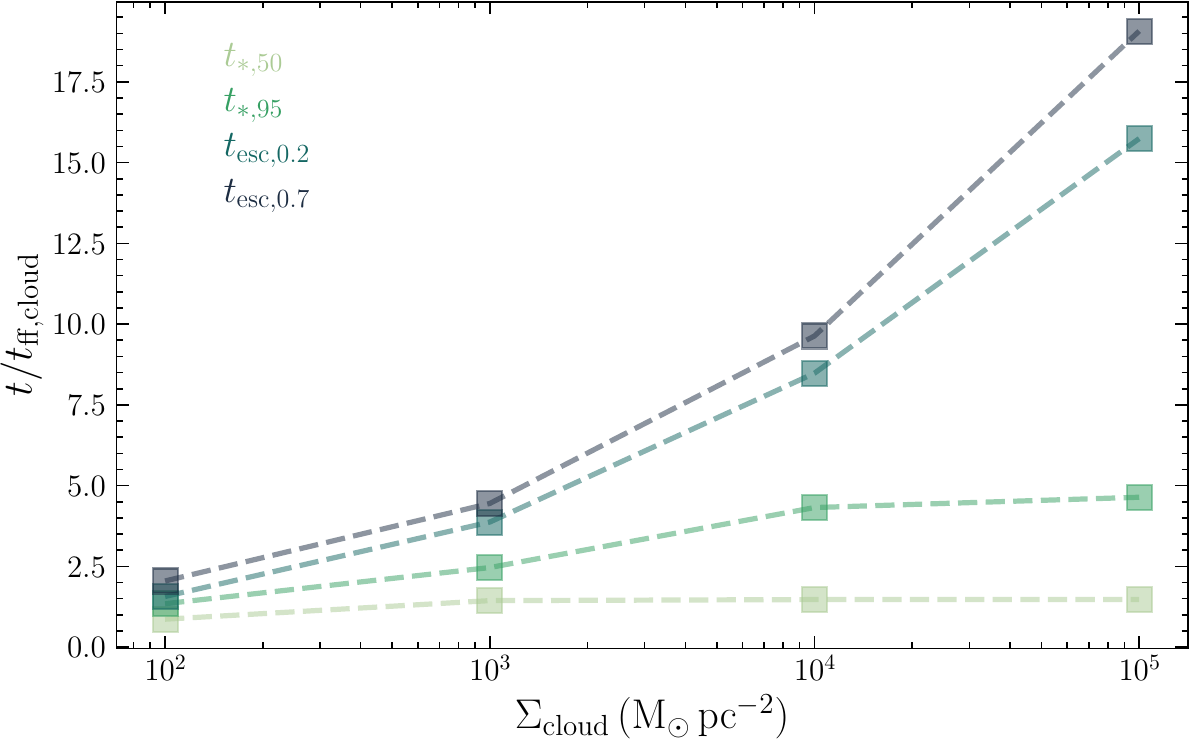}
    \caption{Characteristic timescales (defined in Section~\ref{sec:evacuation}) in units of $t_{\mathrm{ff},\mathrm{cloud}}$ for different $\Sigmacloud$. $t_{*,50}$ and $t_{*,95}$ indicate the times at which 50\% and 95\% of the final saturated stellar mass is achieved. $t_{\mathrm{esc},0.2}$ and $t_{\mathrm{esc},0.7}$ are the times at which $f_{\mathrm{esc},\mathrm{LyC}}$ are 20\% and 70\% respectively. Denser clouds form stars for slightly more dynamical times than less dense clouds. They also take significantly more dynamical times to leak LyC photons. Note, however, that the dynamical times of denser clouds are significantly shorter in absolute time (Table~\ref{tab:Simulations}).}
    \label{fig:Timescales}
\end{figure}

To systematically quantify the distinct temporal behavior across our simulation suite, we define four characteristic timescales that we can use to compare the runs:
\begin{enumerate}
    \item $t_{*,50}$: The timescale to reach 50\% of the final saturated stellar mass. 
    \item $t_{*,95}$: The timescale to reach 95\% of the final saturated stellar mass, therefore representing when star formation more or less ceases.
    \item $t_{\mathrm{esc},0.2}$: The time when an instantaneous LyC escape fraction of 20\% is achieved.
    \item $t_{\mathrm{esc},0.7}$: The time when an instantaneous LyC escape fraction of 70\% is achieved.
\end{enumerate}
We show the projected distribution of gas and stars at these characteristic timescales in Figure~\ref{fig:Projection_1}. We plot these characteristic timescales as a function of $\Sigmacloud$ in Figure~\ref{fig:Timescales}. We can see that star formation occurs for a slightly longer duration at higher $\Sigmacloud$. More evidently, the timescale for achieving LyC escape is significantly longer at larger $\Sigmacloud$. This indicates that the processes of photoevaporating and evacuating channels of dusty gas takes longer (in terms of free-fall timescales) at higher $\Sigmacloud$. 

To understand this behavior more clearly, we look at the distribution of gas columns as seen by the radiating star clusters. To obtain this, we first make the assumption that the set of distributed radiating sources can be effectively replaced with a single source placed at the centre of mass of the system, with the combined luminosity of all the sources. We then interpolate the grid quantities from our Cartesian grid onto a spherical grid centered at this point. The spherical grid is discretized with $[N_r,N_{\theta},N_{\phi}] = [128,128,256]$, equally spaced in $r \in [0,2\Rcloud]$, $\phi \in [0,2\pi]$, and $\cos \theta \in [-1,1]$, ensuring that the grid is discretized equally in solid angle. We use trilinear interpolation for this purpose. We then compute the gas column densities along each line of sight by integrating the gas density in the radial direction for a given $\theta,\phi$. This yields a distribution of the gas column in solid angle space -- i.e. $f_\Omega (\Sigma)$. We calculate these distributions for both the total gas ($\Sigma_{\mathrm{g}}$) and the neutral atomic gas ($\Sigma_{\mathrm{HI}}$) separately. We refer to these as the circumcluster gas column densities. 

\begin{figure*}
    \centering
    \includegraphics[width=\textwidth]{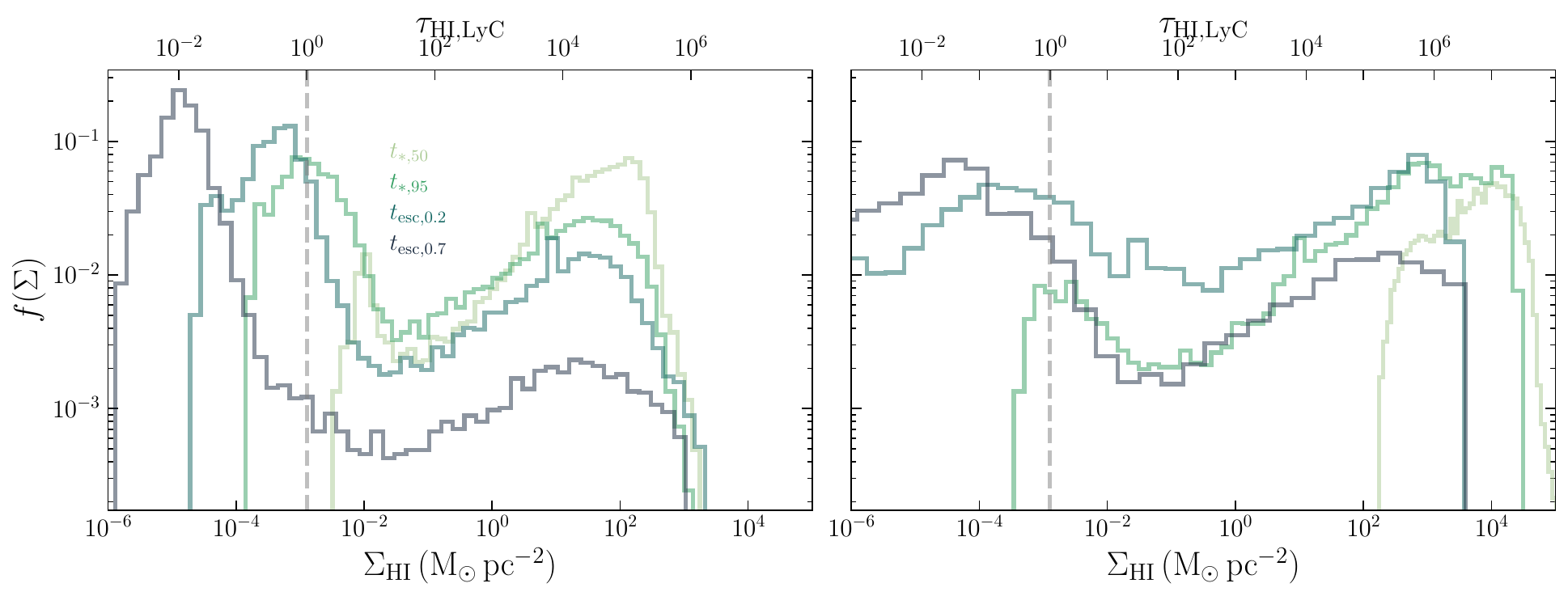}
    \caption{The PDF of neutral HI column densities in different sightlines as seen by the center of mass of the stellar distribution at the characteristic timescales for $\Sigmacloud =  100 \Msolpc$ (left) and $10^4 \Msolpc$ (right). The dashed line indicates the transition from optically thin to thick to LyC photons (i.e. $\tau_{\mathrm{HI},\mathrm{LyC} = 1}$). We can see that the higher $\Sigmacloud$ run contains a significant fraction of high $\Sigma_{\mathrm{HI}}$ sightlines even at later epochs, whereas the lower $\Sigmacloud$ case has already photoionized the bulk of sightlines when star formation has just ceased (i.e. at $t_{*,95}$).}
    \label{fig:SigmaEvolution}
\end{figure*}

To interpret the significant differences in the LyC escape timescale with $\Sigmacloud$, we compare the circumcluster gas column densities between $\Sigmacloud =  100$ and $10^4 \, \Msolpc$ as representative of a relatively diffuse and dense cloud, respectively. In Figure~\ref{fig:SigmaEvolution} we compare the distributions of $\Sigma_{\mathrm{HI}}$ sampled at the characteristic timescales defined above. As expected, at $t_{*,50}$ all sightlines are obscured by the gas, although the higher $\Sigmacloud$ case has significantly higher $\Sigma_{\mathrm{HI}}$ values. However, by $t_{*,95}$, a significant fraction of the gas has been photoionized for $\Sigmacloud =  100 \, \Msolpc$, such that most sightlines are marginally optically thin. This is not the case for the high surface density model with $\Sigmacloud =  10^4 \, \Msolpc$, where most sightlines are still at the mean value, with $\tau_{\mathrm{HI},\mathrm{LyC}} \gtrsim 10^6$. This indicates that the bulk photoevaporation of the gas has simultaneously been occurring along with star formation at lower $\Sigmacloud$, whereas neutral high density regions continue to survive in denser clouds even after star formation more or less ceases. This distinction in outcomes is largely due to the higher densities and higher escape velocities that accompany higher $\Sigmacloud$ (for fixed mass): higher densities make it harder for the ionizing radiation field to compete with recombinations ($\propto n^2$); this is aggravated by the fact that the ionized gas cannot freely expand since the potential well is too deep (i.e. $v_{\mathrm{esc}} > c_{\mathrm{s},\mathrm{ion}}$), and neither can it exert sufficient thermal pressure on the surrounding neutral gas to expand as a D-type front, since the external ram pressure is more significant ($\sigma_v > c_{\mathrm{s},\mathrm{ion}}$). 

Instead, it is the radiation pressure on dust and gas that is responsible for the expansion of HII regions for these denser clouds, consistent with analytic work \citep{Krumholz_Matzner_2009, Murray_2010, Draine_2011} and numerical simulations \citep[e.g.,][]{Kim_2018}. This expansion lowers the ionized gas density, freeing up LyC photons (which earlier were expended balancing recombinations) to ionize additional material. This radiation pressure-driven expansion is the reason that a significant fraction of material becomes ionized by $t_{\mathrm{esc},0.2}$ for the high $\Sigmacloud$ case in Figure~\ref{fig:SigmaEvolution}. That being said, there is still a significant fraction of sightlines with high $\Sigma_{\mathrm{HI}}$, suggesting that radiation pressure is unable to evacuate these sightlines effectively. We verify in Figure~\ref{fig:SigmaEvolution} that the total gas column density ($\Sigma_{\mathrm{g}}$) also shows signatures of surviving dense columns even at $t_{\mathrm{esc},0.2}$ with dust optical depths much greater than unity; in fact, inspection of Figure~\ref{fig:LyCEscapeFraction} reveals that at these times ($\sim 8.5 t_{\mathrm{ff}}$) it is the absorption by dust that dominates LyC absorption rather than the gas, and it is the failure of stellar feedback to eject the existing high-density columns of gas (and dust) that limits the escape fraction of LyC photons. Indeed, semi-analytic models \citep{Thompson_Krumholz_2016} predict that since the radiating sources see a distribution of column densities, there will be a corresponding distribution of outcomes for the competition between gravity and radiation pressure. This competition can be quantified with the Eddington ratio -- the ratio of radiation pressure to gravitational (self- and stellar) forces -- acting through a given column of gas. 

\begin{figure*}
    \centering
    \includegraphics[width=\textwidth]{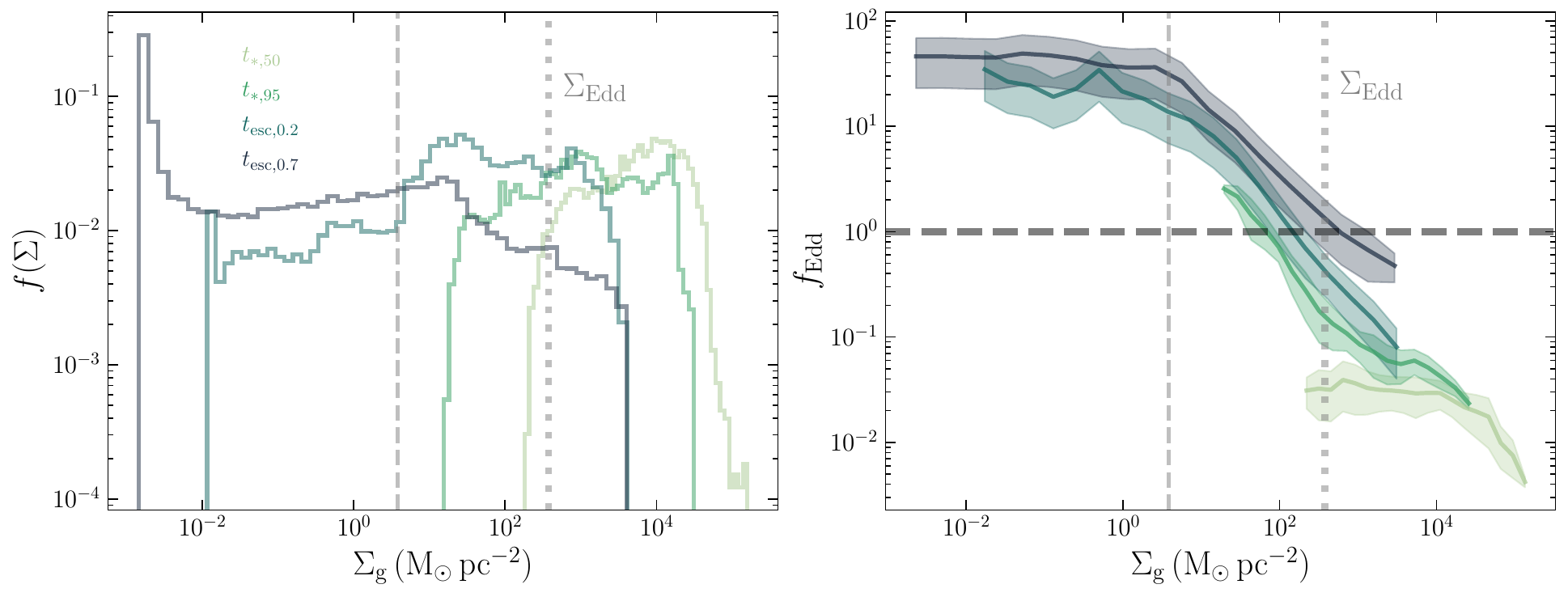}
    \caption{The PDF of total gas column densities ($\Sigma_{\mathrm{g}}$) in different sightlines as seen by the center of mass of the stellar distribution at the characteristic timescales for $\Sigmacloud = 10^4 \Msolpc$ (left) and the median Eddington ratios in each bin of $\Sigma_{\mathrm{g}}$ with the shaded region representing the 1-$\sigma$ variation. Dashed and dotted gray vertical lines represent the points where $\tau_{\mathrm{dust}},\mathrm{LyC} = 1$ and the Eddington surface density respectively. This reveals a scenario where super-Eddington channels are evacuated by radiation pressure on dust, producing the low-column tails in $\Sigma_{\rm g}$ that leak LyC photons at later times. The fraction of sightlines that are super-Eddington increase with time as gas is consumed by star formation and/or locally driven outward by radiative feedback -- both of which systematically lower $\Sigma_{\rm g}$.}
    \label{fig:S4SigmafEdd}
\end{figure*}

We compute the Eddington ratio $f_{\mathrm{Edd}}$ for each sightline by cumulatively integrating the total (UV+IR) radiation pressure forces in that sightline, and dividing by the corresponding cumulative gravitational force. In Figure~\ref{fig:S4SigmafEdd} we look at how $f_{\mathrm{Edd}}$ varies with the corresponding $\Sigma_{\mathrm{g}}$ among different sightlines to understand why some structures survive. We can see that there is no trend of $f_{\mathrm{Edd}}$ with $\Sigma_{\mathrm{g}}$ for $\Sigma_{\mathrm{g}} \lesssim 5 \, \Msolpc$ -- which corresponds to $\tau_{\mathrm{dust}} = 1$ (shown as the dashed line in Figure~\ref{fig:S4SigmafEdd}) -- following which it falls as $\Sigma_{\mathrm{g}}^{-1}$. This is in good agreement with simple back-of-the-envelope estimates of the Eddington ratio for a shell or clump of gas with surface density $\Sigma_{\mathrm{g}}$ \citep{Andrews_2011,Thompson_2015,Thompson_Krumholz_2016,Crocker_2018b,Blackstone_2023}. In the limit where the gas is optically thin in the UV and the gravity is dominated by the stellar component ($M_*$), $f_{\mathrm{Edd}}$ is given by 
\begin{equation}
    \label{eq:fEddOT}
    f_{\mathrm{Edd},\mathrm{OT}} = \frac{\kappa_{\mathrm{UV}}L_{\mathrm{UV}}}{4 \pi GM_*c} \sim 70 \,  \left(\frac{\Psi_{\mathrm{UV}}}{943 L_\sun M_{\sun}^{-1}}\right) \left(\frac{\kappa_{\mathrm{UV}}}{1000 \, \mathrm{cm}^2 \, \mathrm{g}^{-1}}\right),
\end{equation}
where we have used $\Psi_{\mathrm{UV}} = L_{\mathrm{UV}}/M_*$, where $\Psi_{\mathrm{UV}}$ is the total UV luminosity-per-unit mass that we have adopted for our simulations. We have expressed the above scaled by the values we have adopted in our simulations. In the single-scattering (SS) regime -- where the gas is optically thick in the UV yet optically thin in the IR -- the Eddington ratio for a shell/cloud of surface density $\Sigma_{\mathrm{g}}$ is given by \citep{Thompson_Krumholz_2016,Wibking_2018} 
\begin{equation}
    \label{eq:fEddSS}
    f_{\mathrm{Edd},\mathrm{SS}} = \frac{\Psi_{\mathrm{UV}}}{4 \pi Gc \Sigma_{\mathrm{g}}} \sim 1 \,  \left(\frac{\Psi_{\mathrm{UV}}}{943 L_\sun M_{\sun}^{-1}}\right) \left(\frac{\Sigma_{\mathrm{g}}}{500 \, M_\sun \, \mathrm{pc}^{-2}}\right)^{-1},
\end{equation}
which is $\propto \Sigma_{\mathrm{g}}^{-1}$. This expression yields the maximum surface density that can be super-Eddington ($f_{\mathrm{Edd},\mathrm{SS}} > 1$), the Eddington surface density, $\Sigma_{\mathrm{Edd}}$ given by 
\begin{equation}
    \Sigma_{\mathrm{Edd}} \sim 500 \, \Msolpc \left(\frac{\Psi_{\mathrm{UV}}}{943 L_\sun M_{\sun}^{-1}}\right).
\end{equation}
We plot this quantity with dotted lines in Figure~\ref{fig:S4SigmafEdd}; we also denote the point where $\tau_{\mathrm{dust},\mathrm{UV}} >1$ to compare with the expected transition from $f_{\mathrm{OT}} =$constant to $f_{\mathrm{SS}} \propto \Sigma_{\mathrm{g}}^{-1}$ from the above expressions. We can see that both expectations are well satisfied in the obtained trends for our simulations at later times. At earlier times, the trends look different because $f_{\mathrm{Edd},\mathrm{SS}}$ is dependent (linearly) on the fraction of mass in stars ($\epsilon_*$), since gravity has contributions from both the gas and stars, whereas the radiative force is linearly proportional only to the stellar mass -- this lowers $\Sigma_{\mathrm{Edd}}$ at earlier times when $\epsilon_*$ is less than one. Moreover, at earlier times, the radiation is absorbed very close to the sinks, and \citet{Menon_2023} have shown that this leads to cancellation of the forces from individual sinks \citep[see also][]{Socrates_2013}, such that the net radial force from the stellar population is much lower than the one obtained from a single source with the combined luminosity. These additional complexities become weaker at later times, resulting in the strong agreement.

These figures also reveal that the high column density sightlines ($\gtrsim \Sigma_{\mathrm{Edd}}$) are mostly sub-Eddington, at least globally. This does not necessarily mean that the gas in those sightlines is being accreted onto the sink particles. While this is certainly true for certain sightlines, it is possible that gas could still be locally super-Eddington close to the radiating sources, especially considering additional sub-dominant support from the thermal pressure of ionized gas, even if the column as a whole is sub-Eddington -- indeed the mass-weighted velocities as a function of $\Sigma_{\mathrm{g}}$ show that the radial velocities are slightly larger than zero, however, insufficient to escape the cloud ($v < v_{\mathrm{esc}}$). However, the dynamics of high-column density sightlines does reveal why the timescales for evacuating gas are longer for higher $\Sigmacloud$ -- there is a much larger fraction of gas with $\Sigma_{\mathrm{g}} > \Sigma_{\mathrm{Edd}}$ for these cases. The high-density gas in these sightlines is either consumed by star formation, accreted onto existing stars, and/or a small fraction of it could be locally driven outwards by feedback, all of which reduce the $\Sigma_{\mathrm{g}}$ in that sightline. This is unlike the behaviour at lower $\Sigmacloud$ where most of the sightlines are already super-Eddington while star formation is still ongoing.

\subsection{Properties of outflows}

\begin{figure}
    \centering
    \includegraphics[width=0.5\textwidth]{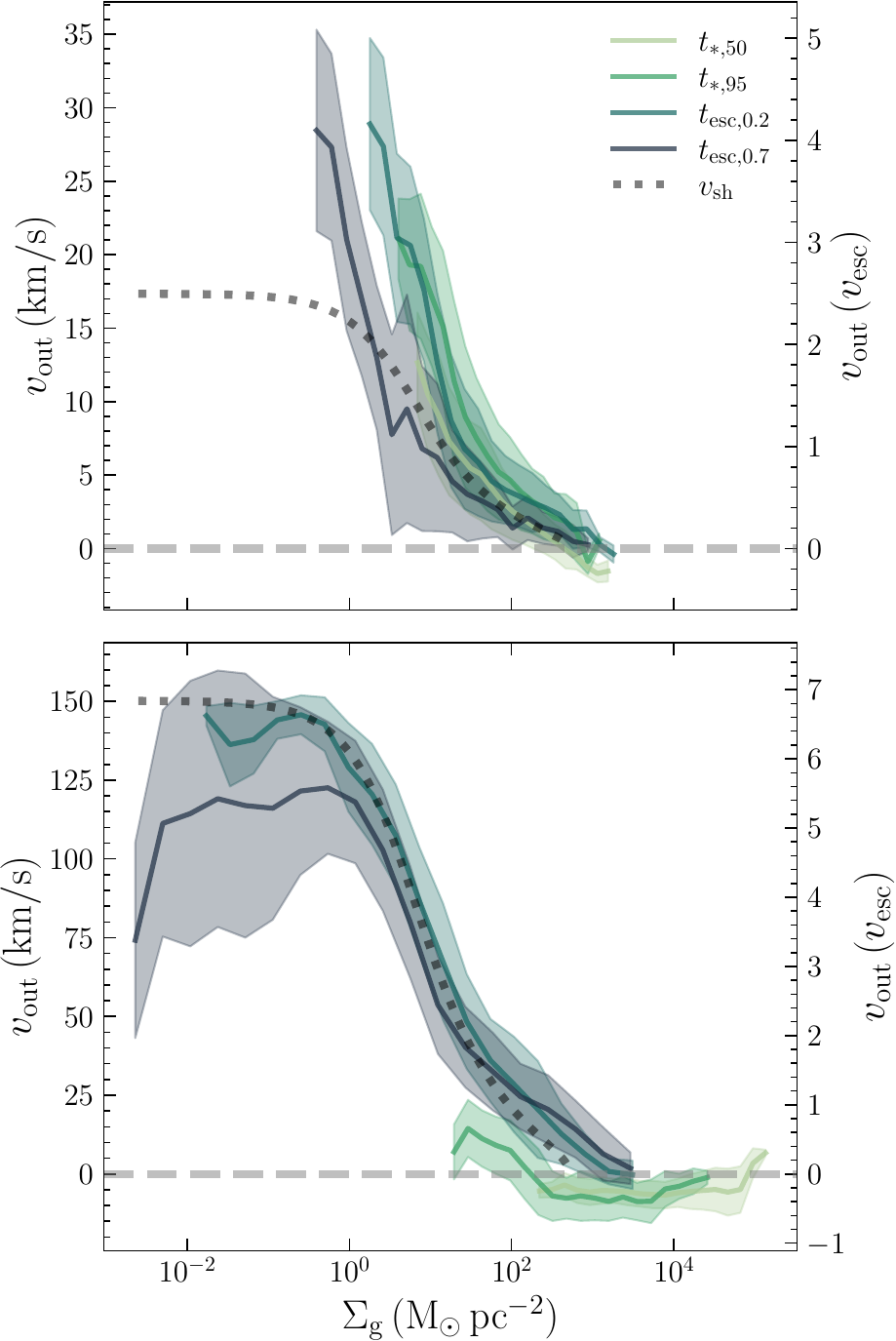}
    \caption{Mass-weighted average radial velocity in different sightlines ($v_{\mathrm{out}}$), binned by the corresponding gas surface densities ($\Sigma_{\mathrm{g}}$) at the characteristic times defined in Section~\ref{sec:evacuation} for $\Sigmacloud = 10^2 \, \Msolpc$ (top) and $10^4 \, \Msolpc$ respectively. Dotted gray lines indicate the analytical solution for a thin shell of surface density ($\Sigma_{\mathrm{g}}$) accelerated purely by radiation pressure on dusty gas (Equation~\ref{eq:shellevol}).}
    \label{fig:radVel}
\end{figure}

The existence of super-Eddington sightlines implies that gas is driven outward by radiative forces in these channels. We show this in Figure~\ref{fig:radVel} for $\Sigmacloud =  100 \Msolpc$ and $10^4 \Msolpc$ for the characteristic times. Specifically, Figure ~\ref{fig:radVel} shows the average mass-weighted radial velocity ($v_{\mathrm{out}}$) along sightlines binned by their corresponding $\Sigma_{\mathrm{g}}$. Comparing the panels, we can see that the lower $\Sigmacloud$ case already has the most sightlines with $v_{\mathrm{out}} >0$ by $t_{*,50}$, consistent with the rapid bulk photoevaporation scenario introduced in the previous section, whereas the higher range of $\Sigma_{\mathrm{g}}$ values for $\Sigmacloud =  10^4 \Msolpc$ ensures that outflows are launched only after most of the star formation has ceased. Lower-column density channels have faster outflow velocities, with the lowest columns accelerated to velocities several times the escape speed of the cloud ($\gtrsim 5 v_{\mathrm{esc}}$). We can also see that this dependence of $v_{\mathrm{out}}$ on $\Sigma_{\mathrm{g}}$ follows analytic estimates based on the scaling of the Eddington ratio with $\Sigma_{\mathrm{g}}$ for radiation pressure on dust \citep{Thompson_2015}. Since there is a linear anti-correlation of the Eddington ratio ($f_{\mathrm{Edd},\mathrm{SS}}$; see Equation~\ref{eq:fEddSS}) with $\Sigma_{\mathrm{g}}$ in the single-scattering regime ($\Sigma_{\mathrm{g}} \gtrsim 5 \, \Msolpc$), the acceleration experienced by the gas, and the resulting outflow velocities, would show a trend of decreasing $v_{\mathrm{out}}$ with increasing $\Sigma_{\mathrm{g}}$ in this regime; this dependence is shown in Figure~\ref{fig:radVel}. On the other hand, no such dependence is expected in the optically-thin regime ($\Sigma_{\mathrm{g}} \lesssim 5 \, \Msolpc$), where we expect optically-thin sightlines to be accelerated to a single characteristic velocity (see below).

To quantify this in further detail, we follow \citet{Thompson_2015} and estimate the dynamics of a spherical geometrically-thin dusty shell of surface density $\Sigma_{\mathrm{sh}}$ exposed to a radiation source. The radial evolution for the velocity of a thin shell accelerated by radiation pressure, $v_{\mathrm{sh}}$ can be written as 
\begin{equation}
    \label{eq:shellevol}
    v_{\mathrm{sh}} \frac{dv_{\mathrm{sh}}}{dr} = \frac{-GM_*}{r^2} + \frac{\Psi M_*}{4 \pi r^2c\Sigma_{\mathrm{sh}}} \left( 1 - e^{-\tau_\mathrm{UV}} + \tau_{\mathrm{IR}} \right),
\end{equation}
where $M_*$ is the stellar mass, which we take to dominate the gravitational potential, and where we have expressed its luminosity in terms of the luminosity-per-unit mass $\Psi$ that we adopt. If we make the additional assumption that the shell expands without entraining a significant fraction of its original mass, then the shell column density is $\propto r^{-2}$ so that $\Sigma_{\mathrm{sh}} = \Sigma_0 r_0^2/r^2$, where $r_0$ is the initial radius from which the shell is driven, and $\Sigma_0$ the column density at that radius. The right hand side of Equation~\ref{eq:shellevol} can now expressed purely as a function of $r$, and integrated to obtain $v_{\mathrm{sh}}$ for a given initial $r_0,\Sigma_0$. We obtain

\begin{equation} \label{eq:vshell}
\begin{split}
\frac{v_{\mathrm{sh}}^2(r,\Sigma_0)}{v_{\mathrm{esc}}^2 (r_0)} & = \frac{r_0}{r} -1 + f_{\mathrm{Edd},0} \left[ \frac{r}{r_0} -1 \right. \\
 & + e^{-\tau_{\mathrm{UV},0}} - \frac{r}{r_0} e^{-\tau_{\mathrm{UV},0} (r_0/r)^2} \\
 & + \sqrt{\pi \tau_{\rm UV,0}} \left(\mathrm{erf}(\sqrt{\tau_{\mathrm{UV},0}}) - \mathrm{erf}(\frac{r_0}{r}\sqrt{\tau_{\mathrm{UV},0}})  \right) \\ 
 & + \left. \tau_{\mathrm{IR},0}\left( 1 - \frac{r_0}{r} \right) \right],
\end{split}
\end{equation}
where $\tau_{\mathrm{UV},0} = \kappa_{\rm UV} \Sigma_0$ and $\tau_{\mathrm{IR},0} = \kappa_{IR} \Sigma_0$ are the initial UV and IR optical depths at $r_0$, $v_{\mathrm{esc}}(r_0) = \sqrt{2GM_*/r_0}$ is the escape velocity measured at $r_0$, and $f_{\mathrm{Edd},0} = \Psi/4 \pi Gc \Sigma_0$ is the single-scattering Eddington ratio (Equation~\ref{eq:fEddSS}) at $r_0$. $\kappa_{\mathrm{UV}}$ and $\kappa_{\mathrm{IR}}$ are the dust UV and IR opacities respectively; we set the former to $1000 \, \mathrm{cm}^2 \, \mathrm{g}^{-1}$ consistent with the FUV band value and the latter to $5 \, \mathrm{cm}^2 \, \mathrm{g}^{-1}$. While $\kappa_{\mathrm{IR}}$ is dependent on the dust temperature in our simulations, we use a constant value for simplicity; the adopted value is irrelevant as long as it is within the range of typical dust IR opacities \citep[$\lesssim 10 \, \mathrm{cm}^2 \, \mathrm{g}^{-1}$; ][]{Semenov_2003}, since the Eddington ratio for IR radiation pressure is less than unity for this range of values and therefore shouldn't affect this calculation \citep{Skinner_2015,Menon_2022b}. We can use Equation~\ref{eq:vshell} to compute the analytical expectation for the velocity of shells with different $\Sigma_{\mathrm{g}}$ accelerated by radiation pressure. Following along the lines of \citet{Raskutti_2017} we assume that the outflow is launched at the cloud radius ($r_0 = \Rcloud$), and the shell has reached the edge of the domain ($r = 2\Rcloud$) -- this choice implies that $\Sigma_{\mathrm{g}} = \Sigma_0/4$. Using these values in Equation~\ref{eq:vshell} allows us to obtain $v_{\mathrm{sh}} (\Sigma_{\mathrm{g}}$), which we overplot as the dashed lines in Figure~\ref{fig:radVel}. We can see that for $\Sigmacloud = 100 \, \Msolpc$, the agreement between the expected velocity from Equation~\ref{eq:vshell} and the simulations is poor, especially at low $\Sigma_{\mathrm{g}}$. This is not surprising, since photoionization is the dominant source of momentum in this regime \citep[e.g.,][]{Kim_2018}, which is not considered in the analytic approximation of Equation~\ref{eq:vshell}. Moreover, the assumption of a single point source of radiation is less appropriate for lower $\Sigmacloud$ since the distribution of star particles is more extended due to the weaker potential. However, for $\Sigmacloud = 10^4 \, \Msolpc$, for which this comparison is more appropriate, the agreement between the estimate from Equation~\ref{eq:vshell} and the simulation at late times (when the outflow is driven) is striking, as is evident from the lower panel of Figure~\ref{fig:radVel}. 

Equations~\ref{eq:shellevol} and \ref{eq:vshell} also demonstrate why radiation pressure is able to accelerate gas to $v_{\mathrm{out}}$ values of several times $v_{\mathrm{esc}}$. The radiation pressure when the dust is optically thick in the UV ($\tau_{\mathrm{UV}} >> 1$) is independent of $r$, whereas the gravitational force drops as $r^{-2}$. This implies that if the shell was super-Eddington at a given radius, the acceleration imposed on the shell is positive and increases with time, until the point where the shell becomes optically thin \citep{Elmegreen_1982,Thompson_2015}.

\begin{figure}
    \centering
    \includegraphics[width=0.48 \textwidth]{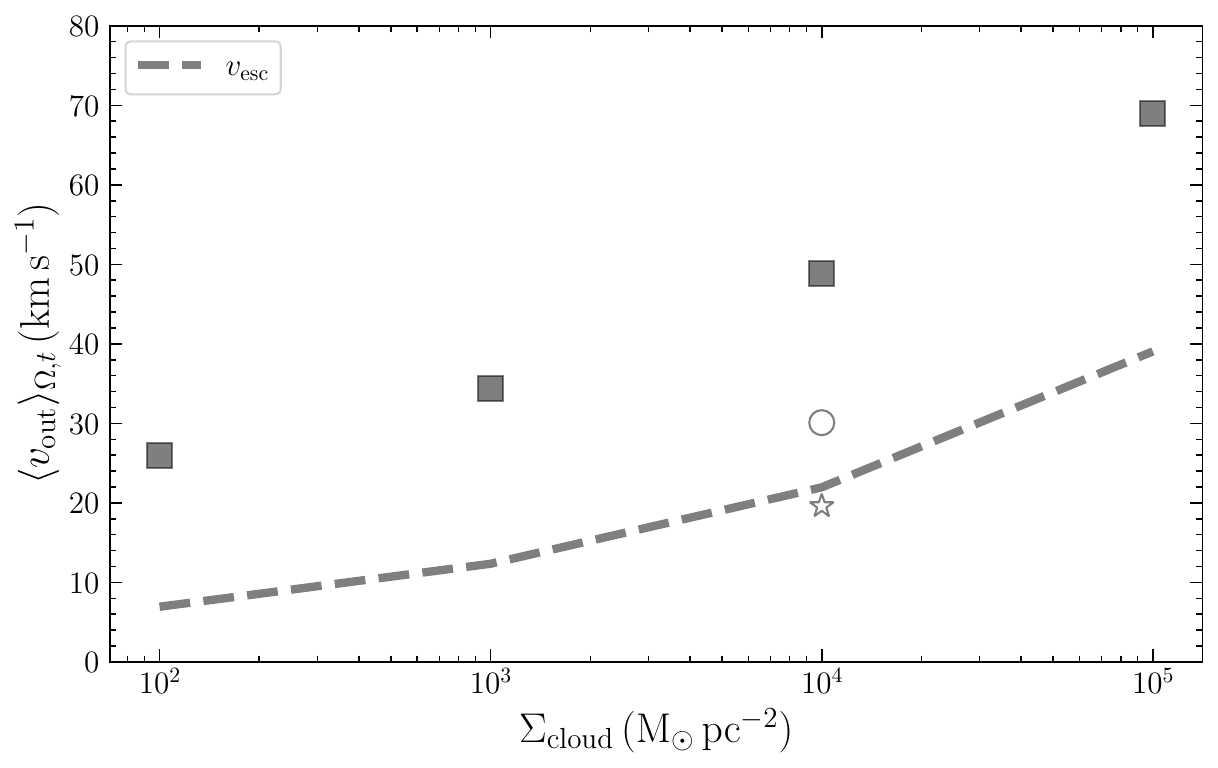}
    \caption{Time-averaged, mass-weighted outflow velocities for the simulations with different $\Sigmacloud$; the points with (unfilled) circle and star markers are the runs with $0.1 Z_{\mathrm{d},\sun}$ and $0.01 Z_{\mathrm{d},\sun}$ respectively. The dashed line indicates the escape velocity for a given $\Sigmacloud$. The magnitude of the outflow velocities is significantly higher at higher $\Sigmacloud$, and smaller for lower $Z_{\mathrm{d}}$ for a given $\Sigmacloud$. This potentially has implications on whether these outflows can facilitate LyC escape through the ISM of the host galaxies of the star clusters.}
    \label{fig:radVelaverage}
\end{figure}

While the outflow velocities are a few times the cloud escape speed ($v_{\mathrm{esc}}$) in both $\Sigmacloud$ cases, the much higher $v_{\mathrm{esc}}$ for higher $\Sigmacloud$ (Table~\ref{tab:Simulations}) implies that the \textit{absolute} outflow velocities are much higher -- up to 100 $\kms$ for the low-column sightlines. To compare the \textit{average} outflow speeds systematically with $\Sigmacloud$, in Figure~\ref{fig:radVelaverage} we plot $\langle v_{\mathrm{out}} \rangle_{\Omega,t}$, the outflow velocity averaged over all sightlines ($\Omega$) and times ($t$) for our different simulations. We can see the trend of increasing $\langle v_{\mathrm{out}} \rangle_{\Omega,t}$ with $\Sigmacloud$; inspection of the overplotted $v_{\mathrm{esc}}$ shows that $\langle v_{\mathrm{out}} \rangle_{\Omega,t} \sim 2$--$3 v_{\mathrm{esc}}$ for all $\Sigmacloud$. The exception are the runs with lower $Z_{\mathrm{d}}$, which have lower $\langle v_{\mathrm{out}} \rangle_{\Omega,t}$. This is because the Eddington ratio for optically thin gas is lower at lower $Z_{\mathrm{d}}$ (see Equation~\ref{eq:fEddOT}). In addition, the single-scattering regime operates at $\Sigma_{\mathrm{g}} \gtrsim 50 \, \Msolpc$ and $500 \, \Msolpc$ for $0.1 Z_{\mathrm{d},\sun}$ and $0.01 Z_{\mathrm{d},\sun}$ respectively; therefore the rapidly accelerating phase associated with this regime lasts over a smaller range in surface density, corresponding to a shorter duration in time at lower $Z_{\mathrm{d}}$. This results in the lower asymptotic velocities.

\subsection{Cumulative LyC escape}

\begin{figure*}
    \centering
    \includegraphics[width=\textwidth]{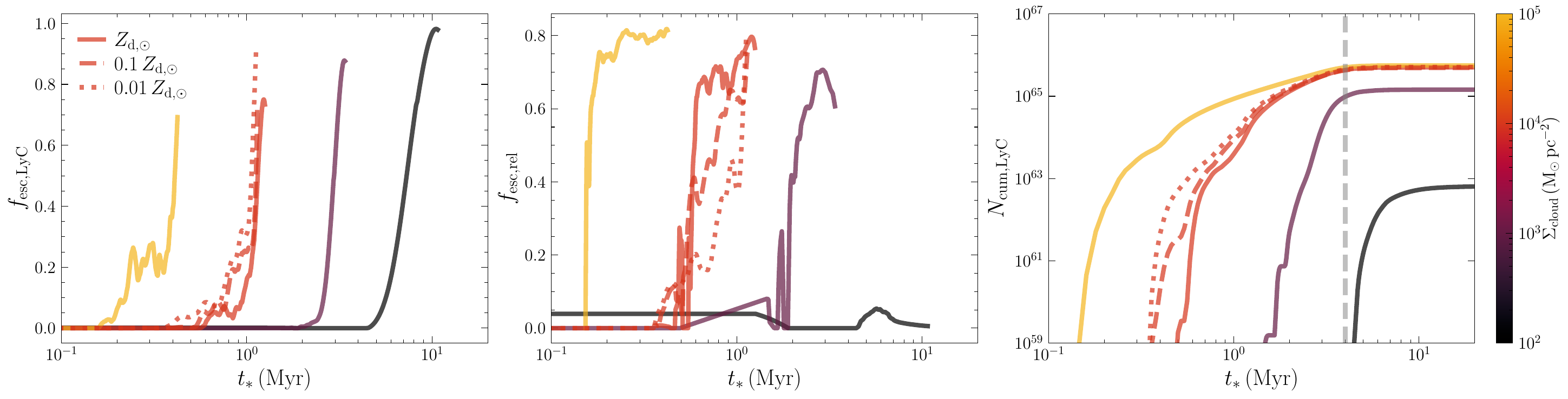}
    \caption{Left: Instantaneous value of $f_{\mathrm{esc},\mathrm{LyC}}$ shown in terms of absolute time in Myr since star formation started ($t_*$) for the different simulations. Middle: $f_{\mathrm{esc},\mathrm{rel}}$, the LyC Escape fraction relative to the escaping bolometric FUV luminosity from a stellar population (see Equation~\ref{eq:fesc_rel}). Right: The cumulative number of LyC photons that escape the cloud over the lifetime of the (instantaneous) burst (see Equation~\ref{eq:Ncum}). The dashed gray vertical line indicates the rough timescale ($\sim 4 \, \mathrm{Myr}$) when massive stars die and the LyC emission rate starts to drop. Denser clouds are able to leak LyC photons well before this timescale due to their rapid evolution governed by their much shorter $t_{\rm ff, cloud}$. This leads to significantly higher values of $N_{\mathrm{cum},\mathrm{LyC}}$ over the lifetime of the stellar population.}
    \label{fig:CumulativeEscape}
\end{figure*}

In the preceding sections we have shown that the higher columns in higher $\Sigmacloud$ cases make it more difficult for feedback to open up channels for LyC escape. In these cases, more dynamical times are required to achieve high $f_{\mathrm{esc},\mathrm{LyC}}$. However, the dynamical timescales for higher $\Sigmacloud$ are also significantly shorter (see Table~\ref{tab:Simulations}), which implies that the phase of clearing out feedback-driven channels of LyC escape occurs when the stellar population is younger  -- so much so that this reverses the trend for the timescales for LyC escape when viewed as a function of the stellar population age. For instance, even though the $\Sigmacloud = 10^4 \, \Msolpc$ requires $\sim 9 t_{\mathrm{ff},\mathrm{cloud}}$ to achieve $f_{\mathrm{esc},\mathrm{LyC}} \gtrsim 50\%$ and $\Sigmacloud = 100 \, \Msolpc$ only requires $\sim 1.5 t_{\mathrm{ff},\mathrm{cloud}}$ (see Figure~\ref{fig:LyCEscapeFraction}), the value of $t_{\mathrm{ff},\mathrm{cloud}}$ is higher in the latter by a factor $\sim 30$ (Table~\ref{tab:Simulations}) -- much more than the factor $\sim 6$ longer period of cloud dispersal and subsequent LyC escape for the denser cloud. This fact results in the trend shown in Figure~\ref{fig:CumulativeEscape} (left panel) for $f_{\mathrm{esc},\mathrm{LyC}}$ in terms of absolute time. In the context of reionization, it is indeed this trend that is relevant, since the LyC emission rate of a stellar population is very strongly dependent on age; this is because as massive stars die on timescales $\gtrsim 3 \, \mathrm{Myr}$, the sources of LyC photons are lost, and the emission rate drops significantly. 

To demonstrate this, we show two quantities that take into account the time dependence of LyC emission in a stellar population in Figure~\ref{fig:CumulativeEscape}. In the middle panel, we show the relative escape fraction, $f_{\mathrm{esc},\mathrm{rel}}$, given by the relation
\begin{equation}
    \label{eq:fesc_rel}
    f_{\mathrm{esc},\mathrm{rel}} = \left( \frac{f_{\mathrm{esc},\mathrm{LyC}}}{f_{\mathrm{esc},\mathrm{FUV}}} \right) \left( \frac{L_{\mathrm{LyC},*}}{L_{\mathrm{FUV},*}} \right),
\end{equation}
where $L_{\mathrm{LyC},*}$ and $L_{\mathrm{FUV},*}$ are the intrinsic luminosities of the stellar population in the LyC and FUV bands, which depends purely on age (and IMF) for the case of a single burst. This is identical to the commonly used definition in the observational literature for galaxies, without the absorption factor due to the IGM along the line of sight to the galaxy (i.e. $\exp({\tau_{\mathrm{HI},\mathrm{IGM}}})$). We can see that $f_{\mathrm{esc},\mathrm{rel}}$ for high $\Sigmacloud$ reaches higher values, and also that such clouds do so at much earlier times. In fact, for the lowest $\Sigmacloud$ case, $f_{\mathrm{esc},\mathrm{rel}}$ barely reaches values of $5 \%$ in the entirety of its evolution; the highest $\Sigmacloud$ on the other hand reaches values $\gtrsim 80\%$ in timescales of $\lesssim 1 \, \mathrm{Myr}$. This trend is primarily because high $f_{\mathrm{esc},\mathrm{LyC}}$ is achieved at timescales much greater (smaller) than 4\,Myr for low (high) $\Sigmacloud$. 

\begin{figure*}
    \centering
    \includegraphics[width=0.8\textwidth]{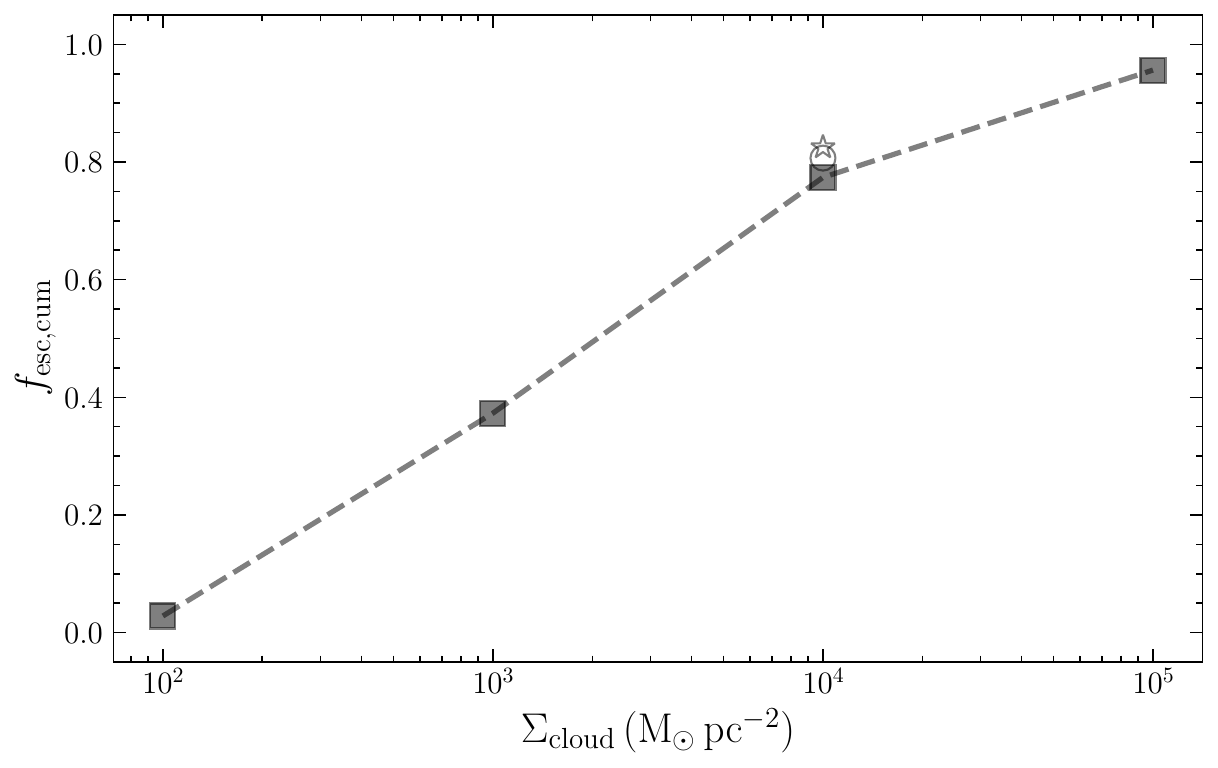}
    \caption{Cumulative LyC escape fractions ($f_{\mathrm{esc},\mathrm{cum}}$) averaged over the lifetime of the stellar population (see Equation~\ref{eq:fEscCum}) for the different simulations in our study. Filled markers are at $Z_{\mathrm{d}} = Z_{\mathrm{d},\sun}$; the unfilled circle and star markers indicate the $0.1 Z_{\mathrm{d},\sun}$ and $0.01 Z_{\mathrm{d},\sun}$ respectively.} 
    \label{fig:fEscCum_Sigma}
\end{figure*}

The effect of the relative timescales for the cloud to disperse and the stellar population to evolve are even more dramatic when comparing the cumulative number of LyC photons that escape the star cluster. This depends on the instantaneous emission rate of LyC photons by the cluster ($\dot{n}_{\mathrm{LyC}}$), instantaneous LyC escape fraction ($f_{\mathrm{esc},\mathrm{LyC}}$) and the duration for which LyC escape occurs, and is given by the relation
\begin{equation}
    \label{eq:Ncum}
    N_{\mathrm{cum},\mathrm{LyC}} = \int_{0}^{t_*} \dot{n}_{\mathrm{LyC}} f_{\mathrm{esc},\mathrm{LyC}} dt,
\end{equation}
where $t_*$ is the age of the young star cluster. We show how $N_{\mathrm{cum},\mathrm{LyC}}$ depends on $\Sigmacloud$ in the right panel of Figure~\ref{fig:CumulativeEscape}. The temporal trends are similar across $\Sigmacloud$ with $N_{\mathrm{cum},\mathrm{LyC}}$ increasing as $f_{\mathrm{esc},\mathrm{LyC}}$ increases, followed by a saturation for $t_* \gtrsim 4 \, \mathrm{Myr}$. This saturation occurs because the emission rate of LyC photons drops significantly beyond these times. We can see that for higher $\Sigmacloud$, the duration of being in a state of high $f_{\mathrm{esc},\mathrm{LyC}}$ at young stellar ages ($t_* \lesssim 4 \, \mathrm{Myr}$) is significantly higher. This results in total $N_{\mathrm{cum},\mathrm{LyC}}$ over the lifetime of the young stellar population that is
orders of magnitude higher than for lower $\Sigmacloud$. We can also use $N_{\mathrm{cum},\mathrm{LyC}}$ to define a cumulative escape fraction, which is the effective fraction of LyC photons emitted by the stellar population over its lifetime that escape; i.e. 
\begin{equation}
    \label{eq:fEscCum}
    f_{\mathrm{esc},\mathrm{cum}} = \frac{N_{\mathrm{cum},\mathrm{LyC}}(t_{\mathrm{life}})}{\int_{0}^{t_{\mathrm{life}}} \dot{n}_{\mathrm{LyC}} dt},
\end{equation}
where we take $t_{\mathrm{life}} = 20 \, \mathrm{Myr}$, since the cumulative LyC emission is well saturated by this point. We show how $f_{\mathrm{esc},\mathrm{cum}}$ varies across our simulations in Figure~\ref{fig:fEscCum_Sigma}. The strong trend of increasing $f_{\mathrm{esc},\mathrm{cum}}$ with $\Sigmacloud$ is evident; the effect of the dust-to-gas ratio is relatively quite minor.

\section{Discussion}
\label{sec:discussion}

\begin{table*}
\caption{Summary of relevant quantities measured across our simulations.}
\centering
\label{tab:Summary}
\begin{threeparttable}
\begin{tabular}{c c c c c c c c c}
\toprule
\multicolumn{1}{c}{$\Sigmacloud$}& \multicolumn{1}{c}{$Z_{\mathrm{d}}$}& \multicolumn{1}{c}{$\epsilon_*$}& \multicolumn{1}{c}{$t_{*,50}$}&  \multicolumn{1}{c}{$t_{*,95}$}& \multicolumn{1}{c}{$t_{\mathrm{esc},0.2}$}& \multicolumn{1}{c}{$t_{\mathrm{esc},0.7}$}& \multicolumn{1}{c}{$\langle v_{\mathrm{out}} \rangle_{\Omega,t}$}& \multicolumn{1}{c}{$f_{\mathrm{esc},\mathrm{cum}}$} \\
$\Msolpc$& $Z_{\mathrm{d},\sun}$ & & $\mathrm{Myr} (t_{\mathrm{ff}})$ & $\mathrm{Myr} (t_{\mathrm{ff}})$& $\mathrm{Myr} (t_{\mathrm{ff}})$& $\mathrm{Myr} (t_{\mathrm{ff}})$& [km/s]& \\
\midrule
$10^2$ &1 &$0.12$ & 3.4 (0.86) &5.3(1.3) &6.2 (1.6) &8 (2) &26 &0.03 \\
$10^3$ &1 &$0.64$ &1 (1.44) &1.7 (2.5) &2.7 (3.9) & 3.1 (4.4) &35 &0.37\\
$10^4$ &1 &$0.88$ &0.2 (1.47) &0.54 (4.3) &1 (8.5) &1.2 (9.6) &49 &0.77\\
$10^5$ &1 &$0.93$ &0.03 (1.47) &0.1 (4.6) &0.35 (16) &0.42 (19) &70 &0.96 \\
$10^4$ &0.1 &$0.86$ &0.18 (1.44) &0.51  (4.1) &0.9 (7.3) &1.2 (9.2)&31 &0.81\\
$10^4$ &0.01 &$0.87$ &0.18 (1.44) &0.52 (4.1) &0.8 (6.5) &1.1 (8.7)&20 &0.84\\
\bottomrule
\end{tabular}
\begin{tablenotes}
\small
\item \textbf{Notes}: Columns in order indicate - $\Sigma_{\mathrm{cloud}}$: mass surface density of the cloud; $Z_{\mathrm{d}}$: dust-to-gas ratio in units of the corresponding value in the solar neighborhood ($Z_{\mathrm{d},\sun}$); $\epsilon_{*,f}$: final saturated value of the (integrated) star formation efficiency given by $\epsilon_{*,f} = M_{*,f}/\Mcloud$; $t_{*,50}$, $t_{*,95}$ : timescale to form 50\% and 95\% of the final stellar mass respectively in Myr and in units of $t_{\mathrm{ff},\mathrm{cloud}}$ in brackets; $t_{\mathrm{esc},0.2}$, $t_{\mathrm{esc},0.2}$: same, but for the times to achieve a LyC escape fraction of 20\% and 70\% respectively; $\langle v_{\mathrm{out}} \rangle_{\Omega,t}$: Time-averaged outflow velocity; $f_{\mathrm{esc},\mathrm{cum}}$: Luminosity-weighted escape fraction averaged over the lifetime of the stellar population (Equation~\ref{eq:fEscCum}). 
\end{tablenotes}
\end{threeparttable}
\end{table*}

\subsection{Can cloud-scale LyC leaking photons also escape the galaxy?}
\label{sec:cloud_to_galaxy}

Our results imply that highly dense and compact young star clusters permit the escape of a significantly larger fraction of their intrinsic LyC emission due to their short dynamical timescales, while also emitting a larger number of LyC photons due to their higher star formation efficiencies; see Figures~\ref{fig:CumulativeEscape} and ~\ref{fig:fEscCum_Sigma}, and Table~\ref{tab:Summary}. In other words, the key parameters determining (cumulative) LyC escape are the dynamical timescale ($t_{\mathrm{ff}} \propto n_{\mathrm{cloud}}^{-1/2}$) and the star formation efficiency ($\epsilon_* \propto \Sigmacloud$) at the cloud-scale. The former permits higher $f_{\mathrm{esc},\mathrm{LyC}}$ while $\dot{n}_{\mathrm{LyC}}$ is high, and the latter promotes higher $\dot{n}_{\mathrm{LyC}}$ for a given initial $\Mcloud$. However, these findings are in the context of the cloud-scale LyC escape in an isolated cloud numerical setup. In order for LyC photons escaping clusters to contribute to cosmic reionization, the LyC photons must then still be able to escape through the turbulent ISM surrounding the dense star-forming clumps. However, high-resolution cosmological simulations have established that LyC escape in galaxies is determined primarily by escape at the scales of GMCs \citep{Paardekooper_2015,Ma_2015,Trebitsch_2017,Kimm_2017}. Therefore, our results already provide insight on the mapping between local ISM properties and galaxy escape fractions. Even so, we caution that the $f_{\mathrm{esc},\mathrm{LyC}}$ values we report should be interpreted as upper limits when thinking about the fraction of LyC photons that escape the host galaxy, and it is the trends with cloud properties ($t_{\mathrm{ff}}$, $\Sigmacloud$) and the physical insights we built (i.e. bursty star formation and radiation-driven outflows) that we would highlight. 

There are two primary considerations in using the results presented here to keep in mind while extrapolating our cloud-scale $f_{\mathrm{esc},\mathrm{LyC}}$ trends to obtain implications on galaxy scales: i) whether high escape fractions can still be achieved if GMCs accrete from their surroundings and ii) whether radiative feedback can also carve pathways for LyC escape from the galaxy and not just in GMCs. 

\subsubsection{Accretion onto GMCs}

GMCs are highly dynamic objects that continually accrete gas from their surroundings, and simultaneously compete with feedback from star formation within them \citep[e.g.,][]{Jeffreson_2024}. In the context of the results presented in this work, these accreting flows would act to increase the gas columns in sightlines around the young stellar population -- i.e. they would affect the PDFs shown in Figure~\ref{fig:SigmaEvolution} -- and thereby lower $f_{\mathrm{esc},\mathrm{LyC}}$. However, we speculate that there are likely two favorable conditions that apply, which would limit the impact of accretion. The first is the expectation that the accretion flow would be highly filamentary -- akin to observed hub-filament systems in the Milky Way \citep{Hacar_2023}. While these flows would be important in the context of star formation, they would only occupy a small fraction of solid angles -- like blinders on a race horse -- and preferentially sightlines with higher columns. Inspection of Figure~\ref{fig:S4SigmafEdd} reveals that LyC escape is facilitated by the ejection of low-column sightlines anyways, and as long as additional accretion does not occur along those sightlines, it shouldn't affect $f_{\mathrm{esc},\mathrm{LyC}}$ strongly. Secondly, how quickly accretion can fill sightlines would depend on the relative timescales for gas consumption due to star formation/feedback-driven ejection and the accretion rate. The smaller spatial and thereby, faster temporal, scales at which the former occur suggest that the visibility of the cluster shouldn't be strongly affected. Indeed observations in the local Universe do reveal that star clusters are visible in the optical/UV on short timescales after formation \citep{Chevance_23} -- although these may not be suitable analogs for higher density clouds that presumable lie in higher density ISM conditions. Therefore, systematic studies of the impact of accretion on cloud-scale properties, and its interaction with feedback, are required to verify these expectations.

\subsubsection{LyC escape through the ISM}

An additional question is whether radiative feedback, which we show carves pathways for LyC escape at the cloud-scale can do so at the galactic scale. This depends on the ISM conditions outside the cloud. For a turbulent ISM, it is reasonable to assume that for the surrounding medium this can be represented as a distribution of column densities (sightlines), and some fraction of these are super-Eddington \citep{Thompson_Krumholz_2016}. These sightlines would continue to be accelerated by photoionization and radiation pressure at low and high $\Sigma$ respectively. Whether these sightlines can continue to be driven to carve density-bounded channels in the galaxy depends on how fast gas is driven outward with respect to the circular/escape velocity of the galaxy, and the timescales over which this occurs with respect to the lifetimes of massive stars. We note that these two conditions make the case for the role of dense star clusters stronger. Outflow velocities are much higher for $\Sigmacloud \gtrsim 10^4 \, \Msolpc$, $\gtrsim 50 \, \kms$ on average (Figure~\ref{fig:radVelaverage}) and even higher along sightlines with lower columns (Figure~\ref{fig:radVel}) -- comparable in magnitude to circular velocities in compact dwarf galaxies \citep[$M_* \sim 60 \, \kms$;][]{Simons_2015}. More importantly, the rapid dynamical evolution in these clouds implies that there is an important epoch where radiatively driven acceleration of super-Eddington channels can occur before the stellar population ages, the SED reddens, and supernovae take over -- a primary limitation in launching radiation-driven outflows of sub-regions in local star-forming galaxies \citep{Murray_2011,Blackstone_2023}. There is scope for using our insights along with semi-analytic models for radiation-driven winds in a galactic potential \citep[e.g.,][]{Thompson_2015} to quantify this possibility in future work. Idealized numerical simulations have demonstrated that radiative feedback can evacuate channels for LyC escape in the ISM before the onset of supernova feedback \citep[e.g.,][]{Kakiichi_2021}.

An additional point to consider in the high $\Sigmacloud$ regime is that these clouds likely form in starburst regions -- for instance, the compact central regions of dwarf galaxies undergoing mergers or inflows, and/or massive galaxies at high redshift. This implies that there would be effects of feedback from other young star clusters simultaneously acting on the ISM in which clouds are embedded. Combined radiative feedback from clustered star formation (and possibly AGN) can drive stronger outflows and make it easier to carve optically thin channels \citep{Fielding_2018,Orr_2022}. Moreover, supernovae from nearby clusters would carve large-scale holes in these regions, which would imply that the \textit{only} bottleneck for LyC escape is at the cloud-scale in such conditions. This supernovae-assisted high LyC escape in starburst regions has been identified in cosmological simulations \citep{Trebitsch_2017,Rosdahl_2022,Yeh_2023} and observations \citep{Micheva_2018,Martin_2024}. 

\subsection{LyC leakage from dense compact star clusters: empirical evidence}

\label{sec:empirical}

We note that there is some (indirect) empirical evidence for the potentially important role of dense compact star clusters in LyC leakage through the scenario we discussed in this study. For instance, LyC leakers in the nearby Universe ($z \lesssim 0.5$) have been found to be quite compact ($< 1$ kpc), and show signs of concentrated star formation \citep[high $\Sigma_{\mathrm{SFR}}$;][]{Jaskot_2013,Nakajima_2014,Borthakur_2014,Schaerer_2016,Izotov_2016,Izotov_2018a,Izotov_2018b}, with even signs of increasing $f_{\mathrm{esc}}$ with $\Sigma_{\mathrm{SFR}}$ \citep{Naidu_2020,Flury_2022b,Jaskot_2024}. They also seem to be characterized by nebular line emission with relatively high contribution from high-ionization lines -- such as a high [\ion{O}{3}]/[\ion{O}{2}] ratio, among others, and show steep UV spectral slopes with relatively low nebular contamination \citep{Izotov_2017,Izotov_2018a}. These can be interpreted to be consistent with nebular emission along density-bounded channels around very young ($\lesssim 3$--4 Myr) stellar populations \citep{Jaskot_2013,Nakajima_2014}. The results of our study seem to suggest the correlation with these spectral features is likely directly related to the compact nature of young super star clusters in these galaxies, whose short dynamical timescales and strong radiative feedback enable LyC leakage while the cluster is still very young. Indeed, nebular signatures seem to suggest very young ages ($\sim 2$--3 Myr) and ionization parameters reminiscent of super star clusters ($\log \mathcal{U} \sim -2$) in local LyC leakers \citep{Izotov_2018a} and $z\sim 2$ Lyman-$\alpha$ emitters \citep[e.g.,][]{Naidu_2022}. The combination of compact star formation with weak gravitational potential wells have also been found to be characteristic of Ly-$\alpha$ emitters at $z\sim 2$ \citep{Pucha_2022}. There are also tentative signs of radiatively-driven outflows from highly dense SSCs in local (low-mass) green-pea galaxies \citep{Komarova21a}. 

More direct evidence for localized LyC escape from pc-scale compact star formation is found in lensed high-redshift galaxies, where lensing has enabled characterization of the physical properties at scales $< 100 \, \mathrm{pc}$ \citep{Bian_2017,Rivera_2019,Vanzella_2020,Kim_2023_escape,Pascale_2023, Vanzella_2023,Messa_2024b}. For instance, in the Sunburst LyC leaker ($z = 2.37$) with an escape fraction $\gtrsim 50\%$, the LyC leakage is highly localized in a region tens of parsecs in scale that dominates the UV output of the galaxy. The inferred stellar surface densities are very high in the LyC leaking region ($\Sigma_* \sim 10^4 \, \Msolpc$) with their spectra closely resembling SSCs in nearby galaxies \citep{Vanzella_2020}. In addition, this region exhibits other indirect signs of density-bounded ionized regions in this localized patch -- such as the presence of high ionization nebular lines and steep UV spectral slopes \citep{Messa_2024b} -- which are partly averaged out over the entire galaxy \citep{Kim_2023_escape}. This raises the intriguing possibility that LyC leakage is highly anisotropic and dominated by localized patches, and observationally determined escape fractions -- when the galaxy is unresolved -- are highly dependent on whether these privileged sightlines are accessible in the observations \citep{Nakajima_2020}.  

\subsection{Implications for reionization by star-forming galaxies}
\label{sec:reionization}

Our results imply that galaxies that can form a significant population of young, compact, dense star clusters could be proficient LyC leakers. The sensitive dependence on the density of the star cluster (Figure~\ref{fig:fEscCum_Sigma}) for attaining high (localized) $f_{\rm esc}$ provides a natural explanation for why $f_{\rm esc}$ is typically low in large samples of star-forming galaxies, especially in the local Universe. For instance, stacked samples of intermediate redshift ($z \sim 2-4$) star-forming galaxies provide upper limits of $f_{\rm esc} \lesssim 10\%$ \citep{Japelj_2017,Wang_2023,Jung_2024} -- much lower than the average $f_{\rm esc} \sim 20\%$ required to match reionization timing constraints \citep[e.g.,][]{Robertson_2013}. However, there are also a small sample of galaxies that show highly elevated $f_{\rm esc} \gtrsim 40\%$ \citep{Vanzella_2012,Shapley_2016,Izotov_2018a,Rivera_2019,Vanzella_2022,Wang_2023,Marques-Chaves_2024}. Apart from line-of-sight anisotropies, it is possible that these galaxies have undergone a recent surge in the formation of these highly dense compact star clusters in high ISM-pressure environments, which may be facilitated via mergers and/or the triggering of central starbursts. The Sunburst LyC leaker does indeed show the presence of a recent bursty event that likely formed the population of young bound star clusters that show evidence of localized LyC leakage \citep{Vanzella_2022,Kim_2023_escape}. This is also consistent with the findings from cosmological simulations on a correlation of high $f_{\rm esc}$ with recent ($\sim$ few Myr) bursts of star formation \citep{Rosdahl_2022}. Therefore, the low $f_{\rm esc}$ in stacks might simply resemble the small fraction of galaxies that happen to have undergone a recent surge of efficient localized bursts of star formation. Along similar lines, the very low LyC escape fractions in star-forming galaxies of the local Universe may simply reflect the relatively lower ISM-pressures ($P \sim 10^4 \, \mathrm{K} \, \mathrm{cm}^{-3}$) encountered, which are not conducive to the formation of dense stellar systems. 

In the early Universe ($z >6$), cosmological evolution of the mean baryon density and sizes ensure that galaxies are compact and have high average densities and short dynamical timescales \citep{FaucherGiguere_2018}. Indeed, JWST has revealed that galaxies at this epoch tend to be compact ($< 0.5 \, \rm pc$) and can achieve stellar surface densities $\Sigma_* \gtrsim 10^4 \, \Msolpc$ over the whole galaxy \citep{Baggen_2023,Casey_2023,Morishita_2023}. Such high densities would enable the prodigious formation of several compact clusters with $\Sigma_*$ (at least) of this order, suggesting that this extreme mode of star formation might be relatively commonplace at these redshifts. This is consistent with semi-analytic models that have empirically constrained that most ($\sim 50-60\%$) star formation at this epoch would be distributed in massive bound clusters \citep{Belokurov_2023}.  Lensed fields have also revealed several young star clusters that achieve these high $\Sigma_*$ values at these redshifts \citep{Adamo_2024,Mowla_2024,Messa_2024,Fujimoto_2024,Messa_2024b} consistent with these expectations. If this is true -- and taking our results at face value -- we would expect that LyC escape fractions from these compact galaxies would be systematically higher, and this would be the case when the stellar populations are still young. JWST has provided clues that this may be occurring with the detection of several galaxies that are characterized by high ionizing efficiencies indicative of very young ages and low dust attenuation \citep[e.g.,][]{Atek_2024,Simmonds_2024}, suggesting that the stellar populations have very quickly cleared out the surrounding gas. In addition, Lyman-$\alpha$ emitting galaxies at extremely high redshift ($z > 10$) seem to be characterized by very high $f_{\rm esc} \sim 80\%$ \citep{Curtis-Lake_2023,Tachella_2023,Hainline_2024}, consistent with expectations from our results when considering that these galaxies would have their star formation primarily distibuted in highly compact dense star clusters \citep[e.g.,][]{Dekel_2023}. These considerations make the case that star-forming galaxies are likely more than capable of reionizing the Universe on their own. In fact, the high ionizing production efficiencies and escape fractions may possibly even lead to a "photon budget crisis" wherein the Universe is reionized too early \citep{Munoz_2024}. There is scope to explore these implications by adopting our cloud-scale escape fractions as subgrid prescriptions in semi-analytic galaxy formation models and/or numerical simulations of cosmological volumes. 

\subsection{Comparison to other numerical simulations}

There have been earlier numerical simulations in the literature that made predictions for the escape fraction and its dependence on cloud properties. A comparison to these is not straightforward in practice, since the numerical setups, radiative feedback implementations, physics included and subgrid sink particle algorithms vary amongst the simulations. Therefore we do not attempt to make a one-to-one comparison of escape fractions obtained across the simulations. However, it is still useful to assess whether there is agreement with the trends of escape fractions with cloud properties. For instance, our study suggests that the key physical parameters determining the cumulative LyC escape are $t_{\mathrm{ff}}$ and $\epsilon_*$, which depend on $n_{\mathrm{cloud}}$ and $\Sigmacloud$ respectively. This is in agreement with the findings of \citet{Dale_2012} who found that dense, compact clouds with short $t_{\mathrm{ff}} \lesssim 3 \, \mathrm{Myr}$ have $f_{\mathrm{esc},\mathrm{cum}} > 0.8$, whereas more massive clouds (for a given size) with longer $t_{\mathrm{ff}}$ have much lower $f_{\mathrm{esc},\mathrm{cum}}$. \citet{Kim_2019_escape} also find a trend of increasing $f_{\mathrm{esc},\mathrm{cum}}$ for more compact clouds with shorter $t_{\mathrm{ff}}$. However, we note that Figure 6 in their paper seems to indicate that $f_{\mathrm{esc},\mathrm{cum}}$ decreases with $\Sigmacloud$ for a given cloud mass; we  stress that this trend only holds for $\Sigmacloud \lesssim 100 \, \Msolpc$ which is lower than the range we probe; for $\Sigmacloud \gtrsim 100 \, \Msolpc$, $f_{\mathrm{esc},\mathrm{cum}}$ shows signs of increasing with $\Sigmacloud$ consistent with our findings. \citet{He_2020} also report a clear trend of increasing $f_{\mathrm{esc},\mathrm{cum}}$ with cloud density; they also show that the total number of ionizing photons escaping the cloud (per unit mass) increases with $\Mcloud$ for compact clouds due to their higher $\Sigmacloud$ (as $\epsilon_*$ is higher). On the other hand, \citet{Kimm_2022} report that the escape fractions for denser and more efficiently star-forming clouds are lower than diffuse clouds -- which is in disagreement with the trend we presented in this work. They attribute their trend to the longer timescales for photoevaporating the cloud and opening up optically-thin channels at higher $\Sigmacloud$. While we too find that higher $\Sigmacloud$ requires more dynamical timescales for LyC photons to start escaping (Figure~\ref{fig:Timescales}), the factor by which this timescale is longer is smaller than the factor by which the dynamical timescales are shorter at higher $\Sigmacloud$, unlike what \citet{Kimm_2022} find, indicating that we need to identify possible reasons for these differences. Before doing so, we note that in \citet{Kimm_2022}, the difference in escape fraction between their denser and diffuse cloud\footnote{Our reference to the denser and diffuse clouds correspond to runs $\texttt{SM6D}\_\texttt{sZ002}$ and $\texttt{SM6}\_\texttt{sZ002}$ respectively in their paper.} -- which vary in $t_{\mathrm{ff}}$ by a factor of $\sim 2$ -- is only 10\%, and therefore shows a relatively weak difference. Regardless, we note that differences can arise due to different treatments for feedback in their study. For instance, \citet{Kimm_2022} assume that most dust (99\%) is destroyed in \ion{H}{2} regions; this would lower the effect of radiation pressure on dust which we find are important to expand the \ion{H}{2} regions in denser clouds, and drive their subsequent photoevaporation. Conversely, for the more diffuse cloud where thermal pressure dominates over radiation pressure, lower dust abundance in ionized gas can permit LyC photons to photoionize material at larger radii. In other words, the impact of destroying dust in ionized gas could work in opposite directions in the regimes where photoionization and radiation pressure dominates, where it aids and hampers LyC escape respectively. \citet{Kimm_2022} also include supernovae, which can help disrupt the more diffuse cloud with longer $t_{\mathrm{ff}}$ earlier. In addition, the $M_1$ numerical implementation for radiative feedback in their study, and associated flux cancellations in optically-thin media intrinsic to the approach \citep[see, for instance, Figure 13 in][]{Menon_2022}, could also lower the ability for radiative feedback to evacuate its surroundings. That being said, their higher dynamic range in numerical resolution and star-by-star Poisson sampling approach could also capture more localized effects of LyC absorption -- although \citet{He_2020}, who also have these advantages agree with our trends. Any of these factors, or a combination thereof could likely explain the differences in trends, motivating a systematic comparison between different numerical setups and physics implementations and their corresponding impacts on LyC escape.

\label{sec:simulations_literature}

\subsection{Caveats}
\label{sec:caveats}

We have made several choices in our modelling approach, some of which may have implications for our results; we briefly discuss some of them here. 

\subsubsection{Binary evolution and radiative outputs}
In our simulations, we have used the LyC and FUV radiative outputs for stellar populations based on the \texttt{SLUG} stellar population synthesis code, which uses the \texttt{MIST} stellar evolution models \citep{Choi_2016}. This model does not account for the presence of massive star binaries and associated physics, or the higher stellar surface temperatures expected at lower metallicities, both of which can increase the intrinsic LyC emission and the timescales up to which LyC photons are emitted. This could possibly have implications on the relative contribution of $\Sigmacloud \lesssim 10^3 \, \Msolpc$ clouds to LyC escape, since we find that the primary reason for their lower $f_{\mathrm{esc},\mathrm{cum}}$ is that their intrinsic emission drops by the time they disperse their clouds -- a longer timescale for LyC production might reduce this effect. Indeed \citet{Ma_2016} find that adopting stellar models that include the effects of binary evolution can significantly boost the luminosity-averaged escape fractions in cosmological simulations. In Appendix~\ref{sec:appendix_bpass} we check whether adopting the Binary Population and Spectral Synthesis (\texttt{BPASS}) model v2.2 \citep{Stanway_2018} would affect our results on the relative importance of dense clouds on LyC escape. We find that adopting \texttt{BPASS} models can indeed increase $N_{\mathrm{cum},\mathrm{LyC}}$ for $\Sigmacloud = 100 \, \Msolpc$ by factors of up to 10 and decrease the gap in the lifetime-averaged $N_{\mathrm{cum},\mathrm{LyC}}$ with denser clouds. However, this is still insufficient to catch up to the $\gtrsim 100$ higher $N_{\mathrm{cum},\mathrm{LyC}}$ for $\Sigmacloud \gtrsim 10^4 \, \Msolpc$, providing reassurance that their disproportionate degree of LyC leakage still holds. 

\subsubsection{Dust in ionized gas}

In our simulations we make the assumption that dust grains survive in photoionized gas, and the dust opacity does not change when the gas becomes ionized. However, this is unlikely to be true at UV wavelengths since it is the presence of small carbanaceous grains and polycyclic aromatic hydrocarbons (PAHs) that dominate the extinction at these wavelengths, and these types of grains are likely destroyed by intense UV radiation \citep{Draine_2001} -- findings with JWST in \ion{H}{2} regions of local galaxies are consistent with this picture \citep[e.g.,][]{Egorov_2023}. This could impact the absorption coefficients at UV wavelengths, which would lower the amount of dust absorption in photoionized gas, and thereby increase escape fractions. However, the lack of dust absorption in photoionized gas would also reduce the impact of radiation pressure. This is unlikely to affect the evolution of lower $\Sigmacloud$ cases, since photoionization is the dominant radiative feedback mechanism anyways. This has been shown by \citet{Kim_2019_escape} where they considered the two extremes of complete dust survival and destruction for a cloud with $\Sigmacloud \sim 80 \, \Msolpc$, and found that the broad evolution is highly similar; the escape fractions are much higher due to lowered dust extinctions. However, for the denser clouds, it is possible that dust destruction would affect the extent to which radiation pressure can impact the evolution of the clouds. For instance, the simulations with lower dust abundances in our study do have lower outflow velocities, hinting at the diminished impact of radiation pressure. However, this is uncertain since the increased transparency of LyC photons through photoionized gas (as opposed to everywhere) might also shift the momentum deposition due to radiation pressure to larger scales, and make it easier to accelerate gas. Future simulations that systematically quantify the impact of PAH destruction on dust UV opacities, and explore the impact of the modified opacities on dense compact clouds will help to clarify these issues.

\subsubsection{Missing feedback mechanisms}

We only model early radiative feedback in our simulations, and do not include protostellar jets, stellar winds and supernovae. Jets are unlikely to affect the bulk evolution of our clouds due to their limited momentum injection when averaged over a stellar population; for instance \citet{Matzner_2015} show that jets can unbind clouds with $v_{\rm esc} \lesssim 1 \, \kms$, reducing their effect to localized clumps\footnote{We note that the effect of jets is likely crucial on scales we do not resolve in our simulations, since they would unbind localized cores where massive stars are forming, permitting the escape of their intrinsic LyC emission \citep{Rosen_2020}. Therefore the effects of jets are \textit{implicitly} assumed through our choice of a subgrid escape fraction of 100\% from our sink particles.}. Stellar winds on the other hand have a momentum contribution of the same order of magnitude as (UV) radiation pressure. Therefore, we expect their inclusion to increase escape fractions by making it easier to disperse the gas. However, the relative importance of winds as a feedback mechanism has been shown to be more prominent for denser clouds \citep{Lancaster_2021c}, which suggests that their inclusion would only amplify the trends with $\Sigmacloud$ we report in this study. On the other hand, the omission of supernovae would primarily affect the escape fractions of the lowest density cloud (i.e. $\Sigmacloud = 100 \Msolpc$), since Figure~\ref{fig:CumulativeEscape} show that these clouds start leaking LyC photons at times comparable to the lifetime of massive stars ($\sim 4 \, \mathrm{Myr}$). Supernovae could quickly remove the surrounding gas when they explode, possibly achieving higher $f_{\mathrm{esc},\mathrm{LyC}}$ slightly earlier than we find in our simulations. However, we emphasize that this should not affect our conclusions significantly since the onset of supernovae is precisely the point where the intrinsic LyC emission starts to drop. Therefore the luminosity-averaged escape fraction ($f_{\mathrm{esc},\mathrm{cum}}$) and its trends with $\Sigmacloud$ (Figure~\ref{fig:fEscCum_Sigma}) is unlikely to be strongly affected.

\section{Summary}
\label{sec:summary}

We conduct idealized radiation hydrodynamic simulations of self-gravitating turbulent clouds with different initial cloud gas surface densities ($\Sigmacloud$) and dust abundances ($Z_{\mathrm{d}}$) to systematically quantify the escape fraction of LyC (and FUV) photons and its dependence on cloud properties. We use the \texttt{FLASH} MHD code with \texttt{VETTAM} -- a highly accurate variable Eddington tensor-closed moment method -- to model the transport of LyC, FUV and IR photon bands, and the associated ionization and radiation pressure forces. We explore clouds with $\Sigmacloud \sim 10^2$--$10^5 \, \Msolpc$, which resembles conditions found in star cluster-forming clumps typical of star-forming galaxies in the local Universe, to the gas-rich, high ISM-pressure conditions that are typical of the epoch of reionization (EoR). We also experiment with $Z_{\mathrm{d}} = 0.01,0.1,1 Z_{\mathrm{d},\sun}$ where $Z_{\mathrm{d},\sun}$ is the dust abundance in the solar neighborhood (see Table~\ref{tab:Simulations}). We summarize relevant quantities measured in our study in Table~\ref{tab:Summary}, and draw the following conclusions. 

\begin{enumerate}
    \item Higher $\Sigmacloud$ results in much higher (integrated) star formation efficiencies ($\epsilon_*$) with $\epsilon_* \sim 15\%$ and $95\%$ for the $\Sigmacloud = 10^2$ and $10^5 \, \Msolpc$ respectively (Figure~\ref{fig:SFE}). 

    \item $f_{\mathrm{esc},\mathrm{LyC}}$ increases with time as star formation proceeds, and the radiative feedback increasingly disperses the cloud (Figure~\ref{fig:LyCEscapeFraction} and \ref{fig:Projection_1}). At higher $\Sigmacloud$, feedback takes a larger number of dynamical times to start allowing the escape of LyC photons due to higher optical depths (Figure~\ref{fig:Timescales}).

    \item Neutral gas absorbs most LyC photons at early times, and $f_{\mathrm{esc},\mathrm{LyC}}$ is high at late times; only for a brief period (when $f_{\mathrm{esc},\mathrm{LyC}} \sim 50\%$) does dust dominate the absorption, and that too only at $Z_\mathrm{d} = Z_\mathrm{d,\sun}$. However, this period is longer for higher $\Sigmacloud$ (Figure~\ref{fig:LyCEscapeFraction}).

    \item The longer duration (in dynamical times) of efficient LyC escape at higher $\Sigmacloud$ can be explained by the competition between UV radiation pressure  and gravity (Eddington limit) and the larger fraction of gas that exceeds this limit at higher $\Sigmacloud$ (Figures~\ref{fig:SigmaEvolution} and ~\ref{fig:S4SigmafEdd}).

    \item Radiative feedback drives super-Eddington sightlines in an ionized outflow, producing density-bounded channels for LyC escape, with faster outflow velocities at lower columns (Figure~\ref{fig:radVel}), and average velocities that are higher for denser clouds (Figure~\ref{fig:radVelaverage}).

    \item While denser clouds ($\Sigmacloud \gtrsim 10^4 \, \Msolpc$) take a larger number of dynamical times to permit LyC escape, their significantly shorter dynamical times cause this trend to reverse when viewed in terms of the stellar population age in absolute time (Figure~\ref{fig:CumulativeEscape}), with $f_{\mathrm{esc},\mathrm{LyC}} >50\%$ in $\lesssim 2 \, \mathrm{Myr}$. This ensures that the phases of high $f_{\mathrm{esc},\mathrm{LyC}}$ and high intrinsic LyC emission from stellar populations ($\lesssim 3 \, \mathrm{Myr}$) are occurring contemporaneously, such that the cumulative number of LyC photons that escape are significantly higher.

    \item This results in a luminosity-weighted average escape fraction $f_{\mathrm{esc},\mathrm{cum}}$ that is much higher for denser clouds:$\gtrsim 70\%$  and $\lesssim 10\%$ for $\Sigmacloud \gtrsim 10^4 \, \Msolpc$ and $\Sigmacloud \sim 10^2 \, \Msolpc$ respectively (Figure~\ref{fig:fEscCum_Sigma}).  
    
\end{enumerate}

We provide a brief discussion on what implications the cloud-scale LyC escape would have on its counterpart at the galaxy scale (Section~\ref{sec:cloud_to_galaxy}), compile some signs of possible evidence in observational studies of the aforementioned scenario playing out (Section~\ref{sec:empirical}), and discuss its implications on cosmic reionization (Section~\ref{sec:reionization}). We compare with results of other numerical simulations in the literature and discuss agreements/disagreements with them (Section~\ref{sec:simulations_literature}) and discuss some caveats associated with our simulations and their possible implications on our findings (Section~\ref{sec:caveats}). Overall, our results seem to suggest a scenario where dense compact bursts of star formation forming bound clusters ($\Sigma > 10^4 \, \Msolpc$) in a galaxy are highly efficient at leaking LyC photons due to their rapid bursts of star formation and radiation-driven outflows, which carve density-bounded channels before massive stars in the cluster die and the intrinsic LyC emission drops. Increasing evidence for clumpy star clusters in (lensed) galaxies at the EoR, approaching (and exceeding) these ``extreme" stellar surface densities \citep{Vanzella_2023,Adamo_2024,Messa_2024,Mowla_2024,Fujimoto_2024,Messa_2024b} hints at their important role as the drivers of reionization of the Universe.

\begin{acknowledgments}
S.~H.~M would like to thank Nicholas Choustikov for extensive discussions on escape fractions in galaxies. S.~H.~M also thanks Eve Ostriker, Rahul Kannan, Aaron Smith, Ulrich~Steinwandel, Lachlan Lancaster, Oleg Gnedin and Crystal Martin for useful discussions in the context of this paper. We acknowledge high-performance computing resources provided by the provided by the Simons Foundation as part of the CCA. B.B. and S.H.M. acknowledge the support of NASA grant No. 80NSSC20K0500 and NSF grant AST-2009679. B.B. also thanks the Alfred P. Sloan Foundation and the Packard Foundation for support. This research was supported in part by grant NSF PHY-2309135 to the Kavli Institute for Theoretical Physics (KITP). T.A.T. is supported in part by NASA grant No. 80NSSC23K1480. This research has made use of NASA's Astrophysics Data System (ADS) Bibliographic Services. 
\end{acknowledgments}

%

\vspace{5mm}


\software{\texttt{SLUG} \citep{daSilva_2012,Krumholz_Slug}, \texttt{PETSc} \citep{PetscConf,PetscRef}, \texttt{NumPy} \citep{numpy}, \texttt{SciPy} \citep{scipy}, \texttt{Matplotlib} \citep{matplotlib}, \texttt{yt} \citep{yt}, \texttt{TurbGen} \citep{Federrath_2010,FederrathEtAl2022ascl}, \texttt{CFpack}.
          }



\appendix


\section{Effect of binary stellar evolution}
\label{sec:appendix_bpass}

\begin{figure}
    \centering
    \includegraphics[width=\linewidth]{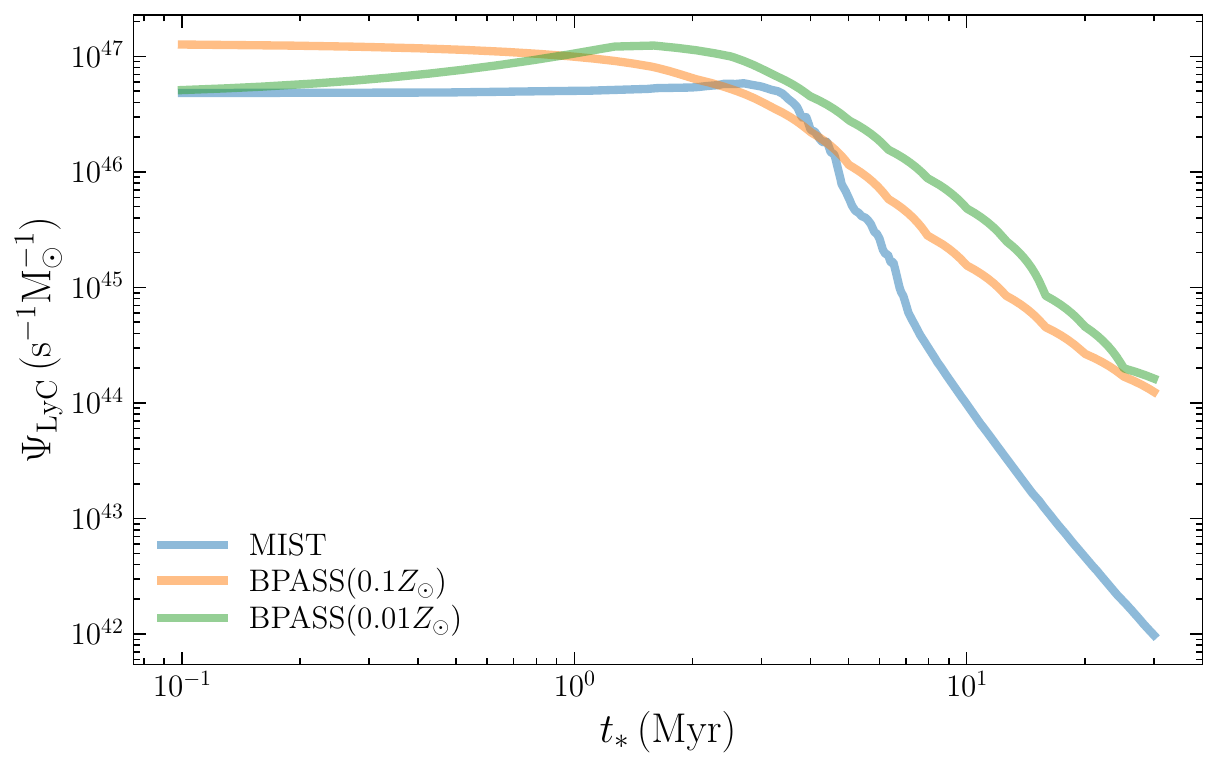}
    \caption{LyC photon emission rate per unit mass ($\Psi_{\mathrm{LyC}}$) emitted by (fully IMF-sampled) stellar populations using the (\texttt{MIST}) v1.0 isochrones \citep{Choi_2016} used in \texttt{SLUG} -- which we adopt in this study -- compared with that obtained with the \texttt{BPASS} v2.2 binary stellar population models \citep{Stanway_2018} at metallicities of $0.1 Z_{\sun}$ and $0.01 Z_{\sun}$. We can see that $\Psi_{\mathrm{LyC}}$ is higher and remains higher for longer in the \texttt{BPASS} models.}
    \label{fig:Bpass_Psi}
\end{figure}

To check whether binary evolution/metallicity would have an impact on our conclusions, we (re)compute and compare the time evolution of $N_{\mathrm{cum},\mathrm{LyC}}$ (Equation~\ref{eq:Ncum}) for $\Sigmacloud = 100 \Msolpc$ and $10^4 \Msolpc$, but with the intrinsic emission ($\dot{n}_{\mathrm{ion}}$) taken from the \texttt{BPASS} model v2.2 \citep{Stanway_2018}. We use the (pre-computed) LyC emission rates for a $10^6 \Msun$ stellar population with a \citet{Chabrier2003} IMF, an upper mass limit of 100$\Msun$, and stellar metallcities of $0.1 Z_{\sun}$ and $0.01 Z_{\sun}$. We use this to compare the LyC emission rate per unit-mass ($\Psi_{\mathrm{LyC}}$) in these models with the adopted \texttt{MIST} v1.0 values \citep{Choi_2016} in our simulations; this is shown in Figure~\ref{fig:Bpass_Psi}. We can clearly see the enhanced LyC outputs at young ages and more importantly, the longer timescales of LyC emission in the \texttt{BPASS} models. We use these to (re)compute the time evolution of $N_{\mathrm{cum},\mathrm{LyC}}$. We do not rerun the simulations, so $f_{\mathrm{esc},\mathrm{LyC}}(t)$ and $M_*(t)$ are identical to those presented in the paper. While the \texttt{BPASS} luminosities would also change this in principle -- permitting quicker cloud disruption due to enhanced UV outputs -- we do not expect it to change significantly since the enhancement of the UV output is only a factor $\lesssim 3$. Moreover, this elevated UV outputs would also slightly decrease the star formation efficiency \citep[by $\lesssim 20\%$;][]{Menon_2024}, so the combined effect on $N_{\mathrm{cum},\mathrm{LyC}}$ remains likely small. As for the longer timescale of intrinsic LyC emission, we did not account for the temporal drop in intrinsic UV emission in our simulations anyways, so are already overestimating the impact of radiative feedback at late times ($t \gtrsim 4 \, \mathrm{Myr}$) at low $\Sigmacloud$ and therefore $f_{\mathrm{esc},\mathrm{LyC}}$. These considerations suggest that sticking with the $f_{\mathrm{esc},\mathrm{LyC}}(t)$ we have obtained is reasonable.

\begin{figure}
    \centering
    \includegraphics[width=\linewidth]{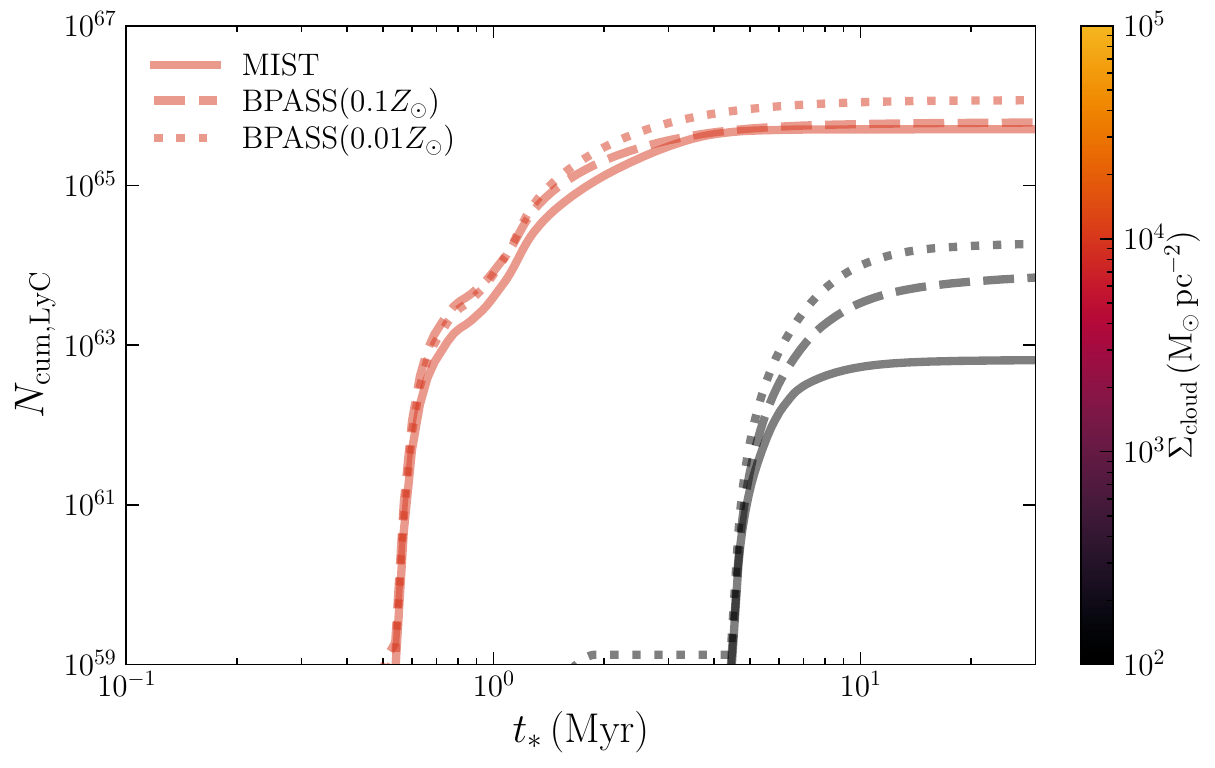}
    \caption{Comparison of the cumulative number of escaping LyC photons ($N_{\mathrm{cum},\mathrm{LyC}}$; Equation~\ref{eq:Ncum}) obtained with our adopted \texttt{MIST} stellar tracks for the LyC emission rate, with the corresponding curves obtained with the \texttt{BPASS} models. We show only two cases, $\Sigmacloud = 100 \, \Msolpc$ and $10^4 \, \Msolpc$, to check if the trends with $\Sigmacloud$ reported in our paper are robust to these changes. To produce these curves we still use $f_{\mathrm{esc},\mathrm{LyC}}$ obtained in our simulations (which use \texttt{MIST}), and only use the \texttt{BPASS} for $\dot{n}_{\mathrm{LyC}}$ in Equation~\ref{eq:Ncum}. We can see that the relative effect of higher and more extended $\Psi_{\mathrm{LyC}}$ in the \texttt{BPASS} models is more prominent at lower $\Sigmacloud$, yet is insufficient to strongly affect the trend of $N_{\mathrm{cum},\mathrm{LyC}}$ with $\Sigmacloud$ that we have reported in this work. } 
    \label{fig:Bpass_Ncum}
\end{figure}

In Figure~\ref{fig:Bpass_Ncum} we compare the time evolution of $N_{\mathrm{cum},\mathrm{LyC}}$ obtained with the \texttt{BPASS} and \texttt{MIST} models for $\Sigmacloud = 100 \, \Msolpc$ and $10^4 \, \Msolpc$ to see its effect on their relative (leaking) LyC outputs. We can see that the impact of \texttt{BPASS} models is more pronounced for the lower $\Sigmacloud$ case -- increasing $N_{\mathrm{cum},\mathrm{LyC}}$ by a factor $\sim 10$; this can be easily understood as arising from the longer intrinsic LyC emission in these models. The impact on the denser cloud is milder since the timescale of LyC leakage is very short anyways, and the additional impact at longer timescales is modest. Regardless, even with the factor $\sim$ 10 increase, $N_{\mathrm{cum},\mathrm{LyC}}$ for the denser cloud is still higher by a factor $\sim 10$. Therefore, adopting \texttt{BPASS} models do not qualitatively change the picture of disproportionate LyC leakage from dense clusters, but only quantitatively change the degree to which the denser clouds are important.

\bibliography{references,federrath,fEscape}{}

\begin{thebibliography}{}
\expandafter\ifx\csname natexlab\endcsname\relax\def\natexlab#1{#1}\fi
\providecommand{\url}[1]{\href{#1}{#1}}
\providecommand{\dodoi}[1]{doi:~\href{http://doi.org/#1}{\nolinkurl{#1}}}
\providecommand{\doeprint}[1]{\href{http://ascl.net/#1}{\nolinkurl{http://ascl.net/#1}}}
\providecommand{\doarXiv}[1]{\href{https://arxiv.org/abs/#1}{\nolinkurl{https://arxiv.org/abs/#1}}}

\bibitem[{{Adamo} {et~al.}(2024){Adamo}, {Bradley}, {Vanzella}, {Claeyssens}, {Welch}, {Diego}, {Mahler}, {Oguri}, {Sharon}, {Abdurro'uf}, {Hsiao}, {Messa}, {Zackrisson}, {Brammer}, {Coe}, {Kokorev}, {Ricotti}, {Zitrin}, {Fujimoto}, {Inoue}, {Resseguier}, {Rigby}, {Jim{\'e}nez-Teja}, {Windhorst}, \& {Xu}}]{Adamo_2024}
{Adamo}, A., {Bradley}, L.~D., {Vanzella}, E., {et~al.} 2024, arXiv e-prints, arXiv:2401.03224, \dodoi{10.48550/arXiv.2401.03224}

\bibitem[{{Andrews} \& {Thompson}(2011)}]{Andrews_2011}
{Andrews}, B.~H., \& {Thompson}, T.~A. 2011, \apj, 727, 97, \dodoi{10.1088/0004-637X/727/2/97}

\bibitem[{{Atek} {et~al.}(2024){Atek}, {Labb{\'e}}, {Furtak}, {Chemerynska}, {Fujimoto}, {Setton}, {Miller}, {Oesch}, {Bezanson}, {Price}, {Dayal}, {Zitrin}, {Kokorev}, {Weaver}, {Brammer}, {Dokkum}, {Williams}, {Cutler}, {Feldmann}, {Fudamoto}, {Greene}, {Leja}, {Maseda}, {Muzzin}, {Pan}, {Papovich}, {Nelson}, {Nanayakkara}, {Stark}, {Stefanon}, {Suess}, {Wang}, \& {Whitaker}}]{Atek_2024}
{Atek}, H., {Labb{\'e}}, I., {Furtak}, L.~J., {et~al.} 2024, \nat, 626, 975, \dodoi{10.1038/s41586-024-07043-6}

\bibitem[{{Baggen} {et~al.}(2023){Baggen}, {van Dokkum}, {Labb{\'e}}, {Brammer}, {Miller}, {Bezanson}, {Leja}, {Wang}, {Whitaker}, {Suess}, \& {Nelson}}]{Baggen_2023}
{Baggen}, J. F.~W., {van Dokkum}, P., {Labb{\'e}}, I., {et~al.} 2023, \apjl, 955, L12, \dodoi{10.3847/2041-8213/acf5ef}

\bibitem[{Balay {et~al.}(1997)Balay, Gropp, McInnes, \& Smith}]{PetscConf}
Balay, S., Gropp, W.~D., McInnes, L.~C., \& Smith, B.~F. 1997, in Modern Software Tools in Scientific Computing, ed. E.~Arge, A.~M. Bruaset, \& H.~P. Langtangen (Birkh{\"{a}}user Press), 163--202

\bibitem[{Balay {et~al.}(2021)Balay, Abhyankar, Adams, Benson, Brown, Brune, Buschelman, Constantinescu, Dalcin, Dener, Eijkhout, Gropp, Hapla, Isaac, Jolivet, Karpeev, Kaushik, Knepley, Kong, Kruger, May, McInnes, Mills, Mitchell, Munson, Roman, Rupp, Sanan, Sarich, Smith, Zampini, Zhang, Zhang, \& Zhang}]{PetscRef}
Balay, S., Abhyankar, S., Adams, M.~F., {et~al.} 2021, {PETSc/TAO} Users Manual, Tech. Rep. ANL-21/39 - Revision 3.16, Argonne National Laboratory

\bibitem[{{Belokurov} \& {Kravtsov}(2023)}]{Belokurov_2023}
{Belokurov}, V., \& {Kravtsov}, A. 2023, \mnras, 525, 4456, \dodoi{10.1093/mnras/stad2241}

\bibitem[{{Bera} {et~al.}(2023){Bera}, {Hassan}, {Smith}, {Cen}, {Garaldi}, {Kannan}, \& {Vogelsberger}}]{Bera_2023}
{Bera}, A., {Hassan}, S., {Smith}, A., {et~al.} 2023, \apj, 959, 2, \dodoi{10.3847/1538-4357/ad05c0}

\bibitem[{{Bian} {et~al.}(2017){Bian}, {Fan}, {McGreer}, {Cai}, \& {Jiang}}]{Bian_2017}
{Bian}, F., {Fan}, X., {McGreer}, I., {Cai}, Z., \& {Jiang}, L. 2017, \apjl, 837, L12, \dodoi{10.3847/2041-8213/aa5ff7}

\bibitem[{{Blackstone} \& {Thompson}(2023)}]{Blackstone_2023}
{Blackstone}, I., \& {Thompson}, T.~A. 2023, arXiv e-prints, arXiv:2302.10136, \dodoi{10.48550/arXiv.2302.10136}

\bibitem[{{Borthakur} {et~al.}(2014){Borthakur}, {Heckman}, {Leitherer}, \& {Overzier}}]{Borthakur_2014}
{Borthakur}, S., {Heckman}, T.~M., {Leitherer}, C., \& {Overzier}, R.~A. 2014, Science, 346, 216, \dodoi{10.1126/science.1254214}

\bibitem[{{Bosman} {et~al.}(2022){Bosman}, {Davies}, {Becker}, {Keating}, {Davies}, {Zhu}, {Eilers}, {D'Odorico}, {Bian}, {Bischetti}, {Cristiani}, {Fan}, {Farina}, {Haehnelt}, {Hennawi}, {Kulkarni}, {Mesinger}, {Meyer}, {Onoue}, {Pallottini}, {Qin}, {Ryan-Weber}, {Schindler}, {Walter}, {Wang}, \& {Yang}}]{Bosman_2022}
{Bosman}, S. E.~I., {Davies}, F.~B., {Becker}, G.~D., {et~al.} 2022, \mnras, 514, 55, \dodoi{10.1093/mnras/stac1046}

\bibitem[{Buntemeyer {et~al.}(2016)Buntemeyer, Banerjee, Peters, Klassen, \& Pudritz}]{Buntemeyer_2016}
Buntemeyer, L., Banerjee, R., Peters, T., Klassen, M., \& Pudritz, R.~E. 2016, New Astronomy, 43, 49, \dodoi{10.1016/j.newast.2015.07.002}

\bibitem[{{Burkhart} {et~al.}(2022){Burkhart}, {Tillman}, {Gurvich}, {Bird}, {Tonnesen}, {Bryan}, {Hernquist}, \& {Somerville}}]{Burkhart_2022}
{Burkhart}, B., {Tillman}, M., {Gurvich}, A.~B., {et~al.} 2022, \apjl, 933, L46, \dodoi{10.3847/2041-8213/ac7e49}

\bibitem[{{Casey} {et~al.}(2023){Casey}, {Akins}, {Shuntov}, {Ilbert}, {Paquereau}, {Franco}, {Hayward}, {Finkelstein}, {Boylan-Kolchin}, {Robertson}, {Allen}, {Brinch}, {Cooper}, {Ding}, {Drakos}, {Faisst}, {Fujimoto}, {Gillman}, {Harish}, {Hirschmann}, {Jin}, {Kartaltepe}, {Koekemoer}, {Kokorev}, {Liu}, {Long}, {Magdis}, {Maraston}, {Martin}, {McCracken}, {McKinney}, {Mobasher}, {Rhodes}, {Rich}, {Sanders}, {Silverman}, {Toft}, {Vijayan}, {Weaver}, {Wilkins}, {Yang}, \& {Zavala}}]{Casey_2023}
{Casey}, C.~M., {Akins}, H.~B., {Shuntov}, M., {et~al.} 2023, arXiv e-prints, arXiv:2308.10932, \dodoi{10.48550/arXiv.2308.10932}

\bibitem[{{Chabrier}(2003)}]{Chabrier2003}
{Chabrier}, G. 2003, \pasp, 115, 763, \dodoi{10.1086/376392}

\bibitem[{{Chevance} {et~al.}(2023){Chevance}, {Krumholz}, {McLeod}, {Ostriker}, {Rosolowsky}, \& {Sternberg}}]{Chevance_23}
{Chevance}, M., {Krumholz}, M.~R., {McLeod}, A.~F., {et~al.} 2023, in Astronomical Society of the Pacific Conference Series, Vol. 534, Protostars and Planets VII, ed. S.~{Inutsuka}, Y.~{Aikawa}, T.~{Muto}, K.~{Tomida}, \& M.~{Tamura}, 1, \dodoi{10.48550/arXiv.2203.09570}

\bibitem[{{Choi} {et~al.}(2016){Choi}, {Dotter}, {Conroy}, {Cantiello}, {Paxton}, \& {Johnson}}]{Choi_2016}
{Choi}, J., {Dotter}, A., {Conroy}, C., {et~al.} 2016, \apj, 823, 102, \dodoi{10.3847/0004-637X/823/2/102}

\bibitem[{{Choi} {et~al.}(2020){Choi}, {Dalcanton}, {Williams}, {Skillman}, {Fouesneau}, {Gordon}, {Sandstrom}, {Weisz}, \& {Gilbert}}]{Choi_2020}
{Choi}, Y., {Dalcanton}, J.~J., {Williams}, B.~F., {et~al.} 2020, \apj, 902, 54, \dodoi{10.3847/1538-4357/abb467}

\bibitem[{{Choustikov} {et~al.}(2024){Choustikov}, {Katz}, {Saxena}, {Cameron}, {Devriendt}, {Slyz}, {Rosdahl}, {Blaizot}, \& {Michel-Dansac}}]{Choustikov_2024}
{Choustikov}, N., {Katz}, H., {Saxena}, A., {et~al.} 2024, \mnras, 529, 3751, \dodoi{10.1093/mnras/stae776}

\bibitem[{{Crocker} {et~al.}(2018){Crocker}, {Krumholz}, {Thompson}, {Baumgardt}, \& {Mackey}}]{Crocker_2018b}
{Crocker}, R.~M., {Krumholz}, M.~R., {Thompson}, T.~A., {Baumgardt}, H., \& {Mackey}, D. 2018, \mnras, 481, 4895, \dodoi{10.1093/mnras/sty2659}

\bibitem[{{Curti} {et~al.}(2023){Curti}, {Maiolino}, {Curtis-Lake}, {Chevallard}, {Carniani}, {D'Eugenio}, {Looser}, {Scholtz}, {Charlot}, {Cameron}, {{\"U}bler}, {Witstok}, {Boyett}, {Laseter}, {Sandles}, {Arribas}, {Bunker}, {Giardino}, {Maseda}, {Rawle}, {Rodr{\'\i}guez Del Pino}, {Smit}, {Willott}, {Eisenstein}, {Hausen}, {Johnson}, {Rieke}, {Robertson}, {Tacchella}, {Williams}, {Willmer}, {Baker}, {Bhatawdekar}, {Egami}, {Helton}, {Ji}, {Kumari}, {Perna}, {Shivaei}, \& {Sun}}]{Curti_2023}
{Curti}, M., {Maiolino}, R., {Curtis-Lake}, E., {et~al.} 2023, arXiv e-prints, arXiv:2304.08516, \dodoi{10.48550/arXiv.2304.08516}

\bibitem[{{Curtis-Lake} {et~al.}(2023){Curtis-Lake}, {Carniani}, {Cameron}, {Charlot}, {Jakobsen}, {Maiolino}, {Bunker}, {Witstok}, {Smit}, {Chevallard}, {Willott}, {Ferruit}, {Arribas}, {Bonaventura}, {Curti}, {D'Eugenio}, {Franx}, {Giardino}, {Looser}, {L{\"u}tzgendorf}, {Maseda}, {Rawle}, {Rix}, {Rodr{\'\i}guez del Pino}, {{\"U}bler}, {Sirianni}, {Dressler}, {Egami}, {Eisenstein}, {Endsley}, {Hainline}, {Hausen}, {Johnson}, {Rieke}, {Robertson}, {Shivaei}, {Stark}, {Tacchella}, {Williams}, {Willmer}, {Bhatawdekar}, {Bowler}, {Boyett}, {Chen}, {de Graaff}, {Helton}, {Hviding}, {Jones}, {Kumari}, {Lyu}, {Nelson}, {Perna}, {Sandles}, {Saxena}, {Suess}, {Sun}, {Topping}, {Wallace}, \& {Whitler}}]{Curtis-Lake_2023}
{Curtis-Lake}, E., {Carniani}, S., {Cameron}, A., {et~al.} 2023, Nature Astronomy, 7, 622, \dodoi{10.1038/s41550-023-01918-w}

\bibitem[{{da Silva} {et~al.}(2012){da Silva}, {Fumagalli}, \& {Krumholz}}]{daSilva_2012}
{da Silva}, R.~L., {Fumagalli}, M., \& {Krumholz}, M. 2012, \apj, 745, 145, \dodoi{10.1088/0004-637X/745/2/145}

\bibitem[{{Dale} {et~al.}(2012){Dale}, {Ercolano}, \& {Bonnell}}]{Dale_2012}
{Dale}, J.~E., {Ercolano}, B., \& {Bonnell}, I.~A. 2012, \mnras, 424, 377, \dodoi{10.1111/j.1365-2966.2012.21205.x}

\bibitem[{{Dayal} \& {Ferrara}(2018)}]{Dayal_2018}
{Dayal}, P., \& {Ferrara}, A. 2018, \physrep, 780, 1, \dodoi{10.1016/j.physrep.2018.10.002}

\bibitem[{{Dayal} {et~al.}(2024){Dayal}, {Volonteri}, {Greene}, {Kokorev}, {Goulding}, {Williams}, {Furtak}, {Zitrin}, {Atek}, {Chemerynska}, {Feldmann}, {Glazebrook}, {Labbe}, {Nanayakkara}, {Oesch}, \& {Weaver}}]{Dayal_2024}
{Dayal}, P., {Volonteri}, M., {Greene}, J.~E., {et~al.} 2024, arXiv e-prints, arXiv:2401.11242, \dodoi{10.48550/arXiv.2401.11242}

\bibitem[{{Dekel} {et~al.}(2023){Dekel}, {Sarkar}, {Birnboim}, {Mandelker}, \& {Li}}]{Dekel_2023}
{Dekel}, A., {Sarkar}, K.~C., {Birnboim}, Y., {Mandelker}, N., \& {Li}, Z. 2023, \mnras, 523, 3201, \dodoi{10.1093/mnras/stad1557}

\bibitem[{{Draine}(2011)}]{Draine_2011}
{Draine}, B.~T. 2011, \apj, 732, 100, \dodoi{10.1088/0004-637X/732/2/100}

\bibitem[{{Draine} \& {Li}(2001)}]{Draine_2001}
{Draine}, B.~T., \& {Li}, A. 2001, \apj, 551, 807, \dodoi{10.1086/320227}

\bibitem[{{Dubey} {et~al.}(2008){Dubey}, {Reid}, \& {Fisher}}]{Dubey_2008}
{Dubey}, A., {Reid}, L.~B., \& {Fisher}, R. 2008, Physica Scripta Volume T, 132, 014046, \dodoi{10.1088/0031-8949/2008/T132/014046}

\bibitem[{{Egorov} {et~al.}(2023){Egorov}, {Kreckel}, {Sandstrom}, {Leroy}, {Glover}, {Groves}, {Kruijssen}, {Barnes}, {Belfiore}, {Bigiel}, {Blanc}, {Boquien}, {Cao}, {Chastenet}, {Chevance}, {Congiu}, {Dale}, {Emsellem}, {Grasha}, {Klessen}, {Larson}, {Liu}, {Murphy}, {Pan}, {Pessa}, {Pety}, {Rosolowsky}, {Scheuermann}, {Schinnerer}, {Sutter}, {Thilker}, {Watkins}, \& {Williams}}]{Egorov_2023}
{Egorov}, O.~V., {Kreckel}, K., {Sandstrom}, K.~M., {et~al.} 2023, \apjl, 944, L16, \dodoi{10.3847/2041-8213/acac92}

\bibitem[{{Elmegreen} \& {Chiang}(1982)}]{Elmegreen_1982}
{Elmegreen}, B.~G., \& {Chiang}, W.~H. 1982, \apj, 253, 666, \dodoi{10.1086/159667}

\bibitem[{Fall {et~al.}(2010)Fall, Krumholz, \& Matzner}]{Fall_2010}
Fall, S.~M., Krumholz, M.~R., \& Matzner, C.~D. 2010, Astrophysical Journal Letters, 710, L142, \dodoi{10.1088/2041-8205/710/2/L142}

\bibitem[{{Fan} {et~al.}(2006){Fan}, {Carilli}, \& {Keating}}]{Fan_2006}
{Fan}, X., {Carilli}, C.~L., \& {Keating}, B. 2006, \araa, 44, 415, \dodoi{10.1146/annurev.astro.44.051905.092514}

\bibitem[{{Faucher-Gigu{\`e}re}(2018)}]{FaucherGiguere_2018}
{Faucher-Gigu{\`e}re}, C.-A. 2018, \mnras, 473, 3717, \dodoi{10.1093/mnras/stx2595}

\bibitem[{{Federrath} {et~al.}(2010{\natexlab{a}}){Federrath}, {Banerjee}, {Clark}, \& {Klessen}}]{Federrath_2010_Sinks}
{Federrath}, C., {Banerjee}, R., {Clark}, P.~C., \& {Klessen}, R.~S. 2010{\natexlab{a}}, \apj, 713, 269, \dodoi{10.1088/0004-637X/713/1/269}

\bibitem[{{Federrath} {et~al.}(2011){Federrath}, {Banerjee}, {Seifried}, {Clark}, \& {Klessen}}]{FederrathBanerjeeSeifriedClarkKlessen2011}
{Federrath}, C., {Banerjee}, R., {Seifried}, D., {Clark}, P.~C., \& {Klessen}, R.~S. 2011, in IAU Symposium, Vol. 270, Computational Star Formation, ed. {J.~Alves, B.~G.~Elmegreen, J.~M.~Girart, \& V.~Trimble}, 425--428, \dodoi{10.1017/S1743921311000755}

\bibitem[{{Federrath} {et~al.}(2010{\natexlab{b}}){Federrath}, {Roman-Duval}, {Klessen}, {Schmidt}, \& {Mac Low}}]{Federrath_2010}
{Federrath}, C., {Roman-Duval}, J., {Klessen}, R.~S., {Schmidt}, W., \& {Mac Low}, M.~M. 2010{\natexlab{b}}, \aap, 512, A81, \dodoi{10.1051/0004-6361/200912437}

\bibitem[{{Federrath} {et~al.}(2022){Federrath}, {Roman-Duval}, {Klessen}, {Schmidt}, \& {Mac Low}}]{FederrathEtAl2022ascl}
---. 2022, {TG: Turbulence Generator}, Astrophysics Source Code Library, record ascl:2204.001.
\newblock \doeprint{2204.001}

\bibitem[{{Fielding} {et~al.}(2018){Fielding}, {Quataert}, \& {Martizzi}}]{Fielding_2018}
{Fielding}, D., {Quataert}, E., \& {Martizzi}, D. 2018, \mnras, 481, 3325, \dodoi{10.1093/mnras/sty2466}

\bibitem[{{Finkelstein} {et~al.}(2019){Finkelstein}, {D'Aloisio}, {Paardekooper}, {Ryan}, {Behroozi}, {Finlator}, {Livermore}, {Upton Sanderbeck}, {Dalla Vecchia}, \& {Khochfar}}]{Finkelstein_2019}
{Finkelstein}, S.~L., {D'Aloisio}, A., {Paardekooper}, J.-P., {et~al.} 2019, \apj, 879, 36, \dodoi{10.3847/1538-4357/ab1ea8}

\bibitem[{{Finkelstein} {et~al.}(2023){Finkelstein}, {Bagley}, {Ferguson}, {Wilkins}, {Kartaltepe}, {Papovich}, {Yung}, {Arrabal Haro}, {Behroozi}, {Dickinson}, {Kocevski}, {Koekemoer}, {Larson}, {Le Bail}, {Morales}, {P{\'e}rez-Gonz{\'a}lez}, {Burgarella}, {Dav{\'e}}, {Hirschmann}, {Somerville}, {Wuyts}, {Bromm}, {Casey}, {Fontana}, {Fujimoto}, {Gardner}, {Giavalisco}, {Grazian}, {Grogin}, {Hathi}, {Hutchison}, {Jha}, {Jogee}, {Kewley}, {Kirkpatrick}, {Long}, {Lotz}, {Pentericci}, {Pierel}, {Pirzkal}, {Ravindranath}, {Ryan}, {Trump}, {Yang}, {Bhatawdekar}, {Bisigello}, {Buat}, {Calabr{\`o}}, {Castellano}, {Cleri}, {Cooper}, {Croton}, {Daddi}, {Dekel}, {Elbaz}, {Franco}, {Gawiser}, {Holwerda}, {Huertas-Company}, {Jaskot}, {Leung}, {Lucas}, {Mobasher}, {Pandya}, {Tacchella}, {Weiner}, \& {Zavala}}]{Finkelstein_2023a}
{Finkelstein}, S.~L., {Bagley}, M.~B., {Ferguson}, H.~C., {et~al.} 2023, \apjl, 946, L13, \dodoi{10.3847/2041-8213/acade4}

\bibitem[{{Flury} {et~al.}(2022){Flury}, {Jaskot}, {Ferguson}, {Worseck}, {Makan}, {Chisholm}, {Saldana-Lopez}, {Schaerer}, {McCandliss}, {Xu}, {Wang}, {Oey}, {Ford}, {Heckman}, {Ji}, {Giavalisco}, {Amor{\'\i}n}, {Atek}, {Blaizot}, {Borthakur}, {Carr}, {Castellano}, {De Barros}, {Dickinson}, {Finkelstein}, {Fleming}, {Fontanot}, {Garel}, {Grazian}, {Hayes}, {Henry}, {Mauerhofer}, {Micheva}, {Ostlin}, {Papovich}, {Pentericci}, {Ravindranath}, {Rosdahl}, {Rutkowski}, {Santini}, {Scarlata}, {Teplitz}, {Thuan}, {Trebitsch}, {Vanzella}, \& {Verhamme}}]{Flury_2022b}
{Flury}, S.~R., {Jaskot}, A.~E., {Ferguson}, H.~C., {et~al.} 2022, \apj, 930, 126, \dodoi{10.3847/1538-4357/ac61e4}

\bibitem[{{Fryxell} {et~al.}(2000){Fryxell}, {Olson}, {Ricker}, {Timmes}, {Zingale}, {Lamb}, {MacNeice}, {Rosner}, {Truran}, \& {Tufo}}]{Fryxell_2000}
{Fryxell}, B., {Olson}, K., {Ricker}, P., {et~al.} 2000, \apjs, 131, 273, \dodoi{10.1086/317361}

\bibitem[{{Fujimoto} {et~al.}(2024){Fujimoto}, {Ouchi}, {Kohno}, {Valentino}, {Gim{\'e}nez-Arteaga}, {Brammer}, {Furtak}, {Kohandel}, {Oguri}, {Pallottini}, {Richard}, {Zitrin}, {Bauer}, {Boylan-Kolchin}, {Dessauges-Zavadsky}, {Egami}, {Finkelstein}, {Ma}, {Smail}, {Watson}, {Hutchison}, {Rigby}, {Welch}, {Ao}, {Bradley}, {Caminha}, {Caputi}, {Espada}, {Endsley}, {Fudamoto}, {Gonz{\'a}lez-L{\'o}pez}, {Hatsukade}, {Koekemoer}, {Kokorev}, {Laporte}, {Lee}, {Magdis}, {Ono}, {Rizzo}, {Shibuya}, {Shimasaku}, {Sun}, {Toft}, {Umehata}, {Wang}, \& {Yajima}}]{Fujimoto_2024}
{Fujimoto}, S., {Ouchi}, M., {Kohno}, K., {et~al.} 2024, arXiv e-prints, arXiv:2402.18543, \dodoi{10.48550/arXiv.2402.18543}

\bibitem[{{Gnedin} \& {Madau}(2022)}]{Gnedin_2022}
{Gnedin}, N.~Y., \& {Madau}, P. 2022, Living Reviews in Computational Astrophysics, 8, 3, \dodoi{10.1007/s41115-022-00015-5}

\bibitem[{Gritschneder {et~al.}(2009)Gritschneder, Naab, Walch, Burkert, \& Heitsch}]{Gritschneder_2009}
Gritschneder, M., Naab, T., Walch, S., Burkert, A., \& Heitsch, F. 2009, Astrophysical Journal, 694, \dodoi{10.1088/0004-637X/694/1/L26}

\bibitem[{{Grudi{\'c}} {et~al.}(2018){Grudi{\'c}}, {Hopkins}, {Faucher-Gigu{\`e}re}, {Quataert}, {Murray}, \& {Kere{\v{s}}}}]{Grudic_2018}
{Grudi{\'c}}, M.~Y., {Hopkins}, P.~F., {Faucher-Gigu{\`e}re}, C.-A., {et~al.} 2018, \mnras, 475, 3511, \dodoi{10.1093/mnras/sty035}

\bibitem[{{Hacar} {et~al.}(2023){Hacar}, {Clark}, {Heitsch}, {Kainulainen}, {Panopoulou}, {Seifried}, \& {Smith}}]{Hacar_2023}
{Hacar}, A., {Clark}, S.~E., {Heitsch}, F., {et~al.} 2023, in Astronomical Society of the Pacific Conference Series, Vol. 534, Protostars and Planets VII, ed. S.~{Inutsuka}, Y.~{Aikawa}, T.~{Muto}, K.~{Tomida}, \& M.~{Tamura}, 153, \dodoi{10.48550/arXiv.2203.09562}

\bibitem[{{Hainline} {et~al.}(2024){Hainline}, {D'Eugenio}, {Jakobsen}, {Chevallard}, {Carniani}, {Witstok}, {Ji}, {Curtis-Lake}, {Johnson}, {Robertson}, {Tacchella}, {Curti}, {Charlot}, {Helton}, {Arribas}, {Bhatawdekar}, {Bunker}, {Cameron}, {Egami}, {Eisenstein}, {Hausen}, {Kumari}, {Maiolino}, {Perez-Gonzalez}, {Rieke}, {Saxena}, {Scholtz}, {Smit}, {Sun}, {Williams}, {Willmer}, \& {Willott}}]{Hainline_2024}
{Hainline}, K.~N., {D'Eugenio}, F., {Jakobsen}, P., {et~al.} 2024, arXiv e-prints, arXiv:2404.04325, \dodoi{10.48550/arXiv.2404.04325}

\bibitem[{{Harikane} {et~al.}(2023){Harikane}, {Ouchi}, {Oguri}, {Ono}, {Nakajima}, {Isobe}, {Umeda}, {Mawatari}, \& {Zhang}}]{Harikane_2023}
{Harikane}, Y., {Ouchi}, M., {Oguri}, M., {et~al.} 2023, \apjs, 265, 5, \dodoi{10.3847/1538-4365/acaaa9}

\bibitem[{Harris {et~al.}(2020)Harris, Millman, van~der Walt, Gommers, Virtanen, Cournapeau, Wieser, Taylor, Berg, Smith, Kern, Picus, Hoyer, van Kerkwijk, Brett, Haldane, del R{\'{i}}o, Wiebe, Peterson, G{\'{e}}rard-Marchant, Sheppard, Reddy, Weckesser, Abbasi, Gohlke, \& Oliphant}]{numpy}
Harris, C.~R., Millman, K.~J., van~der Walt, S.~J., {et~al.} 2020, Nature, 585, 357, \dodoi{10.1038/s41586-020-2649-2}

\bibitem[{{Hassan} {et~al.}(2018){Hassan}, {Dav{\'e}}, {Mitra}, {Finlator}, {Ciardi}, \& {Santos}}]{Hassan_2018}
{Hassan}, S., {Dav{\'e}}, R., {Mitra}, S., {et~al.} 2018, \mnras, 473, 227, \dodoi{10.1093/mnras/stx2194}

\bibitem[{{He} {et~al.}(2020){He}, {Ricotti}, \& {Geen}}]{He_2020}
{He}, C.-C., {Ricotti}, M., \& {Geen}, S. 2020, \mnras, 492, 4858, \dodoi{10.1093/mnras/staa165}

\bibitem[{{Howard} {et~al.}(2018){Howard}, {Pudritz}, {Harris}, \& {Klessen}}]{Howard_2018}
{Howard}, C.~S., {Pudritz}, R.~E., {Harris}, W.~E., \& {Klessen}, R.~S. 2018, \mnras, 475, 3121, \dodoi{10.1093/mnras/stx3276}

\bibitem[{{Hui} \& {Gnedin}(1997)}]{Hui_1997}
{Hui}, L., \& {Gnedin}, N.~Y. 1997, \mnras, 292, 27, \dodoi{10.1093/mnras/292.1.27}

\bibitem[{Hunter(2007)}]{matplotlib}
Hunter, J.~D. 2007, Computing in Science Engineering, 9, 90, \dodoi{10.1109/MCSE.2007.55}

\bibitem[{{Izotov} {et~al.}(2016){Izotov}, {Schaerer}, {Thuan}, {Worseck}, {Guseva}, {Orlitov{\'a}}, \& {Verhamme}}]{Izotov_2016}
{Izotov}, Y.~I., {Schaerer}, D., {Thuan}, T.~X., {et~al.} 2016, \mnras, 461, 3683, \dodoi{10.1093/mnras/stw1205}

\bibitem[{{Izotov} {et~al.}(2018{\natexlab{a}}){Izotov}, {Schaerer}, {Worseck}, {Guseva}, {Thuan}, {Verhamme}, {Orlitov{\'a}}, \& {Fricke}}]{Izotov_2018b}
{Izotov}, Y.~I., {Schaerer}, D., {Worseck}, G., {et~al.} 2018{\natexlab{a}}, \mnras, 474, 4514, \dodoi{10.1093/mnras/stx3115}

\bibitem[{{Izotov} {et~al.}(2017){Izotov}, {Thuan}, \& {Guseva}}]{Izotov_2017}
{Izotov}, Y.~I., {Thuan}, T.~X., \& {Guseva}, N.~G. 2017, \mnras, 471, 548, \dodoi{10.1093/mnras/stx1629}

\bibitem[{{Izotov} {et~al.}(2018{\natexlab{b}}){Izotov}, {Worseck}, {Schaerer}, {Guseva}, {Thuan}, {Fricke}, \& {Orlitov{\'a}}}]{Izotov_2018a}
{Izotov}, Y.~I., {Worseck}, G., {Schaerer}, D., {et~al.} 2018{\natexlab{b}}, \mnras, 478, 4851, \dodoi{10.1093/mnras/sty1378}

\bibitem[{{Japelj} {et~al.}(2017){Japelj}, {Vanzella}, {Fontanot}, {Cristiani}, {Caminha}, {Tozzi}, {Balestra}, {Rosati}, \& {Meneghetti}}]{Japelj_2017}
{Japelj}, J., {Vanzella}, E., {Fontanot}, F., {et~al.} 2017, \mnras, 468, 389, \dodoi{10.1093/mnras/stx477}

\bibitem[{{Jaskot} \& {Oey}(2013)}]{Jaskot_2013}
{Jaskot}, A.~E., \& {Oey}, M.~S. 2013, \apj, 766, 91, \dodoi{10.1088/0004-637X/766/2/91}

\bibitem[{{Jaskot} {et~al.}(2024){Jaskot}, {Silveyra}, {Plantinga}, {Flury}, {Hayes}, {Chisholm}, {Heckman}, {Pentericci}, {Schaerer}, {Trebitsch}, {Verhamme}, {Carr}, {Ferguson}, {Ji}, {Giavalisco}, {Henry}, {Marques-Chaves}, {{\"O}stlin}, {Saldana-Lopez}, {Scarlata}, {Worseck}, \& {Xu}}]{Jaskot_2024}
{Jaskot}, A.~E., {Silveyra}, A.~C., {Plantinga}, A., {et~al.} 2024, arXiv e-prints, arXiv:2406.10171, \dodoi{10.48550/arXiv.2406.10171}

\bibitem[{{Jeffreson} {et~al.}(2024){Jeffreson}, {Semenov}, \& {Krumholz}}]{Jeffreson_2024}
{Jeffreson}, S. M.~R., {Semenov}, V.~A., \& {Krumholz}, M.~R. 2024, \mnras, 527, 7093, \dodoi{10.1093/mnras/stad3550}

\bibitem[{{Jung} {et~al.}(2024){Jung}, {Ferguson}, {Hayes}, {Henry}, {Jaskot}, {Schaerer}, {Sharon}, {Amor{\'\i}n}, {Atek}, {Bayliss}, {Dahle}, {Finkelstein}, {Grazian}, {Guaita}, {{\"O}stlin}, {Pentericci}, {Ravindranath}, {Scarlata}, {Teplitz}, \& {Verhamme}}]{Jung_2024}
{Jung}, I., {Ferguson}, H.~C., {Hayes}, M.~J., {et~al.} 2024, arXiv e-prints, arXiv:2403.02388, \dodoi{10.48550/arXiv.2403.02388}

\bibitem[{{Kakiichi} \& {Gronke}(2021)}]{Kakiichi_2021}
{Kakiichi}, K., \& {Gronke}, M. 2021, \apj, 908, 30, \dodoi{10.3847/1538-4357/abc2d9}

\bibitem[{{Kim} {et~al.}(2023{\natexlab{a}}){Kim}, {Gong}, {Kim}, \& {Ostriker}}]{Kim_2023}
{Kim}, J.-G., {Gong}, M., {Kim}, C.-G., \& {Ostriker}, E.~C. 2023{\natexlab{a}}, \apjs, 264, 10, \dodoi{10.3847/1538-4365/ac9b1d}

\bibitem[{{Kim} {et~al.}(2016){Kim}, {Kim}, \& {Ostriker}}]{Kim_2016}
{Kim}, J.-G., {Kim}, W.-T., \& {Ostriker}, E.~C. 2016, \apj, 819, 137, \dodoi{10.3847/0004-637X/819/2/137}

\bibitem[{{Kim} {et~al.}(2018){Kim}, {Kim}, \& {Ostriker}}]{Kim_2018}
---. 2018, \apj, 859, 68, \dodoi{10.3847/1538-4357/aabe27}

\bibitem[{{Kim} {et~al.}(2019){Kim}, {Kim}, \& {Ostriker}}]{Kim_2019_escape}
---. 2019, \apj, 883, 102, \dodoi{10.3847/1538-4357/ab3d3d}

\bibitem[{{Kim} {et~al.}(2017){Kim}, {Kim}, {Ostriker}, \& {Skinner}}]{Kim_2017}
{Kim}, J.-G., {Kim}, W.-T., {Ostriker}, E.~C., \& {Skinner}, M.~A. 2017, \apj, 851, 93, \dodoi{10.3847/1538-4357/aa9b80}

\bibitem[{{Kim} {et~al.}(2023{\natexlab{b}}){Kim}, {Bayliss}, {Rigby}, {Gladders}, {Chisholm}, {Sharon}, {Dahle}, {Rivera-Thorsen}, {Florian}, {Khullar}, {Mahler}, {Mainali}, {Napier}, {Navarre}, {Owens}, \& {Roberson}}]{Kim_2023_escape}
{Kim}, K.~J., {Bayliss}, M.~B., {Rigby}, J.~R., {et~al.} 2023{\natexlab{b}}, \apjl, 955, L17, \dodoi{10.3847/2041-8213/acf0c5}

\bibitem[{{Kimm} {et~al.}(2022){Kimm}, {Bieri}, {Geen}, {Rosdahl}, {Blaizot}, {Michel-Dansac}, \& {Garel}}]{Kimm_2022}
{Kimm}, T., {Bieri}, R., {Geen}, S., {et~al.} 2022, \apjs, 259, 21, \dodoi{10.3847/1538-4365/ac426d}

\bibitem[{{Kimm} {et~al.}(2017){Kimm}, {Katz}, {Haehnelt}, {Rosdahl}, {Devriendt}, \& {Slyz}}]{Kimm_2017}
{Kimm}, T., {Katz}, H., {Haehnelt}, M., {et~al.} 2017, \mnras, 466, 4826, \dodoi{10.1093/mnras/stx052}

\bibitem[{{Komarova} {et~al.}(2021){Komarova}, {Oey}, {Krumholz}, {Silich}, {Kumari}, \& {James}}]{Komarova21a}
{Komarova}, L., {Oey}, M.~S., {Krumholz}, M.~R., {et~al.} 2021, \apjl, 920, L46, \dodoi{10.3847/2041-8213/ac2c09}

\bibitem[{{Kostyuk} {et~al.}(2023){Kostyuk}, {Nelson}, {Ciardi}, {Glatzle}, \& {Pillepich}}]{Kostyuk_2023}
{Kostyuk}, I., {Nelson}, D., {Ciardi}, B., {Glatzle}, M., \& {Pillepich}, A. 2023, \mnras, 521, 3077, \dodoi{10.1093/mnras/stad677}

\bibitem[{{Krumholz} {et~al.}(2015){Krumholz}, {Fumagalli}, {da Silva}, {Rendahl}, \& {Parra}}]{Krumholz_Slug}
{Krumholz}, M.~R., {Fumagalli}, M., {da Silva}, R.~L., {Rendahl}, T., \& {Parra}, J. 2015, \mnras, 452, 1447, \dodoi{10.1093/mnras/stv1374}

\bibitem[{Krumholz \& Matzner(2009)}]{Krumholz_Matzner_2009}
Krumholz, M.~R., \& Matzner, C.~D. 2009, Astrophysical Journal, 703, 1352, \dodoi{10.1088/0004-637X/703/2/1352}

\bibitem[{{Lancaster} {et~al.}(2021){Lancaster}, {Ostriker}, {Kim}, \& {Kim}}]{Lancaster_2021c}
{Lancaster}, L., {Ostriker}, E.~C., {Kim}, J.-G., \& {Kim}, C.-G. 2021, \apjl, 922, L3, \dodoi{10.3847/2041-8213/ac3333}

\bibitem[{{Leitet} {et~al.}(2013){Leitet}, {Bergvall}, {Hayes}, {Linn{\'e}}, \& {Zackrisson}}]{Leitet_2013}
{Leitet}, E., {Bergvall}, N., {Hayes}, M., {Linn{\'e}}, S., \& {Zackrisson}, E. 2013, \aap, 553, A106, \dodoi{10.1051/0004-6361/201118370}

\bibitem[{{Leitherer} {et~al.}(2016){Leitherer}, {Hernandez}, {Lee}, \& {Oey}}]{Leitherer_2016}
{Leitherer}, C., {Hernandez}, S., {Lee}, J.~C., \& {Oey}, M.~S. 2016, \apj, 823, 64, \dodoi{10.3847/0004-637X/823/1/64}

\bibitem[{{Loeb} \& {Barkana}(2001)}]{Loeb_2001}
{Loeb}, A., \& {Barkana}, R. 2001, \araa, 39, 19, \dodoi{10.1146/annurev.astro.39.1.19}

\bibitem[{{Ma} {et~al.}(2016){Ma}, {Hopkins}, {Kasen}, {Quataert}, {Faucher-Gigu{\`e}re}, {Kere{\v{s}}}, {Murray}, \& {Strom}}]{Ma_2016}
{Ma}, X., {Hopkins}, P.~F., {Kasen}, D., {et~al.} 2016, \mnras, 459, 3614, \dodoi{10.1093/mnras/stw941}

\bibitem[{{Ma} {et~al.}(2015){Ma}, {Kasen}, {Hopkins}, {Faucher-Gigu{\`e}re}, {Quataert}, {Kere{\v{s}}}, \& {Murray}}]{Ma_2015}
{Ma}, X., {Kasen}, D., {Hopkins}, P.~F., {et~al.} 2015, \mnras, 453, 960, \dodoi{10.1093/mnras/stv1679}

\bibitem[{{Madau} {et~al.}(2024){Madau}, {Giallongo}, {Grazian}, \& {Haardt}}]{Madau_2024}
{Madau}, P., {Giallongo}, E., {Grazian}, A., \& {Haardt}, F. 2024, arXiv e-prints, arXiv:2406.18697, \dodoi{10.48550/arXiv.2406.18697}

\bibitem[{{Madau} \& {Haardt}(2015)}]{Madau_2015}
{Madau}, P., \& {Haardt}, F. 2015, \apjl, 813, L8, \dodoi{10.1088/2041-8205/813/1/L8}

\bibitem[{{Marques-Chaves} {et~al.}(2024){Marques-Chaves}, {Schaerer}, {Vanzella}, {Verhamme}, {Dessauges-Zavadsky}, {Chisholm}, {Leclercq}, {Upadhyaya}, {Alvarez-Marquez}, {Colina}, {Garel}, \& {Messa}}]{Marques-Chaves_2024}
{Marques-Chaves}, R., {Schaerer}, D., {Vanzella}, E., {et~al.} 2024, arXiv e-prints, arXiv:2407.18804.
\newblock \doarXiv{2407.18804}

\bibitem[{{Marshak}(1958)}]{Marshak_1958}
{Marshak}, R.~E. 1958, Physics of Fluids, 1, 24, \dodoi{10.1063/1.1724332}

\bibitem[{{Martin} {et~al.}(2024){Martin}, {Peng}, \& {Li}}]{Martin_2024}
{Martin}, C.~L., {Peng}, Z., \& {Li}, Y. 2024, \apj, 966, 190, \dodoi{10.3847/1538-4357/ad34ac}

\bibitem[{{Matzner} \& {Jumper}(2015)}]{Matzner_2015}
{Matzner}, C.~D., \& {Jumper}, P.~H. 2015, \apj, 815, 68, \dodoi{10.1088/0004-637X/815/1/68}

\bibitem[{{McGreer} {et~al.}(2015){McGreer}, {Mesinger}, \& {D'Odorico}}]{Mcgreer_2015}
{McGreer}, I.~D., {Mesinger}, A., \& {D'Odorico}, V. 2015, \mnras, 447, 499, \dodoi{10.1093/mnras/stu2449}

\bibitem[{{McLeod} {et~al.}(2024){McLeod}, {Donnan}, {McLure}, {Dunlop}, {Magee}, {Begley}, {Carnall}, {Cullen}, {Ellis}, {Hamadouche}, \& {Stanton}}]{Mcleod_2023}
{McLeod}, D.~J., {Donnan}, C.~T., {McLure}, R.~J., {et~al.} 2024, \mnras, 527, 5004, \dodoi{10.1093/mnras/stad3471}

\bibitem[{{McQuinn}(2016)}]{Mcquinn_2016}
{McQuinn}, M. 2016, \araa, 54, 313, \dodoi{10.1146/annurev-astro-082214-122355}

\bibitem[{{Menon} {et~al.}(2022{\natexlab{a}}){Menon}, {Federrath}, \& {Krumholz}}]{Menon_2022b}
{Menon}, S.~H., {Federrath}, C., \& {Krumholz}, M.~R. 2022{\natexlab{a}}, \mnras, 517, 1313, \dodoi{10.1093/mnras/stac2702}

\bibitem[{{Menon} {et~al.}(2023){Menon}, {Federrath}, \& {Krumholz}}]{Menon_2023}
---. 2023, \mnras, 521, 5160, \dodoi{10.1093/mnras/stad856}

\bibitem[{{Menon} {et~al.}(2022{\natexlab{b}}){Menon}, {Federrath}, {Krumholz}, {Kuiper}, {Wibking}, \& {Jung}}]{Menon_2022}
{Menon}, S.~H., {Federrath}, C., {Krumholz}, M.~R., {et~al.} 2022{\natexlab{b}}, \mnras, 512, 401, \dodoi{10.1093/mnras/stac485}

\bibitem[{{Menon} {et~al.}(2020){Menon}, {Federrath}, \& {Kuiper}}]{Menon_2020}
{Menon}, S.~H., {Federrath}, C., \& {Kuiper}, R. 2020, \mnras, 493, 4643, \dodoi{10.1093/mnras/staa580}

\bibitem[{{Menon} {et~al.}(2024){Menon}, {Lancaster}, {Burkhart}, {Somerville}, {Dekel}, \& {Krumholz}}]{Menon_2024}
{Menon}, S.~H., {Lancaster}, L., {Burkhart}, B., {et~al.} 2024, \apjl, 967, L28, \dodoi{10.3847/2041-8213/ad462d}

\bibitem[{{Messa} {et~al.}(2024{\natexlab{a}}){Messa}, {Dessauges-Zavadsky}, {Adamo}, {Richard}, \& {Claeyssens}}]{Messa_2024}
{Messa}, M., {Dessauges-Zavadsky}, M., {Adamo}, A., {Richard}, J., \& {Claeyssens}, A. 2024{\natexlab{a}}, \mnras, 529, 2162, \dodoi{10.1093/mnras/stae565}

\bibitem[{{Messa} {et~al.}(2024{\natexlab{b}}){Messa}, {Vanzella}, {Loiacono}, {Bergamini}, {Castellano}, {Sun}, {Willott}, {Windhorst}, {Yan}, {Angora}, {Rosati}, {Adamo}, {Annibali}, {Bolamperti}, {Brada{\v{c}}}, {Bradley}, {Calura}, {Claeyssens}, {Comastri}, {Conselice}, {D'Silva}, {Dickinson}, {Frye}, {Grillo}, {Grogin}, {Gruppioni}, {Koekemoer}, {Meneghetti}, {Me{\v{s}}tri{\'c}}, {Pascale}, {Ravindranath}, {Ricotti}, {Summers}, \& {Zanella}}]{Messa_2024b}
{Messa}, M., {Vanzella}, E., {Loiacono}, F., {et~al.} 2024{\natexlab{b}}, arXiv e-prints, arXiv:2407.20331, \dodoi{10.48550/arXiv.2407.20331}

\bibitem[{{Micheva} {et~al.}(2018){Micheva}, {Oey}, {Keenan}, {Jaskot}, \& {James}}]{Micheva_2018}
{Micheva}, G., {Oey}, M.~S., {Keenan}, R.~P., {Jaskot}, A.~E., \& {James}, B.~L. 2018, \apj, 867, 2, \dodoi{10.3847/1538-4357/aae372}

\bibitem[{{Mitra} {et~al.}(2018){Mitra}, {Choudhury}, \& {Ferrara}}]{Mitra_2018}
{Mitra}, S., {Choudhury}, T.~R., \& {Ferrara}, A. 2018, \mnras, 473, 1416, \dodoi{10.1093/mnras/stx2443}

\bibitem[{{Morishita} {et~al.}(2023){Morishita}, {Stiavelli}, {Chary}, {Trenti}, {Bergamini}, {Chiaberge}, {Leethochawalit}, {Roberts-Borsani}, {Shen}, \& {Treu}}]{Morishita_2023}
{Morishita}, T., {Stiavelli}, M., {Chary}, R.-R., {et~al.} 2023, arXiv e-prints, arXiv:2308.05018, \dodoi{10.48550/arXiv.2308.05018}

\bibitem[{{Mowla} {et~al.}(2024){Mowla}, {Iyer}, {Asada}, {Desprez}, {Tan}, {Martis}, {Sarrouh}, {Strait}, {Abraham}, {Brada{\v{c}}}, {Brammer}, {Muzzin}, {Pacifici}, {Ravindranath}, {Sawicki}, {Willott}, {Estrada-Carpenter}, {Jahan}, {Noirot}, {Matharu}, {Rihtar{\v{s}}i{\v{c}}}, \& {Zabl}}]{Mowla_2024}
{Mowla}, L., {Iyer}, K., {Asada}, Y., {et~al.} 2024, arXiv e-prints, arXiv:2402.08696, \dodoi{10.48550/arXiv.2402.08696}

\bibitem[{{Mu{\~n}oz} {et~al.}(2024){Mu{\~n}oz}, {Mirocha}, {Chisholm}, {Furlanetto}, \& {Mason}}]{Munoz_2024}
{Mu{\~n}oz}, J.~B., {Mirocha}, J., {Chisholm}, J., {Furlanetto}, S.~R., \& {Mason}, C. 2024, arXiv e-prints, arXiv:2404.07250, \dodoi{10.48550/arXiv.2404.07250}

\bibitem[{{Murray} {et~al.}(2011){Murray}, {M{\'e}nard}, \& {Thompson}}]{Murray_2011}
{Murray}, N., {M{\'e}nard}, B., \& {Thompson}, T.~A. 2011, \apj, 735, 66, \dodoi{10.1088/0004-637X/735/1/66}

\bibitem[{{Murray} {et~al.}(2010){Murray}, {Quataert}, \& {Thompson}}]{Murray_2010}
{Murray}, N., {Quataert}, E., \& {Thompson}, T.~A. 2010, \apj, 709, 191, \dodoi{10.1088/0004-637X/709/1/191}

\bibitem[{{Naidu} {et~al.}(2020){Naidu}, {Tacchella}, {Mason}, {Bose}, {Oesch}, \& {Conroy}}]{Naidu_2020}
{Naidu}, R.~P., {Tacchella}, S., {Mason}, C.~A., {et~al.} 2020, \apj, 892, 109, \dodoi{10.3847/1538-4357/ab7cc9}

\bibitem[{{Naidu} {et~al.}(2022){Naidu}, {Matthee}, {Oesch}, {Conroy}, {Sobral}, {Pezzulli}, {Hayes}, {Erb}, {Amor{\'\i}n}, {Gronke}, {Schaerer}, {Tacchella}, {Kerutt}, {Paulino-Afonso}, {Calhau}, {Llerena}, \& {R{\"o}ttgering}}]{Naidu_2022}
{Naidu}, R.~P., {Matthee}, J., {Oesch}, P.~A., {et~al.} 2022, \mnras, 510, 4582, \dodoi{10.1093/mnras/stab3601}

\bibitem[{{Nakajima} {et~al.}(2020){Nakajima}, {Ellis}, {Robertson}, {Tang}, \& {Stark}}]{Nakajima_2020}
{Nakajima}, K., {Ellis}, R.~S., {Robertson}, B.~E., {Tang}, M., \& {Stark}, D.~P. 2020, \apj, 889, 161, \dodoi{10.3847/1538-4357/ab6604}

\bibitem[{{Nakajima} \& {Ouchi}(2014)}]{Nakajima_2014}
{Nakajima}, K., \& {Ouchi}, M. 2014, \mnras, 442, 900, \dodoi{10.1093/mnras/stu902}

\bibitem[{{Nakajima} {et~al.}(2023){Nakajima}, {Ouchi}, {Isobe}, {Harikane}, {Zhang}, {Ono}, {Umeda}, \& {Oguri}}]{Nakajima_2023}
{Nakajima}, K., {Ouchi}, M., {Isobe}, Y., {et~al.} 2023, \apjs, 269, 33, \dodoi{10.3847/1538-4365/acd556}

\bibitem[{{Orr} {et~al.}(2022){Orr}, {Fielding}, {Hayward}, \& {Burkhart}}]{Orr_2022}
{Orr}, M.~E., {Fielding}, D.~B., {Hayward}, C.~C., \& {Burkhart}, B. 2022, \apj, 932, 88, \dodoi{10.3847/1538-4357/ac6c26}

\bibitem[{{Paardekooper} {et~al.}(2013){Paardekooper}, {Khochfar}, \& {Dalla}}]{Paardekooper_2013}
{Paardekooper}, J.~P., {Khochfar}, S., \& {Dalla}, C.~V. 2013, \mnras, 429, L94, \dodoi{10.1093/mnrasl/sls032}

\bibitem[{{Paardekooper} {et~al.}(2015){Paardekooper}, {Khochfar}, \& {Dalla Vecchia}}]{Paardekooper_2015}
{Paardekooper}, J.-P., {Khochfar}, S., \& {Dalla Vecchia}, C. 2015, \mnras, 451, 2544, \dodoi{10.1093/mnras/stv1114}

\bibitem[{{Parsa} {et~al.}(2018){Parsa}, {Dunlop}, \& {McLure}}]{Parsa_2018}
{Parsa}, S., {Dunlop}, J.~S., \& {McLure}, R.~J. 2018, \mnras, 474, 2904, \dodoi{10.1093/mnras/stx2887}

\bibitem[{{Pascale} \& {Dai}(2024)}]{Pascale_2024}
{Pascale}, M., \& {Dai}, L. 2024, arXiv e-prints, arXiv:2404.10755, \dodoi{10.48550/arXiv.2404.10755}

\bibitem[{{Pascale} {et~al.}(2023){Pascale}, {Dai}, {McKee}, \& {Tsang}}]{Pascale_2023}
{Pascale}, M., {Dai}, L., {McKee}, C.~F., \& {Tsang}, B. T.~H. 2023, \apj, 957, 77, \dodoi{10.3847/1538-4357/acf75c}

\bibitem[{{Planck Collaboration} {et~al.}(2020){Planck Collaboration}, {Aghanim}, {Akrami}, {Ashdown}, {Aumont}, {Baccigalupi}, {Ballardini}, {Banday}, {Barreiro}, {Bartolo}, {Basak}, {Battye}, {Benabed}, {Bernard}, {Bersanelli}, {Bielewicz}, {Bock}, {Bond}, {Borrill}, {Bouchet}, {Boulanger}, {Bucher}, {Burigana}, {Butler}, {Calabrese}, {Cardoso}, {Carron}, {Challinor}, {Chiang}, {Chluba}, {Colombo}, {Combet}, {Contreras}, {Crill}, {Cuttaia}, {de Bernardis}, {de Zotti}, {Delabrouille}, {Delouis}, {Di Valentino}, {Diego}, {Dor{\'e}}, {Douspis}, {Ducout}, {Dupac}, {Dusini}, {Efstathiou}, {Elsner}, {En{\ss}lin}, {Eriksen}, {Fantaye}, {Farhang}, {Fergusson}, {Fernandez-Cobos}, {Finelli}, {Forastieri}, {Frailis}, {Fraisse}, {Franceschi}, {Frolov}, {Galeotta}, {Galli}, {Ganga}, {G{\'e}nova-Santos}, {Gerbino}, {Ghosh}, {Gonz{\'a}lez-Nuevo}, {G{\'o}rski}, {Gratton}, {Gruppuso}, {Gudmundsson}, {Hamann}, {Handley}, {Hansen}, {Herranz}, {Hildebrandt}, {Hivon}, {Huang}, {Jaffe}, {Jones}, {Karakci}, {Keih{\"a}nen},
  {Keskitalo}, {Kiiveri}, {Kim}, {Kisner}, {Knox}, {Krachmalnicoff}, {Kunz}, {Kurki-Suonio}, {Lagache}, {Lamarre}, {Lasenby}, {Lattanzi}, {Lawrence}, {Le Jeune}, {Lemos}, {Lesgourgues}, {Levrier}, {Lewis}, {Liguori}, {Lilje}, {Lilley}, {Lindholm}, {L{\'o}pez-Caniego}, {Lubin}, {Ma}, {Mac{\'\i}as-P{\'e}rez}, {Maggio}, {Maino}, {Mandolesi}, {Mangilli}, {Marcos-Caballero}, {Maris}, {Martin}, {Martinelli}, {Mart{\'\i}nez-Gonz{\'a}lez}, {Matarrese}, {Mauri}, {McEwen}, {Meinhold}, {Melchiorri}, {Mennella}, {Migliaccio}, {Millea}, {Mitra}, {Miville-Desch{\^e}nes}, {Molinari}, {Montier}, {Morgante}, {Moss}, {Natoli}, {N{\o}rgaard-Nielsen}, {Pagano}, {Paoletti}, {Partridge}, {Patanchon}, {Peiris}, {Perrotta}, {Pettorino}, {Piacentini}, {Polastri}, {Polenta}, {Puget}, {Rachen}, {Reinecke}, {Remazeilles}, {Renzi}, {Rocha}, {Rosset}, {Roudier}, {Rubi{\~n}o-Mart{\'\i}n}, {Ruiz-Granados}, {Salvati}, {Sandri}, {Savelainen}, {Scott}, {Shellard}, {Sirignano}, {Sirri}, {Spencer}, {Sunyaev}, {Suur-Uski}, {Tauber}, {Tavagnacco},
  {Tenti}, {Toffolatti}, {Tomasi}, {Trombetti}, {Valenziano}, {Valiviita}, {Van Tent}, {Vibert}, {Vielva}, {Villa}, {Vittorio}, {Wandelt}, {Wehus}, {White}, {White}, {Zacchei}, \& {Zonca}}]{Planck_2020}
{Planck Collaboration}, {Aghanim}, N., {Akrami}, Y., {et~al.} 2020, \aap, 641, A6, \dodoi{10.1051/0004-6361/201833910}

\bibitem[{{Pucha} {et~al.}(2022){Pucha}, {Reddy}, {Dey}, {Juneau}, {Lee}, {Prescott}, {Shivaei}, \& {Hong}}]{Pucha_2022}
{Pucha}, R., {Reddy}, N.~A., {Dey}, A., {et~al.} 2022, \aj, 164, 159, \dodoi{10.3847/1538-3881/ac83a9}

\bibitem[{{Qin} {et~al.}(2017){Qin}, {Mutch}, {Poole}, {Liu}, {Angel}, {Duffy}, {Geil}, {Mesinger}, \& {Wyithe}}]{Qin_2017}
{Qin}, Y., {Mutch}, S.~J., {Poole}, G.~B., {et~al.} 2017, \mnras, 472, 2009, \dodoi{10.1093/mnras/stx1909}

\bibitem[{{Ramambason} {et~al.}(2020){Ramambason}, {Schaerer}, {Stasi{\'n}ska}, {Izotov}, {Guseva}, {V{\'\i}lchez}, {Amor{\'\i}n}, \& {Morisset}}]{Ramambason_2020}
{Ramambason}, L., {Schaerer}, D., {Stasi{\'n}ska}, G., {et~al.} 2020, \aap, 644, A21, \dodoi{10.1051/0004-6361/202038634}

\bibitem[{Raskutti {et~al.}(2017)Raskutti, Ostriker, \& Skinner}]{Raskutti_2017}
Raskutti, S., Ostriker, E.~C., \& Skinner, M.~A. 2017, The Astrophysical Journal, 850, 112, \dodoi{10.3847/1538-4357/aa965e}

\bibitem[{{Ricker}(2008)}]{Ricker_2008}
{Ricker}, P.~M. 2008, \apjs, 176, 293, \dodoi{10.1086/526425}

\bibitem[{{Rivera-Thorsen} {et~al.}(2019){Rivera-Thorsen}, {Dahle}, {Chisholm}, {Florian}, {Gronke}, {Rigby}, {Gladders}, {Mahler}, {Sharon}, \& {Bayliss}}]{Rivera_2019}
{Rivera-Thorsen}, T.~E., {Dahle}, H., {Chisholm}, J., {et~al.} 2019, Science, 366, 738, \dodoi{10.1126/science.aaw0978}

\bibitem[{{Robertson} {et~al.}(2023){Robertson}, {Johnson}, {Tacchella}, {Eisenstein}, {Hainline}, {Arribas}, {Baker}, {Bunker}, {Carniani}, {Carreira}, {Cargile}, {Charlot}, {Chevallard}, {Curti}, {Curtis-Lake}, {D'Eugenio}, {Egami}, {Hausen}, {Helton}, {Jakobsen}, {Ji}, {Jones}, {Maiolino}, {Maseda}, {Nelson}, {P{\'e}rez-Gonz{\'a}lez}, {Pusk{\'a}s}, {Rieke}, {Smit}, {Sun}, {{\"U}bler}, {Whitler}, {Willmer}, {Willott}, \& {Witstok}}]{Robertson_2023}
{Robertson}, B., {Johnson}, B.~D., {Tacchella}, S., {et~al.} 2023, arXiv e-prints, arXiv:2312.10033, \dodoi{10.48550/arXiv.2312.10033}

\bibitem[{{Robertson}(2022)}]{Robertson_2022}
{Robertson}, B.~E. 2022, \araa, 60, 121, \dodoi{10.1146/annurev-astro-120221-044656}

\bibitem[{{Robertson} {et~al.}(2015){Robertson}, {Ellis}, {Furlanetto}, \& {Dunlop}}]{Robertson_2015}
{Robertson}, B.~E., {Ellis}, R.~S., {Furlanetto}, S.~R., \& {Dunlop}, J.~S. 2015, \apjl, 802, L19, \dodoi{10.1088/2041-8205/802/2/L19}

\bibitem[{{Robertson} {et~al.}(2013){Robertson}, {Furlanetto}, {Schneider}, {Charlot}, {Ellis}, {Stark}, {McLure}, {Dunlop}, {Koekemoer}, {Schenker}, {Ouchi}, {Ono}, {Curtis-Lake}, {Rogers}, {Bowler}, \& {Cirasuolo}}]{Robertson_2013}
{Robertson}, B.~E., {Furlanetto}, S.~R., {Schneider}, E., {et~al.} 2013, \apj, 768, 71, \dodoi{10.1088/0004-637X/768/1/71}

\bibitem[{{Rosdahl} {et~al.}(2022){Rosdahl}, {Blaizot}, {Katz}, {Kimm}, {Garel}, {Haehnelt}, {Keating}, {Martin-Alvarez}, {Michel-Dansac}, \& {Ocvirk}}]{Rosdahl_2022}
{Rosdahl}, J., {Blaizot}, J., {Katz}, H., {et~al.} 2022, \mnras, 515, 2386, \dodoi{10.1093/mnras/stac1942}

\bibitem[{{Rosen} {et~al.}(2020){Rosen}, {Offner}, {Sadavoy}, {Bhandare}, {V{\'a}zquez-Semadeni}, \& {Ginsburg}}]{Rosen_2020}
{Rosen}, A.~L., {Offner}, S. S.~R., {Sadavoy}, S.~I., {et~al.} 2020, \ssr, 216, 62, \dodoi{10.1007/s11214-020-00688-5}

\bibitem[{{Schaerer} {et~al.}(2016){Schaerer}, {Izotov}, {Verhamme}, {Orlitov{\'a}}, {Thuan}, {Worseck}, \& {Guseva}}]{Schaerer_2016}
{Schaerer}, D., {Izotov}, Y.~I., {Verhamme}, A., {et~al.} 2016, \aap, 591, L8, \dodoi{10.1051/0004-6361/201628943}

\bibitem[{{Semenov} {et~al.}(2003){Semenov}, {Henning}, {Helling}, {Ilgner}, \& {Sedlmayr}}]{Semenov_2003}
{Semenov}, D., {Henning}, T., {Helling}, C., {Ilgner}, M., \& {Sedlmayr}, E. 2003, \aap, 410, 611, \dodoi{10.1051/0004-6361:20031279}

\bibitem[{{Shapley} {et~al.}(2016){Shapley}, {Steidel}, {Strom}, {Bogosavljevi{\'c}}, {Reddy}, {Siana}, {Mostardi}, \& {Rudie}}]{Shapley_2016}
{Shapley}, A.~E., {Steidel}, C.~C., {Strom}, A.~L., {et~al.} 2016, \apjl, 826, L24, \dodoi{10.3847/2041-8205/826/2/L24}

\bibitem[{{Sharma} {et~al.}(2016){Sharma}, {Theuns}, {Frenk}, {Bower}, {Crain}, {Schaller}, \& {Schaye}}]{Sharma_2016}
{Sharma}, M., {Theuns}, T., {Frenk}, C., {et~al.} 2016, \mnras, 458, L94, \dodoi{10.1093/mnrasl/slw021}

\bibitem[{{Simmonds} {et~al.}(2024){Simmonds}, {Tacchella}, {Hainline}, {Johnson}, {McClymont}, {Robertson}, {Saxena}, {Sun}, {Witten}, {Baker}, {Bhatawdekar}, {Boyett}, {Bunker}, {Charlot}, {Curtis-Lake}, {Egami}, {Eisenstein}, {Hausen}, {Maiolino}, {Maseda}, {Scholtz}, {Williams}, {Willott}, \& {Witstok}}]{Simmonds_2024}
{Simmonds}, C., {Tacchella}, S., {Hainline}, K., {et~al.} 2024, \mnras, 527, 6139, \dodoi{10.1093/mnras/stad3605}

\bibitem[{{Simons} {et~al.}(2015){Simons}, {Kassin}, {Weiner}, {Heckman}, {Lee}, {Lotz}, {Peth}, \& {Tchernyshyov}}]{Simons_2015}
{Simons}, R.~C., {Kassin}, S.~A., {Weiner}, B.~J., {et~al.} 2015, \mnras, 452, 986, \dodoi{10.1093/mnras/stv1298}

\bibitem[{{Skinner} \& {Ostriker}(2015)}]{Skinner_2015}
{Skinner}, M.~A., \& {Ostriker}, E.~C. 2015, \apj, 809, 187, \dodoi{10.1088/0004-637X/809/2/187}

\bibitem[{{Socrates} \& {Sironi}(2013)}]{Socrates_2013}
{Socrates}, A., \& {Sironi}, L. 2013, \apjl, 772, L21, \dodoi{10.1088/2041-8205/772/2/L21}

\bibitem[{{Stanway} \& {Eldridge}(2018)}]{Stanway_2018}
{Stanway}, E.~R., \& {Eldridge}, J.~J. 2018, \mnras, 479, 75, \dodoi{10.1093/mnras/sty1353}

\bibitem[{{Tacchella} {et~al.}(2023){Tacchella}, {Johnson}, {Robertson}, {Carniani}, {D'Eugenio}, {Kumari}, {Maiolino}, {Nelson}, {Suess}, {{\"U}bler}, {Williams}, {Adebusola}, {Alberts}, {Arribas}, {Bhatawdekar}, {Bonaventura}, {Bowler}, {Bunker}, {Cameron}, {Curti}, {Egami}, {Eisenstein}, {Frye}, {Hainline}, {Helton}, {Ji}, {Looser}, {Lyu}, {Perna}, {Rawle}, {Rieke}, {Rieke}, {Saxena}, {Sandles}, {Shivaei}, {Simmonds}, {Sun}, {Willmer}, {Willott}, \& {Witstok}}]{Tachella_2023}
{Tacchella}, S., {Johnson}, B.~D., {Robertson}, B.~E., {et~al.} 2023, \mnras, 522, 6236, \dodoi{10.1093/mnras/stad1408}

\bibitem[{{Thompson} {et~al.}(2015){Thompson}, {Fabian}, {Quataert}, \& {Murray}}]{Thompson_2015}
{Thompson}, T.~A., {Fabian}, A.~C., {Quataert}, E., \& {Murray}, N. 2015, \mnras, 449, 147, \dodoi{10.1093/mnras/stv246}

\bibitem[{Thompson \& Krumholz(2016)}]{Thompson_Krumholz_2016}
Thompson, T.~A., \& Krumholz, M.~R. 2016, Monthly Notices of the Royal Astronomical Society, 455, 334, \dodoi{10.1093/mnras/stv2331}

\bibitem[{{Trebitsch} {et~al.}(2017){Trebitsch}, {Blaizot}, {Rosdahl}, {Devriendt}, \& {Slyz}}]{Trebitsch_2017}
{Trebitsch}, M., {Blaizot}, J., {Rosdahl}, J., {Devriendt}, J., \& {Slyz}, A. 2017, \mnras, 470, 224, \dodoi{10.1093/mnras/stx1060}

\bibitem[{Turk {et~al.}(2010)Turk, Smith, Oishi, Skory, Skillman, Abel, \& Norman}]{yt}
Turk, M.~J., Smith, B.~D., Oishi, J.~S., {et~al.} 2010, The Astrophysical Journal Supplement Series, 192, 9, \dodoi{10.1088/0067-0049/192/1/9}

\bibitem[{{Vanzella} {et~al.}(2012){Vanzella}, {Guo}, {Giavalisco}, {Grazian}, {Castellano}, {Cristiani}, {Dickinson}, {Fontana}, {Nonino}, {Giallongo}, {Pentericci}, {Galametz}, {Faber}, {Ferguson}, {Grogin}, {Koekemoer}, {Newman}, \& {Siana}}]{Vanzella_2012}
{Vanzella}, E., {Guo}, Y., {Giavalisco}, M., {et~al.} 2012, \apj, 751, 70, \dodoi{10.1088/0004-637X/751/1/70}

\bibitem[{{Vanzella} {et~al.}(2020){Vanzella}, {Caminha}, {Calura}, {Cupani}, {Meneghetti}, {Castellano}, {Rosati}, {Mercurio}, {Sani}, {Grillo}, {Gilli}, {Mignoli}, {Comastri}, {Nonino}, {Cristiani}, {Giavalisco}, \& {Caputi}}]{Vanzella_2020}
{Vanzella}, E., {Caminha}, G.~B., {Calura}, F., {et~al.} 2020, \mnras, 491, 1093, \dodoi{10.1093/mnras/stz2286}

\bibitem[{{Vanzella} {et~al.}(2022){Vanzella}, {Castellano}, {Bergamini}, {Meneghetti}, {Zanella}, {Calura}, {Caminha}, {Rosati}, {Cupani}, {Me{\v{s}}tri{\'c}}, {Brammer}, {Tozzi}, {Mercurio}, {Grillo}, {Sani}, {Cristiani}, {Nonino}, {Merlin}, \& {Pignataro}}]{Vanzella_2022}
{Vanzella}, E., {Castellano}, M., {Bergamini}, P., {et~al.} 2022, \aap, 659, A2, \dodoi{10.1051/0004-6361/202141590}

\bibitem[{{Vanzella} {et~al.}(2023){Vanzella}, {Claeyssens}, {Welch}, {Adamo}, {Coe}, {Diego}, {Mahler}, {Khullar}, {Kokorev}, {Oguri}, {Ravindranath}, {Furtak}, {Hsiao}, {Abdurro'uf}, {Mandelker}, {Brammer}, {Bradley}, {Brada{\v{c}}}, {Conselice}, {Dayal}, {Nonino}, {Andrade-Santos}, {Windhorst}, {Pirzkal}, {Sharon}, {de Mink}, {Fujimoto}, {Zitrin}, {Eldridge}, \& {Norman}}]{Vanzella_2023}
{Vanzella}, E., {Claeyssens}, A., {Welch}, B., {et~al.} 2023, \apj, 945, 53, \dodoi{10.3847/1538-4357/acb59a}

\bibitem[{Virtanen {et~al.}(2020)Virtanen, Gommers, Oliphant, Haberland, Reddy, Cournapeau, Burovski, Peterson, Weckesser, Bright, {van der Walt}, Brett, Wilson, Millman, Mayorov, Nelson, Jones, Kern, Larson, Carey, Polat, Feng, Moore, {VanderPlas}, Laxalde, Perktold, Cimrman, Henriksen, Quintero, Harris, Archibald, Ribeiro, Pedregosa, {van Mulbregt}, \& {SciPy 1.0 Contributors}}]{scipy}
Virtanen, P., Gommers, R., Oliphant, T.~E., {et~al.} 2020, Nature Methods, 17, 261, \dodoi{10.1038/s41592-019-0686-2}

\bibitem[{{Waagan} {et~al.}(2011){Waagan}, {Federrath}, \& {Klingenberg}}]{Waagan_2011}
{Waagan}, K., {Federrath}, C., \& {Klingenberg}, C. 2011, Journal of Computational Physics, 230, 3331, \dodoi{10.1016/j.jcp.2011.01.026}

\bibitem[{{Wang} {et~al.}(2023){Wang}, {Teplitz}, {Smith}, {Windhorst}, {Rafelski}, {Mehta}, {Alavi}, {Brammer}, {Colbert}, {Grogin}, {Hathi}, {Koekemoer}, {Prichard}, {Scarlata}, {Sunnquist}, {Arrabal Haro}, {Conselice}, {Gawiser}, {Guo}, {Hayes}, {Jansen}, {Ji}, {Lucas}, {O'Connell}, {Robertson}, {Rutkowski}, {Siana}, {Vanzella}, {Ashcraft}, {Bagley}, {Baronchelli}, {Barro}, {Blanche}, {Broussard}, {Carleton}, {Chartab}, {Cheng}, {Codoreanu}, {Cohen}, {Dai}, {Darvish}, {Dave}, {DeGroot}, {De Mello}, {Dickinson}, {Emami}, {Ferguson}, {Ferreira}, {Finkelstein}, {Finkelstein}, {Gardner}, {Gburek}, {Giavalisco}, {Grazian}, {Gronwall}, {Hemmati}, {Howell}, {Iyer}, {Kaviraj}, {Kurczynski}, {Lazar}, {MacKenty}, {Mantha}, {Martin}, {Martin}, {McCabe}, {Mobasher}, {Nedkova}, {Olsen}, {Otteson}, {Ravindranath}, {Redshaw}, {Sattari}, {Soto}, {Yung}, {Zabelle}, \& {the UVCANDELS team}}]{Wang_2023}
{Wang}, X., {Teplitz}, H.~I., {Smith}, B.~M., {et~al.} 2023, arXiv e-prints, arXiv:2308.09064, \dodoi{10.48550/arXiv.2308.09064}

\bibitem[{{Weingartner} \& {Draine}(2001)}]{Weingartner_2001b}
{Weingartner}, J.~C., \& {Draine}, B.~T. 2001, \apjs, 134, 263, \dodoi{10.1086/320852}

\bibitem[{{Wibking} {et~al.}(2018){Wibking}, {Thompson}, \& {Krumholz}}]{Wibking_2018}
{Wibking}, B.~D., {Thompson}, T.~A., \& {Krumholz}, M.~R. 2018, \mnras, 477, 4665, \dodoi{10.1093/mnras/sty907}

\bibitem[{{Wise} {et~al.}(2014){Wise}, {Demchenko}, {Halicek}, {Norman}, {Turk}, {Abel}, \& {Smith}}]{Wise_2014}
{Wise}, J.~H., {Demchenko}, V.~G., {Halicek}, M.~T., {et~al.} 2014, \mnras, 442, 2560, \dodoi{10.1093/mnras/stu979}

\bibitem[{{Xu} {et~al.}(2016){Xu}, {Wise}, {Norman}, {Ahn}, \& {O'Shea}}]{Xu_2016}
{Xu}, H., {Wise}, J.~H., {Norman}, M.~L., {Ahn}, K., \& {O'Shea}, B.~W. 2016, \apj, 833, 84, \dodoi{10.3847/1538-4357/833/1/84}

\bibitem[{{Yeh} {et~al.}(2023){Yeh}, {Smith}, {Kannan}, {Garaldi}, {Vogelsberger}, {Borrow}, {Pakmor}, {Springel}, \& {Hernquist}}]{Yeh_2023}
{Yeh}, J. Y.~C., {Smith}, A., {Kannan}, R., {et~al.} 2023, \mnras, 520, 2757, \dodoi{10.1093/mnras/stad210}

\end{thebibliography}
\bibliographystyle{aasjournal}



\end{document}